\documentclass[prb,aps,psfig,nopacs,amsmath,amssymb,nofootinbib,superscriptaddress]{revtex4}

\pdfoutput=1

\usepackage{graphicx}
\usepackage{hyperref}
\usepackage{doi}
\usepackage{mathtools}
\usepackage{bbm}

\newcommand{\be}[1]{ \begin{eqnarray} \mbox{$\label{#1}$} }

\newcommand{\ee}{\end{eqnarray}}

\newcommand\ie {{\it i.e. }}
\newcommand\eg {{\it e.g. }}

\newcommand\ket [1] {|#1 \rangle }
\newcommand\bra [1] {\langle #1 |}

\newcommand{\av}[1]{\langle #1\rangle}

\newcommand{\orbdis}{\delta\tau}

\def\a{\alpha}
\def\b{\beta}
\def\g{\gamma}

\newcommand{\beq}{\begin{equation}}
\newcommand{\eeq}{\end{equation}}
\newcommand{\eq}[1]{Eq.~\eqref{#1}}

\newcommand{\expn}[1]{\text{exp} \left\{ #1 \right\}}
\newcommand{\norm}[1]{\left\Vert #1 \right\Vert}
\newcommand{\wt}[1]{\widetilde{#1}}
\newcommand{\coll}[1]{\mathbf{#1}}

\newcommand{\cmplx}{\mathbb{C}}
\newcommand{\intg}{\mathbb{Z}}
\newcommand{\proj}{K}

\newcommand{\const}{{\,\delta\tau}}
\newcommand{\lenx}{L}
\newcommand{\onev}{\mathbf{1}}
\newcommand{\coord}{\omega}
\newcommand{\bmu}{{\boldsymbol{\mu}}}

\begin{document}

\title{Matrix product state representation of quasielectron wave functions}

\author{J. Kj\"all}
\affiliation{Department of Physics, Stockholm University, AlbaNova University Center, SE-106 91 Stockholm, Sweden}

\author{E. Ardonne}
\affiliation{Department of Physics, Stockholm University, AlbaNova University Center, SE-106 91 Stockholm, Sweden}

\author{V. Dwivedi}
\affiliation{Institute for Theoretical Physics, University of Cologne, 50937 Cologne, Germany}

\author{M. Hermanns}
\affiliation{Institute for Theoretical Physics, University of Cologne, 50937 Cologne, Germany}
\affiliation{Department of Physics, University of Gothenburg, SE 412 96 Gothenburg, Sweden}

\author{T.H. Hansson}
\affiliation{Department of Physics, Stockholm University, AlbaNova University Center, SE-106 91 Stockholm, Sweden}
\affiliation{Nordita, KTH Royal Institute of Technology and Stockholm University, Roslagstullsbacken 23, SE-106 91 Stockholm, Sweden}

\date{May 3, 2018}

\begin{abstract}
Matrix product state techniques provide a very efficient way to numerically evaluate certain classes of quantum Hall wave functions
that can be written as correlators in two-dimensional conformal field theories.
Important examples are the Laughlin and Moore-Read ground states and their quasihole excitations.
In this paper, we extend the matrix product state techniques to evaluate quasielectron wave functions, a more complex task
because the corresponding conformal field theory operator is not local. 
We use our method to obtain density profiles for states with multiple quasielectrons and quasiholes, and to calculate
the (mutual) statistical phases of the excitations with high precision. 
The wave functions we study are subject to a known difficulty: the position of a quasielectron depends on the presence of other quasiparticles,
even when their separation is large compared to the magnetic length. Quasielectron wave functions constructed using the
composite fermion picture, which are topologically equivalent to the quasielectrons we study, have the same problem.
This flaw is serious in that it gives wrong results for the statistical phases obtained by braiding distant quasiparticles. We analyze this
problem in detail and show that it originates from an incomplete screening of the topological charges, which invalidates the plasma analogy.
We demonstrate that this can be remedied in the case when the separation between the quasiparticles is large, which allows us to obtain
the correct statistical phases.
Finally, we propose that a modification of the Laughlin state, that allows for local quasielectron operators, should have good topological properties
for arbitrary configurations of excitations.

\end{abstract}

\maketitle

\section{Introduction}

The study of the fractional quantum Hall effect~\cite{fqhe} has been of great importance for the understanding of  
many-body states in the extreme quantum regime. It also provides  paradigmatic examples of  topologically
ordered states of matter~\cite{wen}, and the so far only experimentally observed candidate~\cite{willet87,pan99}
for a state with bulk non-abelian excitations.  
Like in all condensed matter systems, the theoretical description of the fractional quantum Hall effect is based on 
constructing various kinds of effective field theories. 
However, it is  also  very special, in that a lot of understanding has been 
gained by the study of various explicit many-body wave functions, the most famous one being the Laughlin wave function~\cite{laugh83}.

In certain cases, as for instance the Laughlin states at filling fractions $\nu = 1/q$ or the non-abelian Moore-Read state~\cite{MR} at
$\nu = 5/2$, the `representative' many-body wave functions are eigenstates of known Hamiltonians with (admittedly singular) short
range interactions. 
The belief is that these idealized Hamiltonians can be adiabatically connected to realistic ones without changing the topological properties of the states.
There are, however, many examples of proposed representative wave functions which are not eigenstates of  any known Hamiltonian. 
The most well-known of these are the composite fermion states~\cite{jaincf,jainbook}, which describe the most prominent members of the hierarchy of abelian states in the lowest Landau level (LLL)  at rational filling fractions $\nu = p/q$, with $q$ odd~\cite{pandata}.
All these wave functions  fit into a theoretical framework based on a deep connection between the topological quantum field theories that provide the long distance description of fractional quantum Hall states and  certain $1+1$ dimensional conformal field theories (CFTs)~\cite{witten}. 
The original works along these lines were by Moore and Read~\cite{MR} and by Wen~\cite{wen91,blokwen}, and it was later generalized to both
abelian~\cite{haldane83,halperinhierarchy} and non-abelian hierarchy states~\cite{bondsling,levhalp,hermanns} (for a review, see Ref.~\onlinecite{hhsv}).

Since the hierarchy states constructed using composite fermions, or more generally by CFT based methods, do not come from a Hamiltonian, adiabatic arguments are not applicable, so other methods must be used to argue that they
are relevant to physics. 
One approach is to show that they, in some approximation, follow from a sound effective field theory, but this has  been achieved only in certain simple cases~\cite{lopezfradkin}. 
In most cases, the physical relevance of the hierarchy states has only been justified by numerical studies. 
Calculating overlaps with states obtained by direct numerical diagonalization of small systems has provided sanity checks for many of the representative wave functions, while the numerical calculation of Berry phases~\cite{arovas,kjonsbergM1999,jeon03,jeon04,MRqh,mpsnashort,wu15} and entanglement entropies~\cite{kitaev-preskill,lw06,Haque07} and spectra~\cite{li08} has allowed for deeper insights into the topological properties of these states. 

A limiting factor for extending these kind of  studies is that in most cases it is computationally very demanding
to evaluate the wave functions, even though they are explicitly known in real space. 
There are several sources of difficulties, starting with the expansion of the wave functions in Slater determinants. 
The need to perform many derivatives and/or anti-symmetrize over a large number of variables is also numerically very costly.
A lot of effort has been put into developing more efficient numerical methods, one of the latest being the adaption of the matrix product state (MPS) technique~\cite{ostlund,perez-garcia} to quantum Hall problems~\cite{dubail12,zm} (see Ref.~\onlinecite{ciracsierra} for a related spin chain model).

The MPS method has its origin in the density matrix renormalization group (DMRG), which has been 
very successful for simulating one-dimensional systems, in particular spin chains~\cite{white,schollwoeck}. 
To explain the basic idea, we consider a lattice model with $N$  sites and attach a Hilbert space $\{\ket{p_l}\}$ with dimension $D_l$ to the site $l$. 
A general state can be written as
\be{MPSstate}
\ket \Psi= \sum_{p_1,p_2\dots p_N} C_{p_1,p_2\dots p_N} \bigotimes_{l=1}^N \ket{p_l},
\ee 
and an MPS representation amounts to expressing the coefficients $C$ as a traces of matrices,
\be{coeff}
C_{p_1,p_2\dots p_N} = {\mathrm Tr} \left[ \mathbf B^{[p_1]} \mathbf B^{[p_2]}\dots \mathbf B^{[p_N]}\right] = B^{[p_1]}_{\a\b} B^{[p_2]}_{\b\g}\dots B^{[p_N]}_{\xi\a} \ ,
\ee
where the Greek variables refer to the {\em auxiliary spaces} which have dimensions $\chi_l$ at bond $l$ (between site $l$ and site $l+1$). 
The physical meaning of this space  can be understood as follows. Imagine dividing the system in two parts at the bond $l$ and note that the only way the two parts depend on each other is via the matrix $\mathbf{B}^{[p_l]}$. If we now concentrate on, say, the left part, the presence of the right part is encoded in the entanglement data, such as the entanglement entropy and the entanglement spectrum, of the divided system. 
This information must be encoded in the matrices $\mathbf{B}^{[p_l]}$, and one would thus think that starting from, say, the leftmost
site, it would require more and more information to encode the entanglement between the parts, as the left part grows bigger.
Consequently, one would expect the dimensions $\chi_l$ of the auxiliary space to grow very quickly. 
The reason for the success of the MPS method is that this does not happen for gapped states of a system described by a local Hamiltonian~\cite{vedral}.
Instead, the entanglement grows only up to a limit, meaning that the state can be accurately described by a \emph{finite-dimensional} matrix. 
The matrices are not uniquely defined, but there is a special representation where the eigenvalues are precisely the entanglement energies, thus providing a precise connection between the original renormalization group ideas of White~\cite{white} and the quantum information viewpoint just described. 
For a translation-invariant state, the matrices $\mathbf{B}^{[p_l]}=\mathbf{B}^{[p]}$ are independent of $l$, and finding a good approximation for the ground state amounts to finding the optimal matrix. 
For a pedagogical review of tensor network states, of which MPS state are a special case, see \eg Ref.~\onlinecite{orus}.

\begin{figure}[th]
		\includegraphics[width=.75\textwidth]{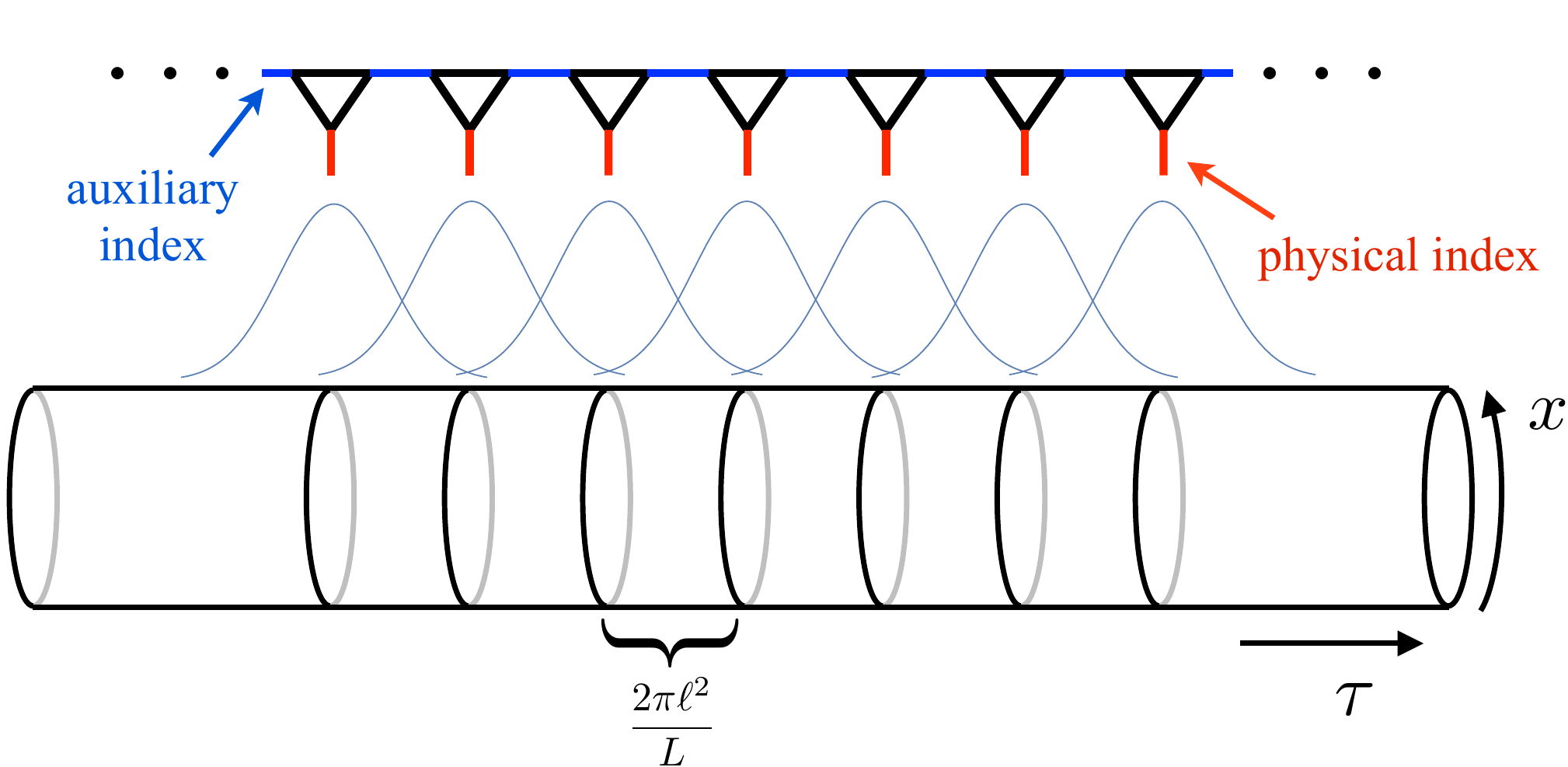}
		\caption{The cylinder geometry, indicating the LLL single-particle orbitals and the corresponding
			MPS structure of the wave function. The physical index is the single-particle occupation number, while the auxiliary index corresponds to the CFT Hilbert space.}
		\label{fig:cylinder}
	\end{figure}

It is far from obvious that the MPS technique can be useful for two-dimensional systems, and in particular for  quantum Hall
liquids. 
As was noted in Ref.~\onlinecite{dubail12}, this is nevertheless the case, because these liquids only occupy a few Landau levels. 
Therefore, it is often sufficient to consider the dynamics in only one of them, the lower ones being completely filled and thus inert.
For this reason, we shall restrict ourselves to states in the LLL. 
For the purpose of calculations, we use periodic boundary conditions in one direction, corresponding to studying the quantum Hall liquid on a cylinder.
Choosing the Landau gauge, the LLL problem is  mapped onto a lattice model as illustrated in Fig.~\ref{fig:cylinder}.
Zaletel and Mong showed how the Laughlin and Moore-Read wave functions can be expressed as MPSs\cite{zm}, which in turn allows for  very efficient computations of topological and entanglement characteristics. 
This method has been applied to study the properties of model wave
functions\cite{zm,estienne-short,estienne-long,mpsnashort}, and adapted to study
Coulomb systems\cite{zmp13,zpm15,mzpp17}.

The starting point is the Moore-Read representation of the QH state as a correlation function in the appropriate CFT,
\be{corrfcn}
\Psi (z_1,z_2,\dots,z_N) = \av{ {\mathcal O}_{\rm bg} V(z_1)V(z_2) \cdots V(z_N)  } \ ,
\ee
where $V(z_i)$ is a primary field of a CFT as a function of the (complex) electron coordinate $z_i$, and ${\mathcal O}_{\rm bg}$ is 
a neutralizing background charge operator that depends on the magnetic length $\ell$. 
In the Hamiltonian picture, the average $\langle \ldots \rangle$ denotes the ground state expectation value of a time (or radial) ordered product of operators. 
The key step is  to insert resolutions of identity, $\mathbbm 1 = \ket{\alpha_i}{\bra{\alpha_i}}$, between the operators by
which the product of operators is turned into a matrix product. The states $\ket{\alpha_i}$ span the Hilbert space
of the CFT, which thus constitutes the auxiliary space\cite{dubail12,zm}.

By inserting `quasihole' operators, $H(\eta)$, into the correlator in Eq.~\eqref{corrfcn}, one obtains quasihole states.
One can also obtain an MPS representations  for these states by introducing extra matrices describing the quasiholes. 
One would think that the generalization to quasielectron states would be straightforward, but this 
has turned out not to be the case. 
The naive guess --- inserting an inverse quasihole $H^{-1}$ in the correlator \eqref{corrfcn}--- does not produce a valid electronic wave function.\footnote{It is interesting to note that the inverse quasihole does provide a valid description of quasielectrons for lattice Laughlin states, as shown in Ref.~\cite{nielsen}.}
It also fails to give excitations with the correct topological properties when implemented as an MPS. 
The underlying reason for this is that while the electron and quasihole operators $V$ and $H$ are both local, the operator describing a quasielectron is \emph{quasi-local}~\cite{hhrv,hhv}.

The starting point of our MPS description of quasielectron excitations of Laughlin states is this quasi-local operator and we review its construction in Section~\ref{qeo}. 
Besides being intrinsically interesting, our results  also point towards a way to construct MPS representations both of hierarchy states, and of quasielectron excitations in the Moore-Read state.
In Section~\ref{sec:mps-laughlin}, we show in detail how to extend the MPS techniques to Laughlin states containing both quasielectrons and quasiholes.
As in the original work by Zaletel and Mong~\cite{zm}, we use a cylinder geometry.
As explained in earlier work (for a detailed review, see Ref.~\onlinecite{hhsv}), to construct the non-local quasielectron operator one must extend the CFT to contain an additional scalar field $\tilde\varphi$, which results in a more complicated matrix structure.
We derive the form of the matrices necessary to obtain the quasielectrons in Section~\ref{sec:mps-qes} and provide some details on the numerical implementation and challenges in Section~\ref{sec:numimp}.
We have checked the validity of our construction by direct comparison with explicit expressions for quasielectron wave functions, which can be obtained  for systems with a small number of particles. 
Going to large systems,  we perform high precision calculations of density profiles and statistical phases for various configurations of quasielectrons and quasiholes. 
These results are presented in Section~\ref{sec:results}.
In Section~\ref{sec:screening}, we discuss a known flaw of the CFT quasielectron wave functions, originally discovered
in the composite fermion picture, which wave functions are topologically equivalent to those we consider in this paper.
We stress that this flaw is not a mere technical glitch but indicates that the wave functions do not encode the topological content of quasielectrons states in a faithful way.
We find that the origin of the difficulties is that the plasma analogy can not be applied.
In the CFT language, this is due to an incomplete screening of a charge associated with the quasielectron operator.

Using this knowledge, we show how one can modify the quasielectron and quasihole operators to obtain full screening when the separation between the excitations is large, and verify that in those cases the statistical phases come out as expected. 
We also suggest an \textit{ad-hoc} modification that numerically appears to have the desired screening even when the separation between the excitations is small. 
We finally discuss an alternative version of the Laughlin wave functions where the quasiparticles are created by local operators and which might have
good screening properties for arbitrary quasiparticle configurations.
We close the paper with a short summary and outlook on future directions in Section~\ref{sec:summary}. 
Some technical background as well as more detailed arguments are given in the Appendices. Appendix~\ref{app:chiralboson} deals with the chiral boson CFT.
In Appendices~\ref{app:matrices-polynomial} and \ref{app:mps-pol-ang-qe}, we provide a detailed derivation of the MPS matrices for the `polynomial part' of the Laughlin state and the quasielectron, while Appendix~\ref{app:matrices-cylinder} deals with the quasielectrons on the cylinder. 
In Appendix~\ref{app:sphereqe}, we discuss the quasielectron wave functions
on the sphere. 
Finally, in Appendix~\ref{app:TTlimit} we give a detailed derivation of the thin-cylinder limit of the quasielectron wave functions in the presence of other quasiholes in the system.

\emph{Notation}: We set $\hbar = c =1$, so the magnetic length is $\ell = 1/\sqrt{eB}$. 
Operators have a hat only where it might otherwise lead to confusion. 
For instance, we use $\hat P$ for the quasielectron operator to distinguish
it from the quantum number $P$, but denote the electron and quasihole operators with $V$ and $H$ respectively.
We use the word `quasiparticle' when the pertinent statement applies to quasielectrons and quasiholes alike.


\section{quasielectron wave functions from CFT} \label{qeo}

In this section, we review how the wave functions for states with quasielectrons can be obtained using CFT techniques.  
Details on  the CFT associated with the compact chiral boson field $\varphi(z)$ are provided in appendix~\ref{app:chiralboson}. 

We start by  recalling  that  the (unnormalized) Laughlin wave function for $N_e$ electrons and $N_{qh}$ quasiholes on a plane can be written as a CFT correlator,
\be{laughlinh}
\Psi_{L,qh}(z_i;\eta_\alpha) &=&
\prod_{\alpha<\beta}^{N_{qh}} (\eta_\alpha-\eta_\beta)^\frac{1}{q}
\prod_{\alpha,i}(\eta_\alpha-z_i) \prod_{i<j}^{N_e}(z_i-z_j)^q  e^{-\frac{1}{4\ell^2} \sum_j |z_j|^2}e^{-\frac{1}{4 q \ell^2} \sum_\alpha |\eta_\alpha|^2}  \\
 &=& \av{{\mathcal O}_{\rm bg}   \prod_{i=1}^{N_e}  V(z_{i}) \prod_{\alpha = 1}^{N_{qh}} H(\eta_\alpha) } \nonumber \, ,
\ee
where the operators $V(z) =\, :e^{i \sqrt{q} \varphi (z)}:$ and $H(\eta) =\, :e^{(i/\sqrt{q})\varphi(\eta)}:$ create an electron at position $z=x+iy$, and a quasihole at position $\eta = x_\eta + i y_\eta$, respectively. 
The background charge operator ${\mathcal O}_{\rm bg}$ ensures that the correlator is charge neutral, so that it does not vanish.
When constructing MPS expressions, we shall use two alternatives for the neutralizing background.
To reproduce the  polynomial part of the wave function \eqref{laughlinh}, we take 
\begin{align}
\label{eq:bg0}
{\mathcal O}_{\rm bg} = e^{-i (q N_e + N_{qh}) \varphi_0/ \sqrt{q}} \ .
\end{align}
Note that  $e^{i\beta\varphi_0}$, where $\varphi_{0}$ is part of the zero mode of $\varphi$, simply creates a charge $\beta\sqrt q$, as explained in appendix \ref{app:chiralboson} [see  Eq.~\eqref{eq:appCharge}]. 
In the absence of quasiholes, the polynomial can be expressed as 
  \begin{equation}
 \label{pol}
 \Psi_{L,{\rm Pol}} (z_i) = \prod_{i<j}^{N_e} (z_i - z_j)^q = \av{ e^{-i N_e \sqrt{q} \varphi_0} V(z_{N_e}) \cdots V(z_1) } \ .
 \end{equation}
Inserting instead a uniform background charge
\begin{align}\label{eq:bg}
\mathcal O_{\rm bg} = :e^{-\frac{i\sqrt{q}}{2\pi q \ell^2} \int d^2z \, \varphi (z)}: ,
\end{align}
as proposed by Moore and Read~\cite{MR}, gives an extra factor $e^{- {|z|^2}/ {4\ell^2}}$ for each electron, up to a gauge transformation. 
Thus,  Eq.~\eqref{laughlinh}  reproduces the Laughlin wave function (in the presence of quasiholes) in a radial gauge.
A corresponding calculation on the cylinder yields the wave function in Landau gauge,
as shown in Ref.~\onlinecite{wu15}, up to a gauge factor%
\footnote{We note the difference in the labeling of the coordinates here and in Ref.~\onlinecite{wu15}.}
\be{eq:gaugefactors}
\Psi_{L,{\rm Landau}} (\tau_i,x_i) =
e^{-i \sum_j \tau_j x_j/\ell^2} \av{ \mathcal O_{\rm bg} V(\tau_1,x_1) \cdots V(\tau_{N_e},x_{N_e})    } \ .
\ee
We use the convention that the $x$-coordinate denotes the position around the circumference of the cylinder and  $\tau$ the position along the cylinder, in order to emphasize the interpretation of the latter direction as  imaginary time (see Fig.~\ref{fig:cylinder}).

At first sight, it looks simple to generalize Eq.~\eqref{laughlinh} to also include quasielectrons.
Since quasiholes are obtained by inserting  $H(\eta)$, one would think
that inserting $H^{-1}(\xi)$ would give a quasielectron at position $\xi = x_\xi +i y_\xi$. 
This is correct from a topological point of view, since this operator has the charge and 
statistics of a quasielectron. 
However, it does not give an acceptable LLL wave function, as the correlators will have poles in the electron coordinates.

In Refs.~\onlinecite{hhrv} and~\onlinecite{hhv}, this problem was overcome as follows.
Instead of inserting an operator that creates the quasielectron excitation at position
$\xi$, one modifies the electrons nearby, by shrinking their correlation
hole. 
This `fusion', which technically amounts to a normal ordering prescription, 
effectively adds the charge of a quasielectron near the position $\xi$.
To properly localize the charge at $\xi$, one weighs the contributions from the
different electrons near $\xi$ with an exponentially decaying factor. 
This procedure is not arbitrary, but is uniquely defined by requiring that the resulting wave function 
resides in the LLL; it in fact amounts to a projection on the LLL.

The operator that creates the `modified' electron {consists} of the usual electron operator to which one `fuses' an `inverse quasihole'.  
As explained in detail in Refs.~\onlinecite{hhrv,hhv} it is not possible to directly fuse $H^{-1}$ with $V$, since the resulting modified electron operator $\hat{P}(z)$ would be anyonic and not give acceptable fermionic wave functions for the electrons. 
The solution is to note, as was first done by Halperin~\cite{halperin84},
that there is a freedom in assigning statistics to the quasihole operators. 
Briefly, the statistics of the operator will determine the `monodromies' of the wave function, but 
the statistics of the quasiparticles, or the `holonomies', will also get a contribution from the Berry phase associated to exchange or braiding. 
The change in the monodromy is compensated by a change in Berry phase, leaving the statistics of the
quasiparticles unchanged.

We choose a fermionic representation of the quasihole operator (for reasons discussed in Ref.~\onlinecite{hhv}), which comes at the expense of introducing an independent scalar field $\tilde\varphi(z)$, with compactification radius $R^2 = q(q-1)$. 
The resulting expression for the quasihole operator is
\be{modhole}
H (\eta) &=& :e^{i/\sqrt{q}\varphi(\eta)}: \, :e^{i(q-1)/\sqrt{q(q-1)}\tilde{\varphi}(\eta)}: \, ,
\ee
which has scaling dimension $h = 1/2$, as appropriate for a fermion. 
The resulting modified quasielectron operator becomes
\be{model}
\hat{P}(z) &=& \partial_z \tilde{V} (z) = \partial_z :e^{i (q-1)/\sqrt{q} \varphi(z)}: \, :e^{-i(q-1)/\sqrt{q(q-1)}\tilde{\varphi}(z)}: \ ,
\ee
where  $\tilde V(z)$ is a primary field with  integer scaling dimension $h = (q-1)/2$ corresponding to a boson, as must be since an electron was fused with a fermionic quasihole. 
Consequently, we cannot just insert the `modified' electron operators to get the quasielectron wave functions, but we have to anti-symmetrize both between the `modified' and the `original' electrons and among the `modified' electrons themselves. 
Recall that the correlator in Eq.~\eqref{laughlinh} directly gives an anti-symmetric electronic wave function since the operators $V(z)$ are fermionic.
 
Replacing one of the operators $V(z_i)$ in Eq.~\eqref{laughlinh} with $\hat{P}(z_i)$ creates a quasielectron at the origin, and 
by multiplying with a factor $z_i^k$ we can put the quasielectron in a state with angular momentum $k$. Explicitly we 
have, 
\be{oneqp}
\Psi_{\rm qe}^{(k)} (z_i) & =&  {\mathcal{A} } \left[ z_1^{k}    \av{ {\mathcal O}_{\rm bg}   
 \hat{P}(z_1) V(z_2)\cdots V(z_{N_e}) }  e^{-\frac{1}{4 q \ell^2} |z_1|^2 } \right] \\ \nonumber
& =& \sum_{i} (-1)^i z_i^k \prod_{i\neq j_1 < j_2 \neq i} (z_{j_1}-z_{j_2})^{q}
\partial_{z_i} \prod_{j_3\neq i} (z_{j_3} - z_i)^{q-1} e^{-\frac{1}{4\ell^2} \sum_j |z_j|^2} \ ,
\ee
where the operator  ${\mathcal O}_{\rm bg} $ must be chosen as to neutralize the correlator with 
respect to both $\varphi$ and $\tilde\varphi$. Here the exponential in the first line is introduced by 
hand, but has a natural interpretation, as explained below in the case of a localized quasielectron. 
$\mathcal A$ denotes anti-symmetrization, which is written out explicitly as a sum in the second line. 
As stressed in Ref.~\onlinecite{hhv}, this wave function is identical to the one obtained using composite fermion techniques~\cite{jainbook}. 

To describe a localized quasielectron at $\xi$,  we multiply the correlator with the kernel
\be{proj1}
K (\xi,z_1) &=& \frac 1{2\pi q\ell^2} e^{-\frac{1}{4 q \ell^2} |z_1- \xi|^2} e^{\frac{1}{4 q \ell^2} (\bar{\xi} z_1 - \xi \bar{z_1})} =\frac 1{2\pi q\ell^2} e^{-\frac{1}{4 q \ell^2} (|z_1|^2 + |\xi|^2 - 2\bar\xi z_1) },  \nonumber
\ee
instead of multiplying it with $z_1^k$. The first expression exhibits the exponential localization around $\xi$ (note that the second factor is only a phase), while the  second expression highlights the analytic structure.
Note that the Gaussian $e^{-\frac{1}{4q\ell^2}|z_1|^2}$, introduced by hand in Eq.~\eqref{oneqp}, follows naturally because of the localization. 
Furthermore, the coefficient $1/(4q\ell^2)$ is necessary to obtain the correct
Gaussian factor associated with a charge $1/q$ particle at position $\xi$ in a magnetic field $eB = \ell^{-2}$.
Also note that $K(\xi,z_1)=e^{-\frac{1}{4 q \ell^2} |z_1|^2 } \delta_h (\xi,z_1)$, where $\delta_h$ is the holomorphic delta function, which is the self-reproducing kernel for LLL wave functions. 

An alternative expression for the localizing kernel is obtained by writing the last exponential as a Taylor series in the angular momentum $k$, i.e. 
\be{proj2}
K (\xi,z_1)=\frac 1{2\pi q\ell^2}   e^{-\frac{1}{4 q \ell^2} (|z_1|^2+|\xi|^2)}
\sum_k \frac{(\bar{\xi} z_1)^k}{(2 q \ell^2)^k k!} =   \sum_{k=0}^{\infty} \bar{\phi}_{k} (\xi) \phi_{k} (z_1) 
\ee
where the second identity follows from the explicit expressions for the normalized single-particle LLL wave functions  $\phi_{k} (z_1)$ in radial gauge (with the modification $\ell^2\rightarrow q \ell^2$).
The second expression in Eq.~\eqref{proj2} shows that the localizing kernel $K (\xi,z_1) $ is nothing but the 
projector on the LLL, while the first expression gives the localized quasielectron as a coherent sum of the angular momentum states in Eq.~\eqref{oneqp}.
This type of explicit form will be used later when we construct the MPS representation for localized quasielectrons on the cylinder. 

Thus, the wave function for a quasielectron, expected to be  localized at $\xi$, is given by
\begin{equation}
\Psi_{L,qe}(z_i; \xi) = 
{\mathcal{A} } \left[
K(\xi,z_1)  \av{ {\mathcal O}_{\rm bg}  \hat{P}(z_1) V(z_2)\cdots V(z_{N_e}) }  \right] \ .
\end{equation}
The generalization to a system with several quasielectrons and quasiholes is straightforward. 
For each quasielectron, there is one (and only one) modified electron operator, and one should anti-symmetrize the result over the coordinates $z_i$. 
In terms of a CFT correlator, this results in the following expression for the wave function with multiple localized
quasiholes and quasielectrons:
\be{full-corr}
\Psi_{L,qp}(z_i; \eta_\alpha; \xi_a) = 
{\mathcal{A} } \left[
\av{ {\mathcal O}_{\rm bg}
\prod_{a=1}^{N_{qe}}K(\xi_a,z_a)
\hat{P}(z_a)
\prod_{i=N_{qe}+1}^{N_e} V(z_i) \prod_{\alpha=1}^{N_{qh}}H(\eta_\alpha)}  \right] \ .
\ee
Note that, since the operators $\hat{P}(z_a)$ are bosonic, the only terms in the sums over the $k_a$'s in the localizing kernels that contribute to the wave function are the ones with all $k_a$ distinct. 
The main goal of this paper is to determine an MPS representation from a general correlator like Eq.~\eqref{full-corr}.


\section{MPS representation for the Laughlin wave functions} 
\label{sec:mps-laughlin}

In their original paper, Zaletel and Mong~\cite{zm} used an elegant field theoretic formulation to find 
an MPS description of the Laughlin wave function in a coherent state representation. In this section we follow an alternative approach~\cite{estienne-long,estienne-short,wu15}. We directly manipulate the expression in Eq.~\eqref{corrfcn} into an MPS form for the Laughlin wave function on the cylinder.

The Laughlin wave function includes (gauge dependent) Gaussian factors characteristic of the Landau problem. The magnetic length, which is set by the size of the Landau orbits and breaks the conformal invariance, is introduced by the spread-out background charge in Eq.~\eqref{eq:bg}.
Having an MPS description on the cylinder, it is a simple matter to find the MPS description for the polynomial part of the
wave function, by taking the large circumference limit. 
This limit is useful, because it allows for an explicit check of the, in our case sometimes involved, expressions for the matrices. 
In this section, we put the emphasis on the conceptual  structure and refer to original papers and Appendices for technical details. 
In particular, we present a direct derivation of the MPS for the polynomial part of the wave functions in Appendices~\ref{app:matrices-polynomial} and \ref{app:mps-pol-ang-qe}.

As mentioned in the introduction, the basic insight that leads to an MPS expression for Eq.~\eqref{laughlinh} is that the auxiliary space, in which the matrices act, \emph{is} the Hilbert space of the CFT. 
This suggests that we should use a Hamiltonian formalism and view the correlator in Eq.~\eqref{laughlinh} as a vacuum expectation value of a time ordered product. 
On the plane, the natural ordering is in the radial direction $r=|z|$, but to get a convenient Hamiltonian formalism it is better to use a cylinder geometry. 
The translation between the two is via the conformal transformation
\be{conftrans}
z \rightarrow \omega = e^{-i \frac {2\pi} L z},
\ee
where $L$ is the circumference of the cylinder (see Fig.~\ref{fig:cylinder}). 
The knowledgeable reader might observe that the operators in Eq.~\eqref{laughlinh}, with conformal dimension $h$, will pick up factors $\omega^h$ under the transformation Eq.~\eqref{conftrans}, but these can be ignored
in the quantum Hall context, since they amount to an uninteresting overall shift of the coordinate system.
The quantization on a cylinder is a standard CFT procedure, but for reference, and
to set the notation, we summarize some important formulas in Appendix A. 

As first shown by Zaletel and Mong~\cite{zm}, it is possible to construct an MPS representation for model
wave functions, such as the Laughlin and Moore-Read states, that directly incorporates the Gaussian factors appropriate for the Landau gauge in the cylinder geometry. 
On the cylinder, the single-particle wave
functions are
\be{cylsp}
\phi_l (\tau,x) =
\frac{1}{\mathcal{N}}\,
e^{ -\frac{ i }{\ell^2}\tau_l x } e^{-\frac{1}{2\ell^2}  (\tau - \tau_l)^2}  =   \frac{1}{\mathcal{N}} \, \omega^l e^{ -\frac{1}{2\ell^2}(\tau^2 + \tau_l^2)} \ ,
\ee
where ${\mathcal{N}} = \sqrt{L \ell \sqrt{\pi}}$ is an $l$-independent normalization constant, and $\tau_l = l \delta\tau$ with $\delta\tau = \frac{2\pi \ell^2}{L}$ the distance in $\tau$ between the centers of two nearest single-particle wave functions.
To derive the MPS description, we follow Refs.~\onlinecite{estienne-long} and~\onlinecite{wu15}, and start with the formal expansion of 
$\Psi_{L,{\rm Landau}}$ in terms of Slater determinants
\begin{equation}
\Psi_{L,{\rm Landau}} = \sum_{\lambda} c_{\lambda} {\rm sl}_\lambda \ ,
\end{equation}
where the partitions $\lambda = (l_{N_e},\ldots,l_2,l_1)$ encode the set of occupied single-particle orbitals for a given Slater determinant.
Thus, the $l_i$ are all distinct and ordered as
$0\leq l_1 < l_2 < \cdots < l_{N_e} \leq N_\phi = q (N_e -1)$, where $N_\phi$ is the highest power of any of the $\omega_{i}$ in Eq.~\eqref{eq:gaugefactors}, or equivalently, the highest power of any of the $z_i$ in Eq.~\eqref{pol}. 

The idea now is to obtain an MPS description of the (Landau gauge) Slater coefficients $c_\lambda$.
This MPS expression can then be used to efficiently calculate physical observables, without having to
compute all the Slater coefficients explicitly.
Following the crucial observation due to Zaletel and Mong, 
one sees that  Eq.~\eqref{cylsp} implies that the single-particle orbitals
$\phi_l(\tau,x)$ simplify if evaluated at the center of the orbital in the $\tau$ direction,
$\phi_l (\tau_l,x) = \frac{1}{\mathcal{N}}e^{-i \tau_l x/\ell^2}$, and we can write the Slater coefficients $c_\lambda$ as
\begin{equation}
\label{eq:slaters}
c_{\lambda} =\left( \prod_{j=1}^{N_e} \int_{-\frac{L}{2}}^{\frac{L}{2}} \frac{dx_j}{L}
e^{i x_j \tau_{l_j}/\ell^2}\right) \Psi_{L,{\rm Landau}} (\tau_j = \tau_{l_j},x_j) \ .
\end{equation}
The phase factors in this expression cancel against the phase factors
in the relation between $\Psi_{L,{\rm Landau}}$ and the CFT correlator, Eq.~\eqref{eq:gaugefactors}, to give the final formula,
\begin{equation}
\label{eq:c-lambda-corr}
c_{\lambda} =\left( \prod_{j=1}^{N_e} \int_{-\frac{L}{2}}^{\frac{L}{2}} \frac{dx_j}{L}\right)
\av{\mathcal  O_{\rm bg} V(\tau_1=\tau_{l_1},x_1) \cdots V(\tau_{N_e} = \tau_{l_{N_e}},x_{N_e})} \ .
\end{equation}
The cancellation of the orbital dependent gauge factors is important:  it implies that the matrices in the MPS  will be orbital-independent, which is one of the reasons for the success of the MPS formalism. 
We should already note, however, that we are forced to deal with orbital dependent matrices when we construct wave functions for systems containing quasielectrons.

To derive the matrix elements, we assume that the $N_e$ electron operators in the correlator in Eq.~\eqref{eq:c-lambda-corr} are ordered in $\tau$, with 
the free `time' evolution given by
$U(\tau'-\tau) = e^{-(\tau'-\tau)H} = e^{-\frac{2\pi}{L}(\tau'-\tau) L_0 }$. 
The Hamiltonian of the 
the CFT is 
\begin{align}
H &= \frac{2\pi}{L} L_0, \qquad L_0 = \frac{1}{2}\pi_0^2 + \sum_{j>0} a_{-j}a_{j},
\end{align}
where $\pi_0$ is part of the zero mode of $\varphi(w)$, and $a_{-j}$ with $j>0$ are the creation
operators corresponding to the non-zero modes. 
We refer to  Appendix \ref{app:chiralboson} for more details, but mention the commutation relation $[\varphi_0,\pi_0] = i$,
while $\varphi_0$ commutes with the other modes $a_j$.
We can write the correlator in Eq.~\eqref{eq:c-lambda-corr} as
\begin{multline}
\av{ \mathcal  O_{\rm bg} V(\tau_1=\tau_{l_1},x_1) \cdots V(\tau_{N_e} = \tau_{l_{N_e}},x_{N_e})}  \\
=
\bra{q-1}
U'(\tau_{N_{\phi}+1}-\tau_{l_{N_e}}) V(0,x_{N_e})
U'(\tau_{l_{N_e}} - \tau_{l_{N_e-1}})
\dots U'(\tau_{l_2} - \tau_{l_1})
V(0,x_1) U'(\tau_{l_1}-0) \ket{0} \ ,
\end{multline}
where the charge mismatch of $q-1$ between the in- and out-state comes about because we consider a finite system with $q N_e-(q-1)$ single-particle orbitals (i.e., the same number of orbitals as one would have on the sphere).
The difference between the free CFT evolution operator $U(\tau-\tau')$ and the operator $U'(\tau-\tau')$ used here is due to the spread-out background charge.
We need to know the form of $U'(\tau-\tau')$ in the case where $\tau$ and $\tau'$ correspond to
the center of two adjacent orbitals. 
The operator creating the background charge associated with one orbital is $e^{-i\varphi_0/\sqrt{q}}$.
Because the actual background charge is spread out homogeneously,
 we split the operator $e^{-i \varphi_0/\sqrt{q}}$ into
$n$ `slices', and act with the time evolution
$U (\delta\tau/n)$ in between these slices (recall that $\delta\tau$ is the distance between
neighboring orbitals).
Thus, we write
$U'(\delta\tau) =  \lim_{n\rightarrow\infty} e^{- \frac{2\pi\delta\tau}{nL} L_0}
e^{-\frac{i}{n\sqrt{q}} \varphi_0}
= e^{-\frac{2\pi\delta\tau}{L}  L_0 -\frac{i}{\sqrt{q}} \varphi_0}$.
Using the Campbell-Baker-Hausdorff formula,
the combined effect of the spread-out background charge and the time evolution results in~\cite{estienne-long}
\begin{equation}
\label{eq:time-evolution}
U' (\delta\tau) =  e^{- \frac{2\pi\delta\tau}{L} \bigl( L_0 + \frac{1}{2\sqrt{q}} \pi_0 + \frac{1}{6q} \bigr)}e^{-i\varphi_{0}/\sqrt{q}}
\equiv U'' (\delta\tau) e^{-i\varphi_{0}/\sqrt{q}} \ .
\end{equation}

The operators can now be associated with the orbitals as follows. 
The operator $U' (\delta\tau)$  takes care of the free time evolution from one orbital to the next in the presence of the homogeneous background charge, and corresponds to an empty orbital%
\footnote{%
We note that the time evolution used in Ref.~\onlinecite{zm}, which is simply $U$,
differs from $U''$ used here by the last two terms in the exponential.
However, by making use of the Campbell-Baker-Hausdorff formula, on finds that
$$
U''(\delta\tau) e^{-i\varphi_0/\sqrt{q}} = e^{-\frac{2\pi\delta\tau}{6qL}} e^{-i\varphi_0/(2\sqrt{q})}
U(\delta\tau)e^{-i\varphi_0/(2\sqrt{q})} \ .
$$
Since the electron operator in Ref.~\onlinecite{zm} uses this symmetric expression, we see that both descriptions are equivalent (up to boundary terms and unimportant factors).
}. 
On an occupied orbital, the operator $U' (\delta\tau)$ needs to be multiplied with $V(\tau=0,x)$, which creates the electron.
 
We can now calculate the matrix elements associated with these operators in the auxiliary Hilbert space, which is the Hilbert space of the chiral boson CFT (see App.~\ref{app:chiralboson} for details). 
We insert resolutions of identity $\mathbbm{1}=\sum_{Q, P, \mu}|Q, P, \mu\rangle\langle Q, P, \mu|$ between all the orbitals and use that the matrix elements of general vertex operators are given by 
\be{master}
\langle Q', P', \mu' | :e^{i \beta \varphi (w)}: | Q, P, \mu \rangle =
\delta_{Q',Q + \sqrt{q} \beta}\,
e^{-\frac{2\pi i}{L}(x+i \tau) \bigl( \frac{\beta Q}{\sqrt{q}}+P'-P \bigr)}
A^{\beta}_{\mu',\mu} \, ,
\ee
with $A^{\beta}_{\mu',\mu}$ given by Eq.~\eqref{eq:Avalues} in Appendix~\ref{app:chiralboson}.
Finally, the matrix elements needed for  the MPS description Eq.~\eqref{coeff} become
\begin{eqnarray}
B^{[0]} &=& \bra{ Q',P',\mu'} \hat U''(\delta\tau)e^{-i\varphi_0/\sqrt{q}}  \ket{ Q,P,\mu}  =
e^{-\frac{2\pi\delta\tau}{L} \bigl( \frac{(Q')^2}{2q} + P' + \frac{Q'}{2q} +\frac{1}{6q} \bigr)}
\delta_{Q',Q-1}\delta_{P',P}\prod_{j} \delta_{m'_j,m_j} \ , \label{noel} \\ \nonumber
B^{[1]} &=& \int_{-\frac{L}{2}}^{\frac{L}{2}} \frac {dx} {L}  \bra{ Q',P',\mu'} \hat U''(\delta\tau) e^{-i\varphi_0/\sqrt{q}} V(\tau=0,x) \ket{ Q,P,\mu}  \\
 &=& e^{-\frac{2\pi\delta\tau}{L} \bigl( \frac{(Q')^2}{2q} + P' + \frac{Q'}{2q} +\frac{1}{6q} \bigr)}
\delta_{Q',Q+q-1} \delta_{P',P-Q} A^{\sqrt{q}}_{\mu',\mu} \, . \label{elpres}
\end{eqnarray}
In the matrix elements of $B^{[1]}$, the $\delta$-function relating $P'$ to $P$ comes from the integral over $x$,
which is to be evaluated at $\tau=0$. For the electron matrix elements, the integral becomes
$\int_{-L/2}^{L/2} \frac{dx}{L} e^{-\frac{2\pi i x}{L} (Q+P'-P)} = \delta_{P',P-Q}$, which is well defined (i.e., it does not
depend on how we choose the limits on the integral) because $Q+P'-P$ is always an integer.
We again emphasize that these matrix elements do not depend on the partition labels  $l$.

It is straightforward  to get an MPS representation for the cylinder version of Eq.~\eqref{laughlinh} for an arbitrary number of quasiholes $N_{qh}$, by inserting $N_{qh}$ operators $H(\eta)$.
In order not to clutter the notation, we use $\eta=(x_\eta + i\tau_\eta)$ for the  complex coordinate on the cylinder and write $\omega_\eta = e^{- i (2\pi/L) (x_\eta + i\tau_\eta)}$ in the following.
Note, the correlator in Eq.~\eqref{laughlinh} is by definition radially ordered, so it does not matter
in what order we choose to write the operators. 
In the MPS formulation, one can also choose the points at which to insert the quasihole matrices. 
Nevertheless, one should insert the operator between the matrices corresponding to the orbitals closest to the quasihole location, to ensure fast convergence as the size of the auxiliary Hilbert space is increased.

To obtain the MPS matrices for the quasiholes, we must take into account the anti-commutation of the electron  $V(\omega)$ and the quasihole operator $H(\omega_\eta)$, which is reflected in the anti-symmetric factor $(\omega-\omega_\eta)$ present in the wave function Eq.~\eqref{laughlinh}.
Therefore, we must include an additional sign in the matrices for the quasiholes. This sign is
$(-1)^{\#V}$ where $\#V$ is the number of matrices $B^{[1]}$, corresponding to occupied orbitals that occur before
the position of the quasihole operator. We denote this position by $l$ if the corresponding matrix is inserted in
between the matrices corresponding to the orbitals $l-1$ and $l$ (where the first orbital has $l=0$).
$\#V$ can be written in terms of the quantum number $Q$ at the location $l$, which is the
number of orbitals that come before the quasihole matrix. For the $\alpha^{\rm th}$ quasihole
(i.e., we already acted with $\alpha-1$ quasihole matrices), $Q$ is given by
$Q = - l + q (\#V) + (\alpha - 1)$, where we assumed that the charge of the in-state is zero.
The term $-l$ comes from the distributed background charge. This leads to the sign $(-1)^{(Q+l-(\alpha-1))/q}$, which
needs to be taken into account in the matrix elements for the quasiholes.

Finally, one must be careful with the time evolution when dealing with the quasiholes.
The $\tau$ coordinate of the quasihole is $\tau_\eta$, and its matrix is inserted between
orbitals $l-1$ and $l$. Since the matrix corresponding to orbital $l-1$ includes the time evolution
from orbital $l-1$ to orbital $l$, we must ``evolve back'' by an amount $l \delta\tau - \tau_\eta$,
then act with the quasihole operator (with its $\tau$ coordinate set to zero), and finally evolve
forward again by $l \delta\tau - \tau_\eta$. In addition, because the correlator gives the
Landau wave functions up to a gauge factor as explained above, there is
an additional contribution of $e^{-\frac{2\pi i x_\eta \tilde\tau_\eta}{q L}}$, where
$\tilde\tau_\eta = \tau_\eta/(\delta\tau)$ is the $\tau$ coordinate of the quasihole in units of the
distance between neighboring orbitals (we note that this is a constant factor).
Putting all the pieces together, the matrices $B^{[p_l=0]}$ and $B^{[p_l=1]}$ of the MPS on orbital
$l$ will be multiplied with a quasihole matrix
$B^{[p_l]}\rightarrow \tilde{B}^{[p_l]}=B^{[p_l]}H_l(\omega_{\eta_\alpha})$ for the
$\alpha^{\rm th}$ quasihole, with the following matrix elements,
\begin{align}\label{mQH}
H_l (\omega_{\eta_\alpha}) =&
(-1)^{(Q+l-(\alpha-1))/q}
e^{+\frac{2\pi}{L} (l \delta\tau - \tau_{\eta_\alpha})
\bigl( \frac{(Q)^2}{2q} + P + \frac{Q}{2q} +\frac{1}{6q} \bigr)}
e^{-\frac{2\pi}{L} (l \delta\tau - \tau_{\eta_\alpha})
\bigl( \frac{(Q')^2}{2q} + P' + \frac{Q'}{2q} +\frac{1}{6q} \bigr)}
\\\nonumber&\times
e^{- \frac{2\pi i x_{\eta_\alpha}}{L} \bigl( P' - P + Q/q +\tilde\tau_{\eta_\alpha}/q \bigr)}
\delta_{Q',Q+1} A^{(1/\sqrt{q})}_{\mu',\mu} \ .
\end{align}
This concludes our review of the MPS description of the Laughlin states
on the cylinder in the presence of quasiholes.


\section{MPS representation for the quasielectron states} \label{sec:mps-qes}

In this section, we give an MPS representation for Laughlin states with quasielectrons on the cylinder.
We consider localized quasielectron states, as well as  angular momentum quasielectrons, which are used to construct the localized ones, as explained in Sec.~\ref{qeo}.
Most of the discussion below applies to both types of quasielectrons and where we need to distinguish them we do so explicitly.  
The insertion of a quasielectron is a non-local procedure (see Eq.~\eqref{oneqp}), since the (single) quasielectron can be placed on any orbital $l$, although with a very small weight when the orbital center is far from the quasielectron position. 

To explain precisely how all 
$
B^{[p]}\rightarrow\tilde{B}^{[p_l]}
$
matrices need to be updated is the main goal of this section.
Because the wave functions can be formulated as a CFT correlator~\eqref{full-corr}, one can find an MPS representation of the Slater coefficients, just as in the previous section, except that the procedure becomes more complicated. 
We therefore only highlight the differences, and provide the details of the derivation as well as the explicit form of the matrix elements and the wave functions in  Appendices~\ref{app:mps-pol-ang-qe} and \ref{app:matrices-cylinder}.

The most obvious difference with the previous section is that the vertex operators for the electrons, the modified electrons and the quasiholes
now depend on two chiral boson fields $\varphi(\omega)$ and $\tilde{\varphi}(\omega)$. In the case of an infinite system, they are given by
\begin{align}
V(\omega) &=\, : e^{i \sqrt{q} \varphi(\omega)}: \\
\tilde{V}^k (\omega) &= \omega^k \partial_\omega \bigl(
\, :e^{i (q-1)/\sqrt{q} \varphi(\omega)}: \, :e^{-i(q-1)/\sqrt{q(q-1)}\tilde{\varphi}(\omega)}: \bigr) \label{eq:tildeV}\\
H (\omega_\eta) &=\, :e^{i/\sqrt{q}\varphi(\omega_\eta)}: \, :e^{i(q-1)/\sqrt{q(q-1)}\tilde{\varphi}(\omega_\eta)}: \ .
\end{align}
We now outline how to calculate the matrix elements of the matrices corresponding to the modified electrons $\tilde{V}^k (\omega)$,
focusing on the differences with the previous section.
We start with the matrix elements of the empty orbitals, the `ordinary' electrons, and the quasiholes. 
Then we provide some details for the `modified' electrons necessary for the quasielectrons, but refer to App.~\ref{app:matrices-cylinder} for the actual derivations.

The presence of the additional field $\tilde{\varphi}(\omega)$ implies that the matrix elements corresponding to empty orbitals and orbitals occupied by `ordinary' electrons will have additional $\delta$-functions for the quantum numbers associated with $\tilde{\varphi}$. 
The factor describing the free time evolution is modified as well. 
The explicit expressions are given in Eqs.~\eqref{eq:two-fields-empty} and \eqref{eq:two-fields-electron}. 
The modifications to the matrices corresponding to the quasiholes are straightforward, and are given in Eq.~\eqref{eq:two-fields-qhole}.

In calculating the matrix elements associated with the modified electron operators, there are several differences compared to the previous section. First, a derivative $\omega^k \partial_\omega$ is present in $\tilde{V}^k (\omega)$. 
The easiest way of taking this into account is by performing a partial integration in the expression for the Slater determinants~\eqref{eq:slaters}, keeping in mind that the integral is performed at $\tau_l$, where $l$ is the orbital on which the modified electron resides. 
Thus, the derivative also acts on the factor $e^{i x \tau_l/\ell^2}$ in Eq.~\eqref{eq:slaters}.
The second difference is that the charge (associated with $\varphi(\omega)$) of the vertex operator in  $\tilde{V}^k(\omega)$ is $q-1$ instead of $q$.
This means that the factor $e^{i x \tau_l/\ell^2}$ present in Eq.~\eqref{eq:slaters} does
not completely cancel  the factor coming from the difference in phase between the Landau gauge wave functions, and the correlators in Eq.~\eqref{eq:gaugefactors}. 
Instead, we are left with an additional factor $e^{-\frac{2\pi i x}{L} (-l/q)}$, where $l$ is the orbital on which the modified electron operator resides. 
This factor is important, because to calculate the matrix elements for the modified electron operators, we have to calculate the integral $\int_{-L/2}^{L/2} \frac{dx}{L} e^{-\frac{2\pi i x}{L} f}$, where $f$ depends on the various quantum numbers (see the discussion below Eq.~\eqref{elpres}). 
For this integral to be well defined, $f$ has to be integer, and the additional factor $(-l/q)$ precisely makes this happen. 
In the end, this factor shows up in the $\delta$-function for the momenta.
 
The third difference concerns the contributions coming from the factors describing the free time evolution in the presence of the background charge. 
At the end of the day, these factors conspire to give the correct cylinder normalization of the wave functions. 
In the present case, they also give rise to factors that depend on both $k$, the angular momentum of the quasielectron, and $l$, the orbital associated which the modified electron operator.
The easiest way to deal with such factors is to calculate them explicitly from the form of the time evolution, and compensate for them by hand. 
The details are presented in Appendix~\ref{app:matrices-cylinder}.

Finally, one has to properly anti-symmetrize the wave functions.
This anti-symmetrization can be split in two parts. 
To begin with, the modified electron operators have to be anti-symmetrized with respect to the ordinary electrons, because $V(\omega)$ and $\tilde{V}^{k}(\omega)$ are bosonic with respect to one another. 
The same is true for the $\tilde{V}^{k}(\omega)$ amongst themselves.

The anti-symmetrization of the modified and ordinary electrons can be taken into account by inserting the factor $(-1)^{\#V}$ in the matrix elements for the modified electron operators. 
Here, $\#V$ denotes the number of ordinary electrons present in the system when acting with the current operator. 
This number can be expressed in terms of the various quantum numbers. 
To perform the anti-symmetrization between the modified electrons, one can not simply change the factor $(-1)^{\#V}$ to $(-1)^{\#V+\#\tilde{V}}$ where $\#\tilde{V}$ is the number of modified electrons already in the system. 
Such a change only leads to an overall sign of the wave function, and not to an actual anti-symmetrization between the modified electron operators. 
We postpone the solution of this problem to the end of this section.

Putting together the results so far, we obtain the matrix elements of the modified electron operator on orbital $l$, which we denote by $E_{k_a,l}$ [see Eq.~\eqref{eq:btilde}], for the $a^{\rm th}$ angular momentum quasielectron, with angular momentum $k_a$.

To obtain the matrix elements for a localized quasielectron on the cylinder, we need to use a localizing kernel on the cylinder, as discussed in Section~\ref{qeo} in the case of the disk geometry.
We denote the position of the quasielectron by $\omega_\xi = e^{-\frac{2\pi i}{L} (x_\xi + i \tau_\xi)} = e^{- \frac{2\pi i}{L} \xi}$.
The localizing kernel basically is the lowest Landau level projector, but with the substitution $\ell^2 \rightarrow q \ell^2$, because we are projecting a particle with charge $1/q$. 
On the cylinder, we have 
\begin{equation}
\label{eq:kernel-cylinder}
K(\omega_\xi, \omega) = \sum_k \bar{\phi}(x_\xi,\tau_\xi) \phi(x,\tau) =
\frac{1}{\sqrt{\pi q \ell^2}  L } e^{-\frac{(\tau^2+\tau_\xi^2)}{2 q \ell^2}} \sum_k e^{-\bigl(\frac{2 \pi}{L}\bigr)^2 q \ell^2 k^2} (\bar{\omega}_\xi)^k \omega^k \ .
\end{equation}
To show that $K(\omega_\xi,\omega)$ really is a localizing kernel, one can rewrite Eq.~\eqref{eq:kernel-cylinder} by means of the Poisson summation formula as
\begin{equation}
\label{eq:kernel-cylinder-real-space}
K(\omega_\xi, \omega) =
\frac{1}{2 \pi q \ell^2} \sum_{n\in\mathbb{Z}} e^{-\frac{1}{4 q \ell^2} | \xi - z + L n|^2}
e^{\frac{1}{8 q\ell^2}\bigl( (\xi-\bar{z})^2 - (\bar{\xi}-z)^2  + 2 L n (\xi +z - \bar{\xi}-\bar{z})\bigr)} \ .
\end{equation}
The second exponential is a pure phase, while the first exponential is the appropriate localizing Gaussian on the cylinder.
In the MPS description, we use the form of the localizing kernel as given in Eq.~\eqref{eq:kernel-cylinder}, noting that the factor $\omega^k$ is already incorporated in the operator $\tilde{V}^k(\omega)$ in Eq.~\eqref{eq:tildeV}.
The background charge gives rise to an incomplete Gaussian factor
$e^{-\frac{\tau^2}{2\ell^2}+\frac{\tau^2}{2q\ell^2}}$, because $\tilde{V}^k(\omega)$ has charge $q-1$, and the factor $e^{-\frac{\tau^2}{2 q \ell^2}}$ in the kernel precisely provides the missing factor.

To sum up, the matrix associated with the $a^{\rm th}$ localized quasielectron $E_l(\xi_a)$, is the
weighted sum of the matrix elements for the angular momentum quasielectrons $E_{k_a,l}$,
\be{qe-matrix}
E_l(\xi_a) = e^{-\frac{\tau_\xi^2}{2 q \ell^2}} \sum_{k_a} 
e^{-\bigl(\frac{2 \pi}{L}\bigr)^2 q \ell^2 k_a^2} e^{\frac{2 \pi}{L} k_a (i x_\xi + \tau_\xi)} E_{k_a,l} \ .
\ee

So far, we have not ensured that for each quasielectron, one and only one electron is modified for each Slater determinant. 
Moreover, this modified electron should be able to occupy an arbitrary orbital. 
To ensure this, we introduce an additional `quantum number' that  keeps track of precisely which modified electron operators have already acted.
This increases the Hilbert space dimension by a factor of $2^{N_{qe}}$, where $N_{qe}$ is the total number of quasielectrons in the system.
Thus, enforcing the right number of modified electron operators comes at a rather high price, which is why we only consider states with a few quasielectrons. 
However, using the enlarged Hilbert space it is easy to ensure that the modified electron operators anti-commute amongst themselves.

To explain the structure, we give the enlarged matrices corresponding to an orbital that is occupied by either an ordinary or a modified
electron. For the case of a single quasielectron (where there is no explicit anti-symmetrization needed), localized at $\xi_1$,
the enlarged matrix reads
\be{B1-1qe}
\tilde{B}^{[p_l=1]} =
\begin{pmatrix}
B^{[1]} & 0 \\
E_l(\xi_1) & B^{[1]} 
\end{pmatrix} .
\ee
Angular momentum quasielectron are obtained by replacing $E_l(\xi_1)$ with $E_{k_1,l}$. 
The first diagonal block in Eq.~\eqref{B1-1qe} corresponds to the operators for which we did not yet act with the modified electron operator.
The in-state has non-zero elements only in the first block, while the out-state only has non-zero elements in the second block. 
This enforces that each Slater determinant is a sum of terms that contain precisely one $E_l({\xi_1})$.
The matrix corresponding to the empty orbitals is simply block-diagonal,
\be{B0-1qe}
\tilde{B}^{[p=0]} =
\begin{pmatrix}
B^{[0]} & 0 \\
0 & B^{[0]} 
\end{pmatrix} .
\ee
On the orbitals with an inserted quasihole-operator, we need to multiply the $\tilde{B}^{[p=0]}$ and $\tilde{B}^{[p_l=1]}$ matrices with a block-diagonal matrix with $2^{N_{qe}}$ $H_l(\eta_\alpha)$ matrices on the diagonal.

For two quasielectrons, the enlarged matrix structure is given by
\be{B1-2qe}
\tilde{B}^{[p_l=1]} =
\begin{pmatrix}
B^{[1]} & 0 & 0 & 0\\
E_l(\xi_1) & B^{[1]} & 0 & 0 \\ 
-E_l(\xi_2) & 0 & B^{[1]} & 0 \\ 
0 & E_l(\xi_2) & E_l(\xi_1) & B^{[1]} \\ 
\end{pmatrix}. 
\ee
We included an explicit sign for the case when $E_l(\xi_2)$ acts before $E_{l'}(\xi_1)$, (i.e., when $l<l'$), which takes care of the
anti-symmetrization between the modified electron operators. The enlarged matrix structure for a system with three quasielectrons is shown in Eq.~\eqref{B_3qe} and the generalization to the cases with more quasielectrons is
straightforward.
We note that the matrix elements $\tilde{B}^{[p_l]}$ for the modified electron operator are orbital dependent, due to the various
factors described above. This is another reason why the MPS calculation of the quasielectron states is more costly compared to the states with quasiholes only.


\section{Numerical implementation} \label{sec:numimp}

All the matrices needed to numerically implement the MPS representation of the Laughlin state with an arbitrary number of quasiholes and quasielectrons were derived in the previous sections. However, there are some important technical issues that need to be dealt with to get an efficient numerical implementation.
In this section we discuss the auxiliary Hilbert space and its truncation, and how to deal with both finite and infinite system sizes. 
We also introduce the observables we calculate within the MPS framework in our study of the Laughlin state with quasielectrons.

\subsection{Auxiliary space cut-off}
The auxiliary space required for the most general wave function Eq.~\eqref{full-corr} containing quasielectron and quasihole excitations is  $|Q,P,\mu,\tilde{Q},\tilde{P},\tilde{\mu},\Gamma\rangle$, where $\Gamma$ labels the different blocks of the enlarged matrices, discussed in the previous sections. 
For pedagogical reasons, we first discuss the three quantum numbers $Q,P,\mu$ associated with the $\varphi$-field. 
These are the only quantum numbers needed in a system without quasielectrons or if a single quasielectron is the only excitation in the system. 
On each orbital, the matrix elements $B^{[0]}$, Eq.~\eqref{eq:two-fields-empty} or $B^{[1]}$, Eq.~\eqref{eq:two-fields-electron}, connect the left auxiliary space $\langle Q',P',\mu'|$ with the right auxiliary space $| Q,P,\mu\rangle$. 
Most of the matrix elements are zero, but there are still in principle infinitely many non-zero elements, with $Q\in\mathbb{Z}$, $P\in\mathbb{N}$ and $\mu$ an integer partition of $P$. 
However, the contribution to the wave function decreases exponentially with increasing $|Q|$ and $P$, because of the exponential factors originating from the free (imaginary) time evolution (see for instance Eqs.~\eqref{noel} and \eqref{elpres}). 
One can therefore truncate the auxiliary Hilbert space by introducing a cut-off, $P\leq P_\text{max}$ and $|Q|\leq Q_\text{max}(P_\text{max})$. 
The observables then converge to their thermodynamic values upon increasing $P_\text{max}$ and $Q_\text{max}$. 
We note that for larger circumference $L$, the convergence is slower, so that a larger cut-off is necessary.

One can reduce the dimension of the auxiliary Hilbert space by noting that the matrix elements come in  
$q$ independent sets which are called sectors. Each sector corresponds to one of the $q$ degenerate ground states on an infinite cylinder.
For a system without quasiparticle excitations, the quantum number $Q$ changes by $(q-1) \bmod q$ when going from one auxiliary Hilbert space to the next, which is enforced by the Kronecker delta's in the matrix elements. One can thus choose a sector by restricting the `incoming' quantum numbers $Q$ for the first orbital to have a definite value modulo $q$. We label the orbitals by $l$, with $l\in \mathbb{Z}$. The sector is then determined by $(Q - l) \bmod q$, which is constant throughout the system if no quasiparticles are present.
For physical observables, it is sufficient to analyze one sector, leading to a decrease in the dimensions of the matrices by a factor of $q$. However, one does need $q$ different versions of the matrices $B^{[0]}$ and $B^{[1]}$, dependent on $l \bmod q$. 
Insertion of a quasihole changes the sector by (plus) one. For the quasielectron, the situation is more complicated, because the quasielectron is non-local. The block structure in the $\Gamma$ quantum number (see for example Eq.~\eqref{B1-2qe}) explicitly keeps track of which quasielectrons already have been inserted. From this, one can determine which sector the block belongs to, since that only depends on the number of quasielectrons (and quasiholes) that were previously inserted.

We now turn our attention to the quantum numbers associated with the field $\tilde{\varphi}$, i.e. $\tilde{Q}$, $\tilde{P}$ and $\tilde{\mu}$.
Because we only consider a limited number of quasiparticles in our system, we do not impose any additional cut-off on $|\tilde{Q}|$.
We do impose a cut-off $\tilde{P}_{\rm max}$ on $\tilde{P}$ in a similar way as for the field $\varphi$. In practice we often use a larger $P_\text{max}$ than $\tilde{P}_\text{max}$, since the field $\tilde{\varphi}$ is only present in the operators for the quasiparticles, which are
typically placed far apart from each other (in $\tau$). Indeed, in the case of a single quasielectron, we can set $\tilde{P}_{\rm max} = 0$ without making any approximation.

\subsection{Finite vs. infinite cylinder}
The differences between an MPS description for a finite and an infinite fractional quantum Hall system are small and the discussion up to this point applies to both cases. 
The main difference between the two is their respective boundary condition.

To simulate a finite cylinder, one can simply start with an in-state that has a specified value of $Q_{\rm in}$ (we often take $Q_{\rm in}=0$) and $P_{\rm in}=0$. 
From this, we can construct the possible $Q$ and $P$ quantum numbers (subjected to the cut-off) on the neighboring orbitals  using the Kronecker delta's present in the matrix elements of the matrices $B^{[p]}$ (and the matrices corresponding to the quasiholes and quasielectrons, if present). 
In this way, we can construct the full auxiliary Hilbert space for the sectors we need. 
For a finite system, we label the orbitals as $l=0,1,\ldots,N_\phi$, and use the same number of orbitals as on a sphere, namely $N_\phi+1$, where $N_\phi = q (N_e -1 ) + N_{\rm qh} - N_{\rm qe}$.

To avoid edge effects that are necessarily present for a finite system, we can take advantage of the translational invariance of the ground state along the cylinder, which allows us to effectively simulate an infinite cylinder. 
In calculating observables, we still consider a finite number of orbitals (the simulation area), but one chooses the in and out states corresponding to an infinite system without quasiparticles. These can be obtained from the translational invariant matrices describing the ground state [Eqs.~\eqref{eq:two-fields-empty} and~\eqref{eq:two-fields-electron}]. 
To obtain the correct in-state for a given sector, we take the product of $q$ transfer matrices of neighboring orbitals corresponding to the sector we are interested in and compute the eigenvector corresponding to the largest eigenvalue~\cite{schollwoeck,orus}. 
Finally, for computational reasons, it is advantageous to bring the MPS of the simulation region to canonical form~\cite{schollwoeck}.

\subsection{Observables}
The density profiles of Laughlin quasiholes, and their braiding statistics was first calculated by Zaletel and Mong~\cite{zm}. Here, we generalize their approach to also include quasielectrons. 
The real space density is given by
\begin{align}
\rho(\vec{r})&=\int d\vec{r}_2d\vec{r}_3...d\vec{r}_{N_e}
\langle\Psi|\vec{r},\vec{r}_2,...,\vec{r}_{N_e}\rangle
\langle\vec{r},\vec{r}_2,...,\vec{r}_N|\Psi\rangle\nonumber\\
&=\displaystyle\sum_{n,m} e^{ix(n-m)\frac{2\pi}{L}}e^{-((\frac{2\pi n}{L}-\tau)^2+(\frac{2\pi m}{L}-\tau)^2)}
\frac{\langle\Psi|c_n^\dagger c_m|\Psi\rangle}{L\ell\sqrt{\pi}}, 
\end{align}
where the position $\vec{r}=(x,\tau)$, and the sum runs over the orbitals in the simulation region.
The correlation matrix $\langle\Psi|c_n^\dagger c_m|\Psi\rangle$ is easy to calculate in the MPS formulation, especially if it is brought to canonical form. 
Then we only need to contract $2(|n-m|+1)$ $B^{[p_l]}$-tensors and the left and right environment~\cite{schollwoeck,orus}. 
The correlation matrix is Hermitian and its elements fall off exponentially away from the diagonal, so that elements corresponding to large values of $|n-m|$ can be neglected. Examples of different density profiles for various quasiparticle constellations are shown and investigated in the next two sections. 

\begin{center}
	\begin{figure}
		\includegraphics[width=\columnwidth]{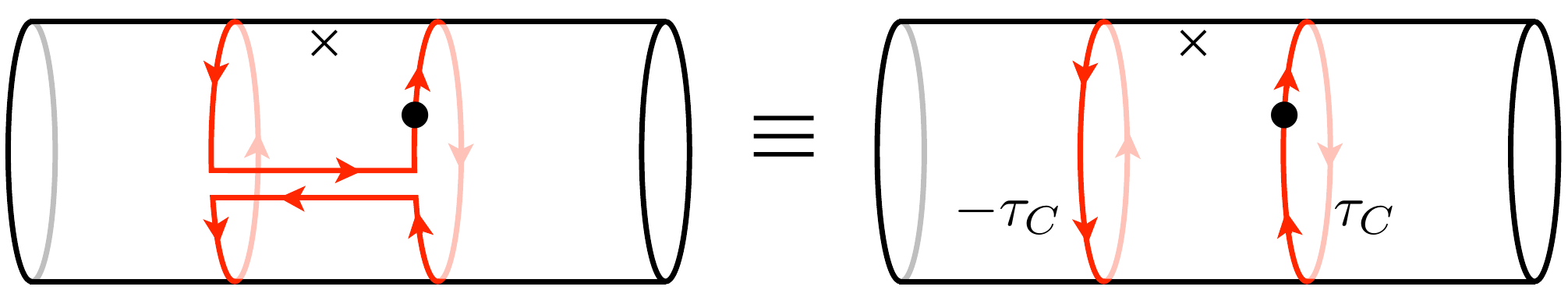}
		\caption{
			A sketch of the exchange path used in the calculation of the statistical phase. 
			The stationary quasiparticle is located at $(\tau_1,x_1)=(0,0)$ ($\times$ marks the spot).
			For the path depicted in the left panel, it is clear that the quasiparticle marked by the
			dot moves around the stationary quasi-particle. This path is equivalent to the path depicted
			on the right . The latter was used for the actual calculations, because it is convenient to
			move the quasiparticles at constant $\tau = \pm \tau_C$.}
	\label{Fig:exchangePath}
	\end{figure}
\end{center}

The braid statistics of quasiparticles is evaluated by calculating the Berry phases associated with various exchange paths.
A quasiparticle tracing out a closed path parametrized by $\zeta$ acquires a phase given by the Berry connection
\begin{align}
\theta=\oint dA=\oint d\zeta\langle \zeta|(-\partial_\zeta)|\zeta\rangle.
\end{align}
There are two contributions to this phase: the Aharonov-Bohm phase (the charged quasiparticle is moved in a magnetic field) and the statistical phase that depends on the quasiparticles that are enclosed by the paths. 
To obtain the statistical phase associated with the process of moving one quasiparticle around another, we take the difference $\Delta\theta$ of two Berry phases.
The first is associated with the process of moving one quasiparticle around the other along some path $C$, while the second amounts to follow the same path, albeit without the other quasiparticle present.
The path $C$ we use in actual calculations is depicted in Fig.~\ref{Fig:exchangePath}. 
To obtain the correct statistical phase, the quasiparticles must be separated sufficiently far from one another. 

In the limit of large separation $\tau_C\rightarrow\infty$ (see Fig.~\ref{Fig:exchangePath} for the definition of $\tau_C$)
it is easy to argue which statistical phases are
possible by using the structure of the MPS description and assuming that the system is in a
screening phase.
In this limit, the quasiparticles have no overlap and the only impact they can have on each another (assuming screening) is to shift the sector the other is in. 
That is, the circular path  at $-\tau_C$ gives the same phase contribution regardless of whether there is a quasiparticle at $\tau = 0$ or not.
For the path at $\tau_C$, the sector differs by $\pm 1$ depending on whether there is a quasihole or quasielectron at $\tau =  0$ or not. 
We know from general arguments that encircling $q$ quasiparticles of the same kind (this amounts to a difference of $q$ sectors) has a trivial statistical phase, i.e. $\Delta \theta_C(\tau_C\rightarrow \infty)=2\pi n$, with $n$ an integer.
As the quasiparticles are indistinguishable, all sectors must be equivalent and give the same phase contribution. 
Hence, the possible values for the statistical phase when encircling a single quasiparticle are 
$\Delta \theta_C(\tau_C\rightarrow \infty)=2\pi n/q$. 
For $q=3$, this includes the analytically known statistical phases $\Delta\theta= 2\pi/3$ for braiding  quasiholes in a Laughlin system. 
In the next two sections we calculate $\Delta\theta_C(\tau_C)$ numerically along the path $C$ as a function of $\tau_C$ for various combinations of quasiparticles.


\section{Properties of the quasielectron} \label{sec:results}

In this section, we study the properties of the quasielectrons in the Laughlin state using the MPS formulation we developed in the last section.
We first check the MPS description by comparing the Slater determinant coefficients it generates with those of the exact quasielectron wave functions.
We then plot the density profiles of various states with quasielectrons.
Here, we observe that in some cases, the quasielectrons are not localized at the expected position $(\tau_\xi,x_\xi)$, but are shifted in the $\tau$ direction if other quasiparticles are present at smaller $\tau$ values. 
Evidence for this effect has previously been seen in the numerical studies of Refs.~\onlinecite{jeon03,jeon04}, but has not been investigated in detail.
We show that this shift is a fundamental problem of the quasielectron wave functions, which is also present in the
angular momentum quasielectron states, and hence not caused by the projector that localizes the quasielectrons,
nor by the MPS description we use to investigate these states.
Because of this shift, the statistical phase associated with the exchange of quasiparticles is incorrect, if computed by moving a quasielectron. 

\subsection{Validating the MPS description}

Before using the MPS description of the quasielectron for calculating observables, it is good to explicitly verify that the wave functions are indeed correctly represented. 
To do this we generated all the Slater coefficients of the polynomial part of the wave functions from the MPS formulation, for small system sizes (up to six electrons), and checked those against the ones obtained by explicitly expanding the polynomials in Eq.~\eqref{eq:angmomqe}. 
For the cases with up to one quasihole, and an arbitrary number of angular momentum quasielectrons, we find exact agreement (including the overall factor) between the two formulations, provided that the cutoff in $P$ and $\tilde{P}$ is large enough.

When more than one quasihole is present, the coefficients are not identical, which is due to the cutoff in $P$ and $\tilde{P}$.
The difference disappears in the limit of large $P_{\rm max}$ and $\tilde{P}_{\rm max}$.
We note that the original formulation of quasiholes, as given by Zaletel and Mong, has the same issue. 
In their case, one needs to go to large $P_{\rm max}$ to faithfully represent factors of the type $(\eta_1-\eta_2)^{1/q}$.
Thus, we conclude that the MPS representation of the angular momentum quasielectrons wave functions given in Eq.~\eqref{eq:angmomqe} is indeed correct. 

In the same way, we explicitly verified the MPS description of these wave functions on the cylinder. 
Again, the Slater coefficients obtained from the MPS description are in exact agreement (for small system sizes and large enough cutoff $P_{\rm max}$ and $\tilde{P}_{\rm max}$) with the coefficients one obtains by explicitly expanding the cylinder wave functions.

\subsection{Density profiles}
\label{sec:density-profiles}
\begin{figure}[t]
	\begin{center}
		\includegraphics[width=\textwidth]{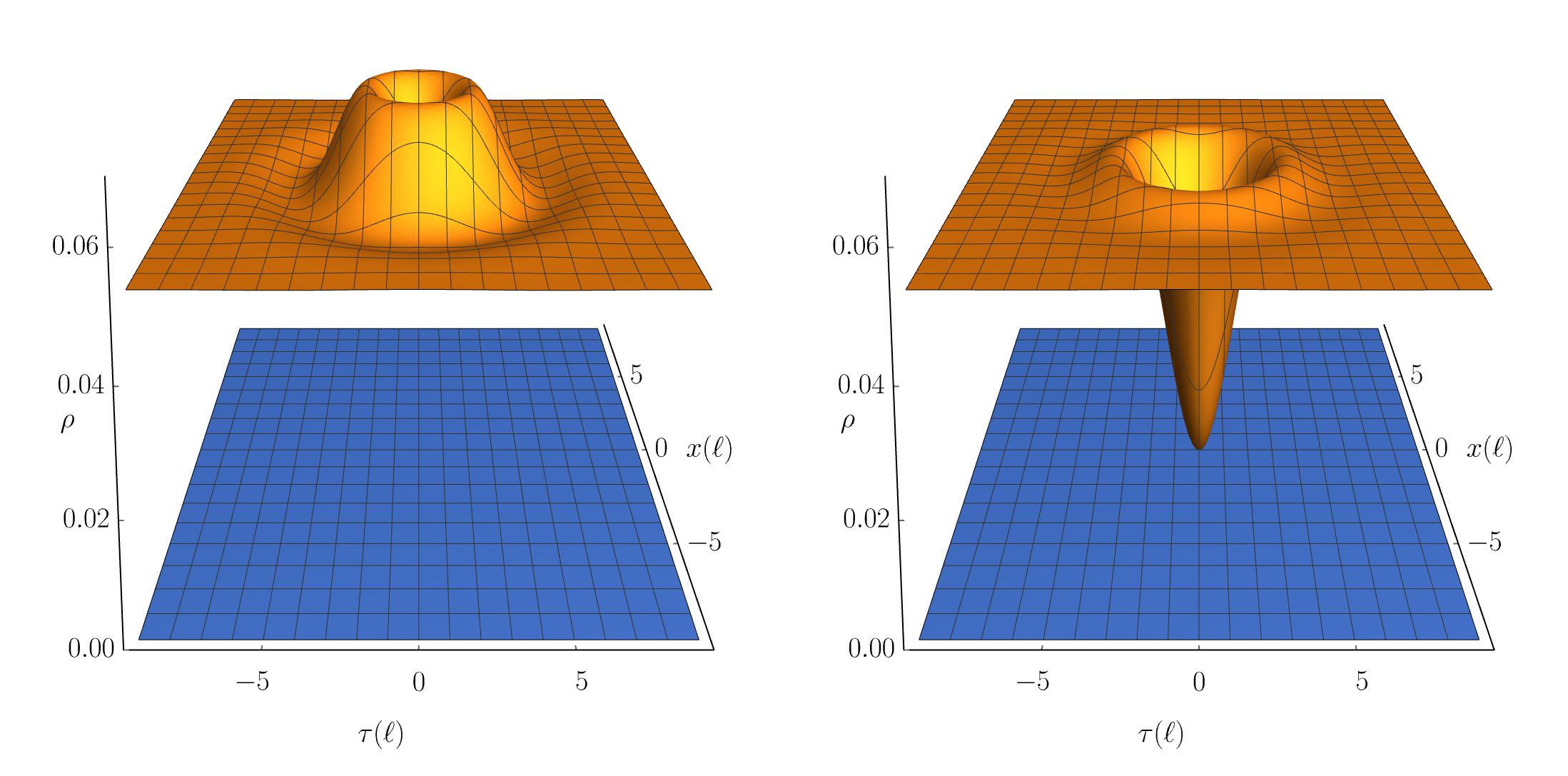}
	\end{center}
	\caption{The density profile of the $q=3$ Laughlin state with a quasielectron (left panel) and a quasihole (right panel) on an infinite cylinder with circumference $L=18\ell$ and cut-off $P_{\rm max} = 10$.}
	\label{fig:QE-3d}
\end{figure}

We start our investigation of the properties of the quasielectron by considering the density profile of a single quasielectron in a $q=3$ Laughlin state (see the left panel of Fig.~\ref{fig:QE-3d}). 
For comparison, the right panel depicts the density profile for a single quasihole. 
The ground state density is given by $\rho=\frac 1 q \frac 1{\delta\tau L}=1/(2\pi\ell^2 q)=0.053\ell^{-2}$, where  $\delta\tau=2\pi\ell^2/L$ is the distance between orbitals. 
Both the quasielectron and the quasihole are cylindrically symmetric around their center at $(\tau,x) = (0,0)$. 
The density at the center of the quasihole approaches zero with increasing $P_{\rm max}$, and is $\rho = 3.6\times10^{-6}\ell^{-2}$ for $P_{\rm max}=10$. 
Although the charges of the quasiparticles are fixed by the charge $Q$ of the vertex operators creating them, we have checked explicitly that they are given by $q_{\rm qh} = e/3$ and $q_{\rm qe} = -e/3$.

\begin{figure}[t]
	\begin{center}
		\includegraphics[width=7.5cm]{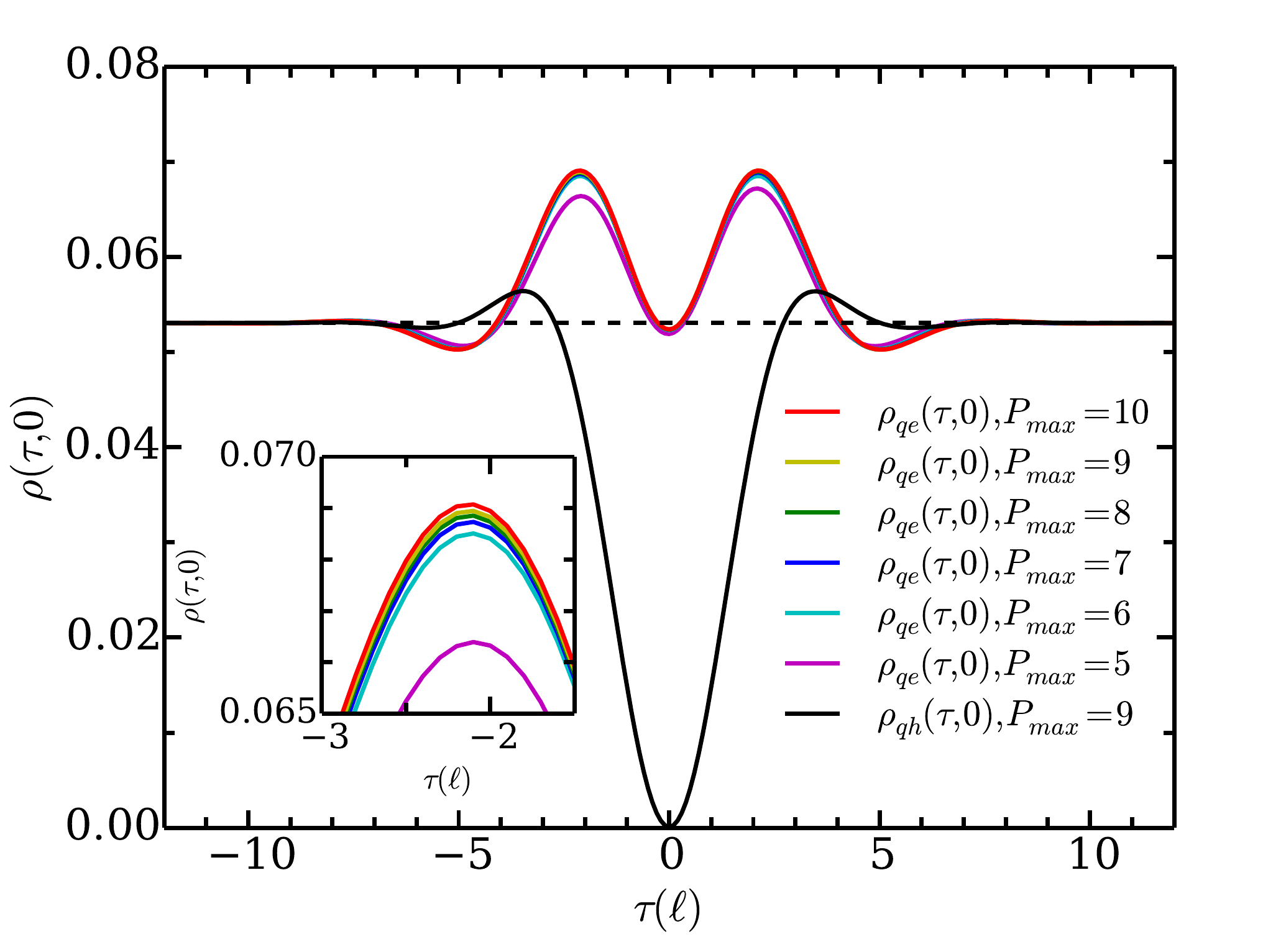}
	\end{center}
	\caption{The density cross section, through the center of a quasielectron for different values of $P_\text{max}$ (colored solid lines), through the center of a quasihole (black solid line) with $P_\text{max}=9$ and for the $q=3$ Laughlin ground state (black dashed line) as a reference, all on an infinite cylinder with circumference $L=20\ell$. The inset is a magnification of the left peak.}
	\label{fig:QE}
\end{figure}

We have studied the convergence of the density as a function of $P_{\rm max}$. 
In Fig.~\ref{fig:QE}, we plot the cross section of the charge density of the quasielectron as a function of $\tau$ through its center. 
For comparison we also include the corresponding cross section of a quasihole. 
For $P_\text{max}=6$ the profile (and other data) is well converged. 
In later more complex simulations with multiple quasiparticles (requiring the $\tilde{\varphi}$-field) $P_\text{max} = 6$ will be used, unless otherwise stated. 
The data is also well converged in the circumference $L$ (not shown) and we conclude that a single localized quasielectron excitation in a thermodynamic Laughlin ground state can be well described by the MPS.

We next consider systems with several quasiparticles, both quasiholes and quasielectrons. 
As long as the quasiparticles are well separated in $\tau$, the shape of both the quasielectrons and quasiholes are identical to those plotted in Fig.~\ref{fig:QE-3d}.
However, the position of a quasielectron is shifted from $(\tau_\xi,x_\xi)$ to $(\tau_\xi+\Delta\tau_\xi,x_\xi)$, where 
\begin{align}\label{shift}
\Delta\tau_\xi = 2(q-1)(n_{\rm qe}-n_{\rm qh})\delta\tau , 
\end{align}
and $n_{\rm qh}$/$n_{\rm qe}$ is the number of quasiholes/quasielectrons that is {\em located at smaller $\tau$ coordinates}. 
That is, each quasielectron is shifted $-2(q-1)$ {\em orbitals} in the $\tau$ direction for each quasihole at smaller $\tau$ coordinate, and 
shifted by $+2(q-1)$ {\em orbitals} in the $\tau$ direction for each quasielectron at smaller $\tau$ coordinate.
This shift persists even when all other quasiparticles are well separated  in the $\tau$-direction.
In contrast, the position of a quasielectron is not influenced by either quasielectrons or quasiholes at larger $\tau$ coordinates. 
In addition, the $\tau$ coordinates of the quasiholes is not influenced at all by the presence of other quasiparticles. 

The result that only quasiparticles at smaller $\tau$ influence the position of a quasielectron is not an inherent asymmetry in the setup, but rather a choice. 
It can be changed by an {\em overall} shift of all the quasielectron coordinates by changing the in quantum number $\tilde{Q}_{\rm in}$. 
For example, on an infinite cylinder it is natural to choose a symmetric prescription where the $\tau$ position of the quasielectrons is shifted $q-1$ orbitals towards every quasihole and $q-1$ orbitals away from every other quasielectron.

\begin{figure}[t]
	\begin{center}
		\includegraphics[width=7.5cm]{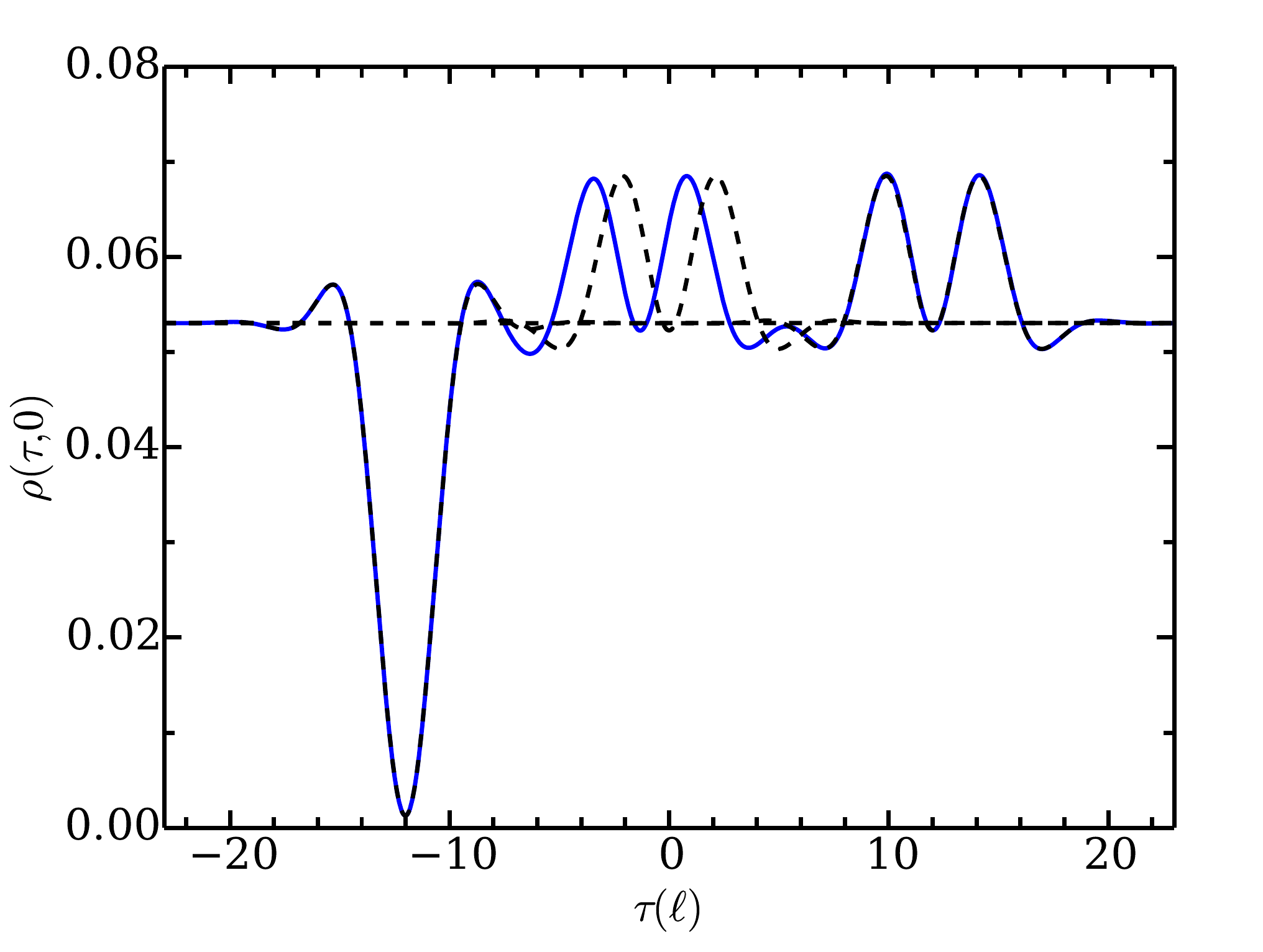}
	\end{center}
	\caption{The cross-section of the density for the $q=3$ Laughlin state with quasiparticles on an infinite cylinder with circumference $L=20\ell$. Blue solid line: two quasielectrons and a quasihole, where one of the quasielectrons is shifted 4 orbitals compared to its coordinate position ($\tilde{P}_\text{max}=1$). 
		The black dashed lines are plotted as references to show the density profile when only the quasihole or one of the quasielectrons is present.}
	\label{fig:shift}
\end{figure}

It is important to emphasize that the shift in the quasielectron coordinate is an additive effect, and not a small, i.e. modulo $q$, effect due to the different sectors. Introducing for instance more and more quasiholes at smaller $\tau$ coordinates of a quasielectron, will cause a shift proportional to the number of such quasiholes. 
In Fig.~\ref{fig:shift} we show an example of a system with two quasielectrons and one quasihole.
The intended location (i.e., the parameters used in the matrices associated with these quasiparticles) is $(\tau_\eta,x_\eta)=(-12\ell,0)$ for the quasihole and $(\tau_{\xi_1},x_{\xi_1})=(0,0)$ and $(\tau_{\xi_2},x_{\xi_2})=(12\ell,0)$ for the quasielectrons. The blue line shows the density of this system as a function of $\tau$, for $x=0$.
As a reference, the three dashed lines show the density as a function of $\tau$ for $x=0$, for systems with one quasihole at $(\tau_\eta,x_\eta)=(-12\ell,0)$, one quasielectron at $(\tau_{\xi},x_{\xi})=(0,0)$ or one quasielectron at $(\tau_{\xi},x_{\xi})=(12\ell,0)$, indicating the expected positions of the quasiparticles.
The quasielectron with coordinate $\tau_{\xi_2} = 12\ell$ is indeed located at the intended position, because there is both a quasielectron and a quasihole at smaller $\tau$, and the shifts caused by them cancel.
The quasielectron with intended coordinate $\tau_{\xi_1} = 0$ is shifted by $\Delta \tau_\xi=-2(q-1)\delta \tau = -4\delta\tau$  in the $\tau$ direction, because only the quasihole has a smaller $\tau$ coordinate.

We should stress that the observed shift in the location of the quasielectrons is not due to an error in our MPS representation of the quasielectron states.
As we reported above, we thoroughly checked our MPS representation.
Indeed, this shift was first observed in Ref.~\onlinecite{jeon03} (see also Refs.~\onlinecite{jeon04} and~\onlinecite{hansson07a}), where the electron density for composite fermion quasielectrons was calculated in the disk geometry by means of Monte Carlo (these composite fermion quasielectrons are the disk versions of the cylinder quasielectron states we consider).
Later the same shift was seen in Ref.~\onlinecite{hansson07b} by an analytical calculation relying on a random phase approximation.
We thus conclude that the observed shift is an actual feature of the states we study in terms of an MPS description.
In section~\ref{sec:screening} below, we study this shift in more detail, and propose a way to correct it.   


\subsection{Statistical phases}
\begin{figure}[th]
	\begin{center}
		\includegraphics[width=7.5cm]{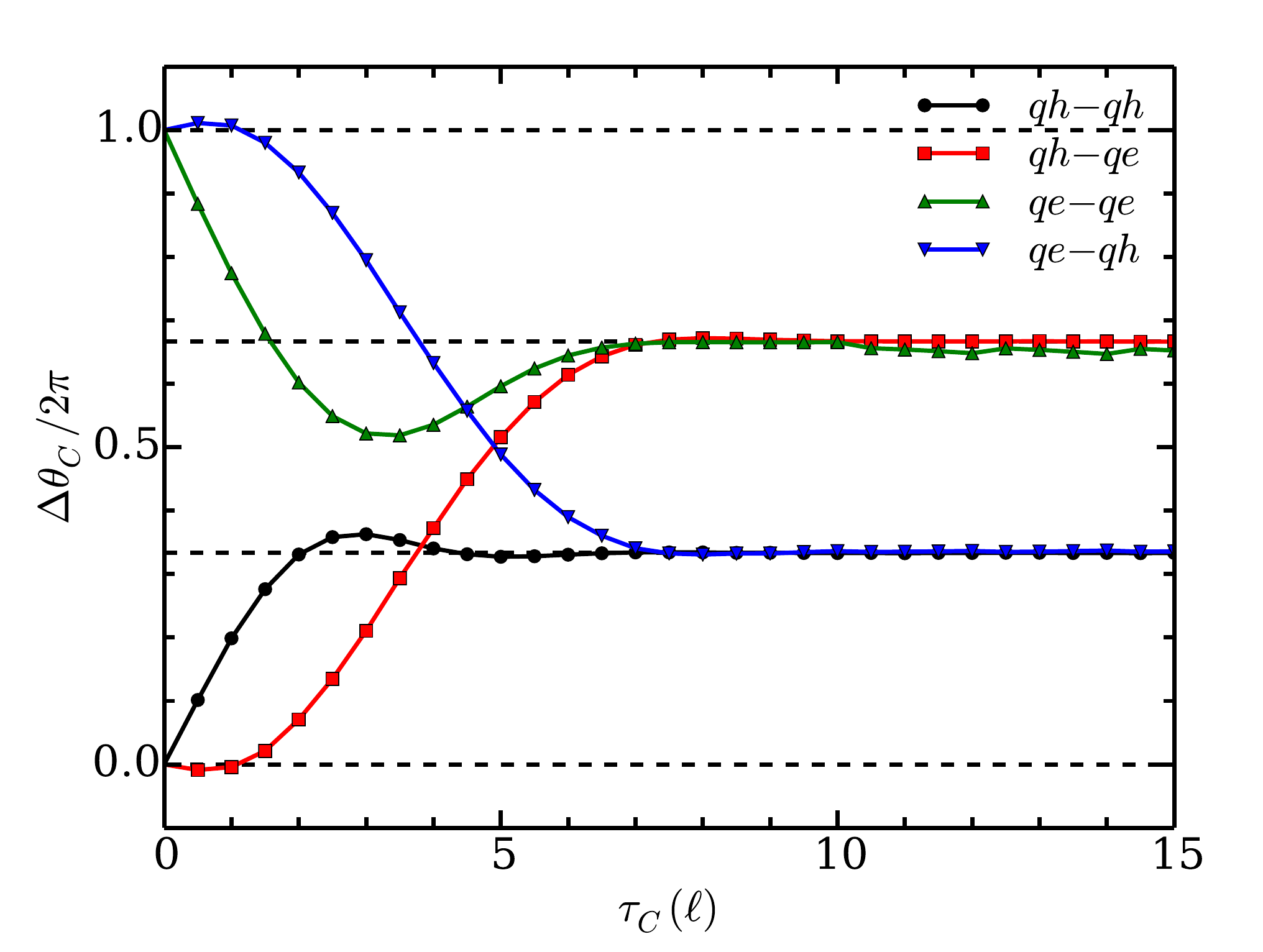}
	\end{center}
	\caption{The statistical phases $\Delta\theta_C/(2\pi)$ for the four ways $q=3$ Laughlin quasielectrons and quasiholes can be braided around each other as a function of $\tau_C$. The calculations are performed on an infinite cylinder with circumference $L=16\ell$. The data is converged in $P_\text{max}$ and $\tilde{P}_\text{max}$ for all data points, except those with quasielectron(s) and $\tau_C\lesssim 4\ell$, where $\tilde{P}_\text{max}=5$ is used ($\tilde{P}_\text{max}=3$ for the qe-qe case). 
	}
	\label{fig:Phase_shifted}
\end{figure}

With the shift detected in the quasielectron position, we expect some errors in the calculations of the statistical phases.
If we move a quasielectron around another quasiparticle, the location of the quasielectron will be shifted from the intended location.
However, when we calculate the contribution of the Aharonov-Bohm phase, the quasielectron will be at the intended location, because no other quasiparticles are present. 
Thus, in the first step, one does not pick up the correct Aharonov-Bohm contribution, leading to an error in the statistical phase.
We nevertheless proceed and plot the statistical phases $\Delta\theta_C(\tau_C)$ along the path $C$, defined in Fig.~\ref{Fig:exchangePath}, for the four different ways of braiding $q=3$ Laughlin quasiholes and quasielectrons (see Fig.~\ref{fig:Phase_shifted}).
The black (red) curve shows the result if a quasihole is moved around another quasihole (quasielectron) at a distance $\tau_C$, whereas 
for the blue (green) curve a quasielectron is braided around a quasihole (quasielectron), instead.
These results agree for large $\tau_C$ with previous numerical studies of Refs.~\onlinecite{kjonsbergL1999,jeon03,jeon04},
but not with what is expected from analytical arguments.
If a quasihole is moved around another quasihole, or a quasielectron around another quasielectron, the statistical phase is given by $e^{2 \pi i/q}$,
while if a quasihole is moved around a quasielectron, or the other way around, the statistical phase is given by $e^{-2 \pi i/q}$ for a $\nu=\frac{1}{q}$ Laughlin state.
The results we obtained are correct if a quasihole is moved around a quasiparticle, but we obtain the wrong sign for  the phase if a quasielectron is moved around a quasiparticle.
This is  consistent with the observation that the location of the quasielectrons is shifted if another quasiparticle is present at smaller $\tau$. 
For the incorrect cases, i.e. when a quasielectron is moved, we also observe some small deviations from $\Delta\theta_C/2\pi = \pm 1/3$ at large $\tau_C$, which we suspect originates from the numerical calculation not being fully converged. 


\subsection{Angular momentum states}

The shift in the position of the quasielectrons is at first glance quite surprising, given that the exponential factor in Eq.~\eqref{proj1} should localize the surplus charge related to the modified electron operator Eq.~\eqref{model} at position $\xi$.
Let us however stress again that the MPS representation faithfully reproduces the CFT wave functions, which are equivalent to the composite fermion construction, and  the problem is inherent already in the wave function.
In order to get a better understanding of the origin of this shift, we have calculated the density profiles for various constellations of quasiparticles in angular momentum states. 
We use a finite system, and only present the numerical results. The necessary formalism is given in Appendix \ref{app:matrices-cylinder}.

In Fig.~\ref{fig:angqe} we show two  examples of density profiles as a function of $\tau$
for angular momentum quasielectrons on a finite cylinder. 
A single angular momentum quasielectron appears in the expected location and is included as a reference. 
We observe that the density profiles of the angular momentum quasielectrons are shifted $q-1$ orbitals towards the quasihole, if there is a quasihole at smaller $\tau$ (i.e., the shift is in the negative $\tau$ direction) and $q-1$ orbitals away from the quasielectron if there is a quasielectron at smaller $\tau$ (i.e., in the positive $\tau$ direction).
Thus, the shift in the location of the angular momentum quasielectrons is half of the shift for the localized quasielectrons in Sec.~\ref{sec:density-profiles}.  
As before, the shift is proportional to the difference in the number of quasiholes and quasielectrons that are located at smaller $\tau$ values,
$ \Delta\tau_\xi = (q-1)(n_{qe}-n_{qh})\delta\tau$
and the position of the angular momentum quasielectrons is not affected by quasiparticles that are located at larger $\tau$ values. 

These results hold as long as the separation between the quasiparticles is sufficiently large. They show that already the angular momentum quasielectron wave functions, which are used to construct the localized quasielectron states, are `deficient' in the sense that the angular-momentum quasielectrons are influenced by the other quasiparticles, even if they are far away.
In the next section, we will argue that this is due to the quasielectrons not being properly screened. 

\begin{figure}[t]
	\begin{center}
		\includegraphics[width=7.5cm]{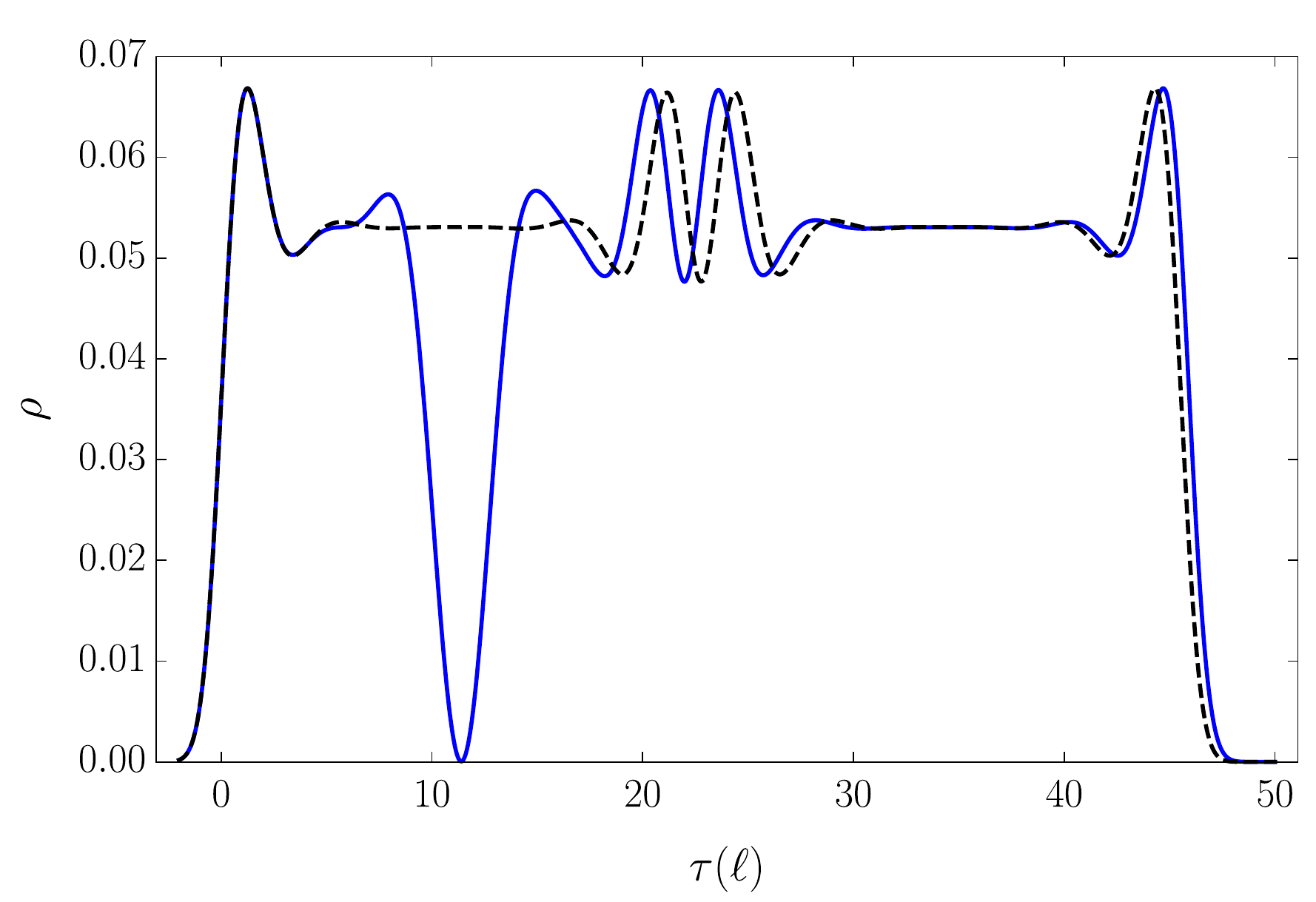}
		\includegraphics[width=7.5cm]{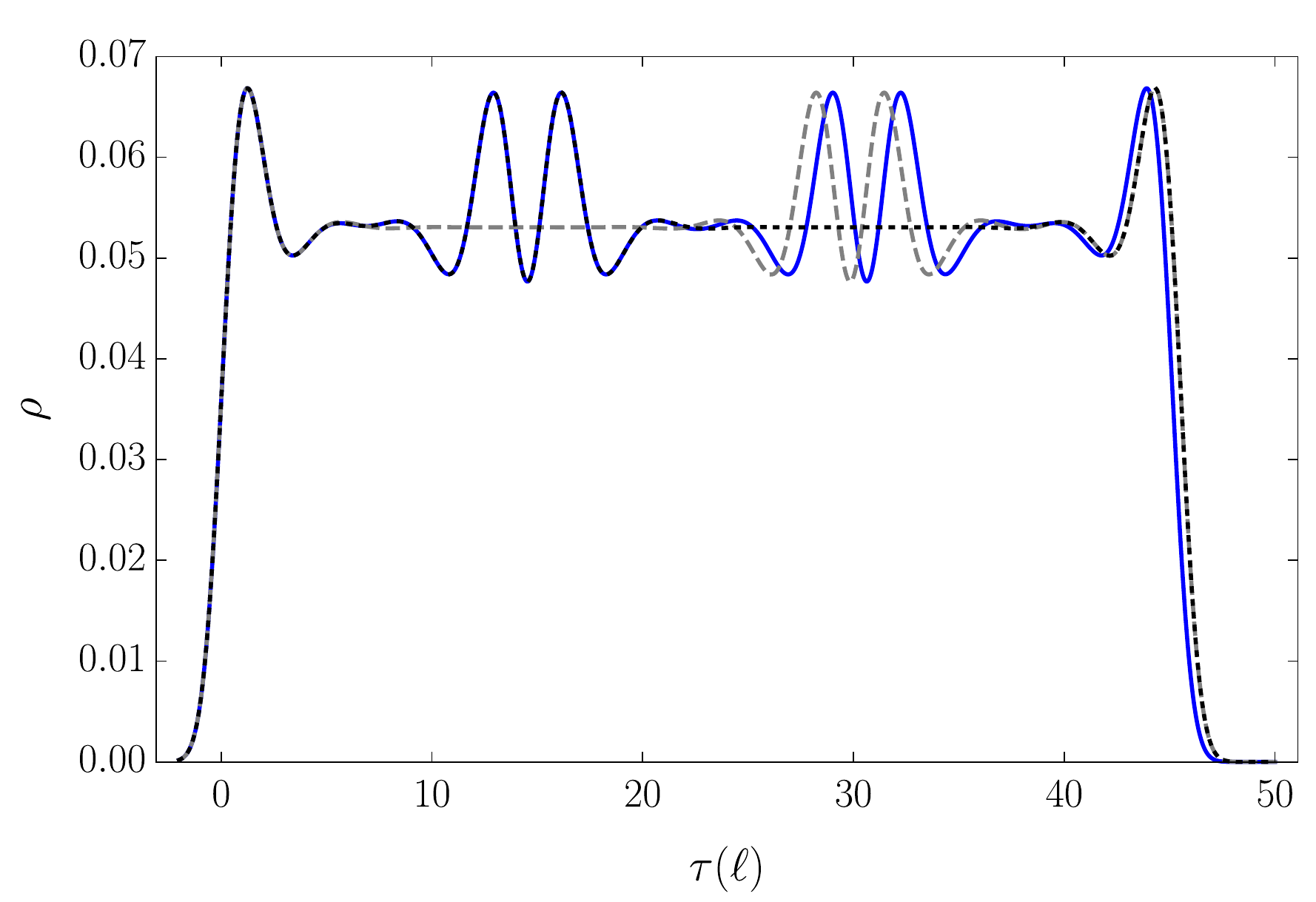}
	\end{center}
	\caption{
		Cross section of the density profile for a $q=3$ Laughlin state with quasiparticles on a finite cylinder, $N_e = 40$, $L=16\ell$ and $\tilde{P}_{\rm max} = 3$.
		Left panel: a quasihole at $\tau = 29 \delta\tau$ and an angular momentum quasielectron with $k = 20$ (solid blue line).
		The density profile of an angular momentum quasielectron with $k = 20$ without quasihole is shown  as reference (dashed black line).
		Right panel: two angular momentum quasielectrons with $k = 13,26$ (solid blue line).
		The density profile of a single angular momentum quasielectron is shown as reference ($k = 13$ dashed black, $k = 26$ dotted black line).
		Note that adding or subtracting a flux quantum changes the size of the droplet, which is the origin of the deviation between the dashed and solid curves at large $\tau$.}
	\label{fig:angqe}
\end{figure}


\section{Screening the quasielectrons}
\label{sec:screening}

In the previous section, we learned that the problem of the shift in the position of the quasielectrons is neither due to the MPS implementation, nor to the particular projection that construct localized states from the angular momentum states, but is a deficiency in the original wave functions.

In this section we show that the problem can be traced back to the improper screening of the modified electron operators.
We start by briefly recalling the meaning and significance of the plasma analogy, then use results in the so called thin-cylinder (or Tao-Thouless)
limit (see for instance Ref.~\onlinecite{bergholtz08}) to highlight the shift problem in an analytically accessible setting and show how it can be cured in the case of widely separated quasielectrons. 
Next we verify the conclusions from the Tao-Thouless limit in a full MPS calculation, and present a modification of the quasiparticle operators which do localize the quasielectrons at the expected positions regardless of quasiparticle configuration. 
This also results in the correct statistical phase for all ways of braiding quasiparticles around one another, showing that the topological properties are as expected. 
The modification of the quasiparticle operator  has a minor drawback, namely that the density profile of a quasielectron is distorted, when other quasiparticles are located at similar values of $\tau$ (but arbitrary separation in $x$). 
This distortion can be cured by an additional, ad-hoc modification of the quasiparticle matrices in the MPS formulation, as discussed below.   
Finally, we discuss the theoretical significance of the failure of the screening and suggest an alternative CFT construction that is likely to localize the quasielectrons at their correct positions.   


\subsection{The plasma analogy - a primer}
\label{sec:plasma}
Laughlin's plasma analogy is based on the observation that the modulus squared of the wave function in Eq.~\eqref{laughlinh}, can be written as 
\be{laughlinp}
|\Psi_{L,qh}(z_1\dots z_{N_e}; \eta_1\dots \eta_{N_{qh}})|^2 &=&
\prod_{\alpha<\beta}^{N_{qh}} |\eta_\alpha-\eta_\beta|^\frac{2}{q}
\prod_{\alpha,i} |\eta_\alpha-z_i|^2 \prod_{i<j}^{N_e} |z_i-z_j|^{2q}  e^{-\frac{1}{2\ell^2} \sum_j |z_j|^2}  
= e^{-\beta H} \nonumber \, ,
\ee 
with $\beta=2/q$ and
\be{plasmaham}
H =     -q^2 \sum_{i<j}^{N_e} \ln |z_i-z_j|  -q  \sum_{\alpha,i} \ln |\eta_\alpha-z_i|   
- \sum_{\alpha<\beta}^{N_{qh}}   \ln |\eta_\alpha-\eta_\beta|  + \frac{q}{2\ell^2} \sum_j |z_j|^2 \ .
\ee
This is the Hamiltonian of a two dimensional Coulomb plasma with unit charges at the positions $\eta_\alpha$ and charge $q$ particles at the positions $z_i$ in a homogeneous neutralizing background charge density $\rho=1/(2\pi\ell^2)$. 
The normalization factor $\mathcal N$ of the wave function  is given by
\be{laughnorm}
{\mathcal N}^2 = \int \prod_{i<j}^N d^2z\, e^{-\beta H} = e^{-\beta F} \ ,
\ee
where $F$ is the free energy of the plasma with unit charge impurities at the positions $\eta_\alpha$. For $q\lesssim 70$, the plasma is in a screening phase \cite{screeningplasma,deleeuw82}, and we can conclude that $\mathcal N$ is independent of the quasihole positions $\eta_\alpha$, as long as they are sufficiently separated. From this it is fairly easy to show that there are no Berry phases associated with quasiparticle braidings, so that the statistical phases can be directly read from the wave functions in Eq.~\eqref{laughlinh}~\cite{arovas}. 
The quasiholes have charge $e/q$, since putting $q$ of them at the same position corresponds to one missing electron. Using the plasma analogy, one can also show that the charges of widely separated (compared to the magnetic length) quasiholes are quantum mechanically sharp~\cite{kjonsbergL1999}. 
From this it should be clear that the plasma analogy is at the heart of the successful phenomenology of the Laughlin wave functions. 
For the following discussion it is important to keep in mind the physical reason for why a plasma screens: it is due to the combination of an  energy cost for deviations from charge neutrality, and the presence of itinerant charges, or in a field theory language, a fluctuating charge density.  
Thus, for the plasma to be in a screening phase even in the more complicated cases where there are several components, it is important to have fluctuating charges for all components.

Turning to the  realization of QH wave functions in terms of CFT correlators, we first notice that for the Laughlin states the electric charge of the quasiholes is directly given by the $U(1)$ charge current $J = i\partial_z \varphi(z)/\sqrt q$. 
This charge is indeed fluctuating (with respect to the constant background charge density), because the electrons are itinerant. 
The situation is different for the field $\tilde \varphi$ that is needed for the modified electron operators, which build up the quasielectrons.
The field $\tilde\varphi$ does not carry {\em electric} charge, but nevertheless has an associated $U(1)$ charge, as encoded by the quantum number $\tilde{Q}$. The problem lies in that this $\tilde{Q}$ charge does not fluctuate, while the $Q$ charge does.
In more technical terms, when inserting a $\tilde V$ operator, the incoming $\tilde Q$ charge is fixed by the charge of the quasiparticles that are located at smaller $\tau$, when they are sufficiently far apart.
Thus, the $\tilde\varphi$ field is not screened, and consequently, the quasielectron positions are shifted depending on the positions of the other quasiparticles.

Note, however, that while the shift in the positions of the quasielectrons indicates that at least one of the fields is not screened, the reverse conclusion does not hold. 
In particular, the quasiholes are located at the correct positions, even though they also have a component in the (unscreened) $\tilde\varphi$ field. 
One can understand this by noting that the quasihole operator is localized at $\eta$ and enforces a vanishing density at this position, at least in the limit where all the quasielectron excitations are very far from the quasihole. The quasielectron, on the other hand, is build from the itinerant, modified electrons around $\xi$, which are not properly screened.

One may wonder what this implies for the hierarchical states, where one inserts $O(N_e)$ quasiparticles and integrates
over their positions~\cite{haldane83,halperin83,halperin84}. 
In this case, we would expect also the $\tilde \varphi$ field to screen, as the corresponding charges have become itinerant in the daughter state. 
Indeed, while there is no rigorous proof of this, there are several compelling heuristic arguments for why the plasma analogy should hold for the hierarchical/composite fermion states, as reviewed in Ref.~\onlinecite{hhsv}.


\subsection{Screening in the Tao-Thouless limit} 

We believe that the shift observed in the quasielectron positions is due to the absence of local charge fluctuations related to the field $\tilde\varphi$. To substantiate this claim, we consider the thin cylinder limit where the shift observed in the numerical calculations can be reproduced using analytical methods. We note that in the thin cylinder limit, the wave function reduces to a single Slater determinant.
Consequently, in this limit screening, if present, is classical screening that occurs for configurations that minimize the Coulomb energy.

In the Tao-Thouless (TT)-limit of a  thin cylinder~\cite{TaoThouless}, a QH wave function simply becomes a charge density wave that minimizes the repulsive static Coulomb energy (see for instance Ref.~\onlinecite{bergholtz08}). For the simple example of the $q=3$ Laughlin state with filling $\nu = 1/3$, the occupation pattern of the ground state (often referred to as the TT pattern) is $...1001001001...$ , i.e. there are $q-1$ empty orbitals in between the occupied orbitals. A quasihole amounts to adding an extra zero to get a pattern like $...1001\underline{000}1001...$ while a quasielectron amounts to removing one zero $...100\underline{101}001...$ . The horizontal line indicates the position of the quasihole/quasielectron. 

These patterns are reproduced by taking the TT-limit of the CFT wave functions, but (as shown below) the quasielectron ``motif" $101$ is displaced by a distance $2(q-1)(n_{qe} - n_{qh})$ precisely as seen in the full MPS calculation.
By introducing screened operators (or rather, operators that do not carry $\tilde{Q}$ charge) this shift goes away and the quasielectrons appear at their  expected positions. We now illustrate this with the simplest example of a single quasielectron and a number of quasiholes (all placed at the same position at a smaller $\tau$ value), first for the original (unscreened) operators, and afterwards for the screened versions. 

Usually, one derives the TT-limit by identifying the dominant component of the wave function when the circumference $L\rightarrow 0$. 
For the sake of completeness, such a calculation is presented in appendix \ref{app:TTlimit}. 
Here, we use an alternative approach that is both simpler and more closely related to the MPS representation --- the TT-limit wave function is reproduced by considering only the zero modes of the chiral fields. 
At least for the ground state, it is straightforward to see this from the form of the matrices in Eqs.~\eqref{eq:b0pol} and \eqref{eq:b1pol}, as the $L$ dependent exponential becomes maximal (in the $L\rightarrow0$ limit) when choosing the charge $Q$ cyclically in $0,\ldots, q-1$ and the momentum $P=0$ and $\tilde{P}=0$ throughout. As the momentum originates solely from the non-zero modes, we can  ignore these in the TT-limit. 

Taking into account only the zero modes and putting the quasielectron at position $\omega_\xi \sim (\tau_\xi,x_\xi)$ and the $n_{qh}$ quasiholes at $\omega_0 \sim (\tau_0,x_0)$, with $\tau_0 \ll \tau_\xi$, we  have to evaluate the correlator
\begin{align}\label{eq:corr}
\Psi_{qe}=\sum_{\alpha} K(\omega_{\xi},w_\alpha) \langle\mathcal{O}_{\rm bg} V(\omega_{N_e}) \ldots V(\omega_{\alpha+1})\tilde V(\omega_\alpha) V(\omega_{\alpha-1}) \ldots V(\omega_{1}) H(\omega_0)^{n_{qh}}\rangle \, ,
\end{align}
where the kernel $K$ is defined in Eq.~\eqref{proj1}, 
and where only the zero modes are kept in the vertex operators, {\it e.g.}
$V(\omega_j) = : e^{i\sqrt q (\varphi_0+i \pi_0 \frac{2\pi}{L}(ix_j-\tau_j))}:$.
To evaluate Eq.~\eqref{eq:corr}, we assume that all coordinates are ordered such that $\tau_0<\tau_1<\ldots <\tau_N$ and use the following formula to normal order two vertex operators containing only the zero modes 
\begin{align}
\label{eq:normalordering}
:e^{i\gamma_1 (\varphi_0+i \pi_0\frac{2\pi}{L}(iz_j))}:\,: e^{i\gamma_2(\varphi_0+i\pi_0\frac{2\pi}{L}(iz_k))}:&=\omega_j^{\gamma_1\gamma_2}   :e^{i(\gamma_1+\gamma_2)\varphi_0+i \pi_0\frac{2\pi}{L}(\gamma_1 iz_j+\gamma_2 iz_k)}:\,.
\end{align}
Normal ordering with respect to the background charge reproduces the Gaussian factor needed for a LLL wave function. 

Up to unimportant phases and overall factors that we ignore, 
the wave function becomes
\begin{align}\label{eq:product state}
\Psi_{qe}&\sim\sum_{\alpha,s} \bar{\omega}_\xi^s e^{-\frac{1}{2q\ell^2}[\tau_\xi^2+2(\orbdis q s)^2]}
\underbrace{ \prod_{j=1}^{N_e} e^{\frac{1}{2\ell^2}(\orbdis\mu_j^\alpha)^2}}_{\equiv W(\alpha,s)}
\times \prod_{j=1}^{N_e} e^{-i\orbdis x_j}e^{-\frac{1}{2  \ell^2}(\tau_j-\orbdis\mu_j^\alpha)^2},
\end{align}
with 
\begin{align}
\mu_j^\alpha&=\left\{\begin{array}{cc}
q(j-1)+n_{qh} & \mbox{ for } j<\alpha\\
(q-1)(j-1)+s-1 & \mbox{ for } j=\alpha\\
q(j-1)+n_{qh}-1 & \mbox{ for } j>\alpha,
\end{array}\right.	
\end{align}
where  $\orbdis=\frac{2\pi\ell^2}{L}$ again denotes the separation of two single-particle orbitals on the cylinder. The extra contribution of $s-1$ for $\mu_{\alpha}^{\alpha}$ originates from the kernel and the derivative. 
We can interpret the summands of Eq.~\eqref{eq:product state} as a product state of single-particle orbitals, where the $\mu_j^\alpha$ are nothing but the momentum labels of the occupied orbitals. 
In the thin cylinder limit, only the product state with the maximal weight $\exp[-\frac{1}{2q\ell^2}(\tau_\xi^2+2(\orbdis \, q s)^2)] \,W(\alpha,s)$ survives.

In order to maximize the sum over $\alpha$ (for any given $s$), we note that  $W(\alpha,s)$ can be written as (up to overall factors that are independent of $\alpha$ and $s$ and that are ignored in the following)
\begin{align}
W(\alpha,s)\sim\exp\left[-(q-1)\frac{\orbdis^2}{2 \ell^2}\left\{\alpha^2-2\alpha(s-n_{qh}+\frac{1}{2})\right\}\right]\exp\left[\frac{\orbdis^2}{2 \ell^2}(s^2-2qs)\right],
\end{align}
which becomes maximal at $\alpha_{0}=s-n_{qh}$ or $\alpha_{0}=s-n_{qh}+1$ (both choices lead to the same thin cylinder pattern --- in fact, they will yield the position of the right/left `1' of the `101' quasielectron motif respectively). Reinserting $\alpha_0$ in $W(\alpha,s)$ yields 
\begin{align}
W(\alpha_{0},s)\sim&\exp\left[\frac{\orbdis^2}{2\ell^2}\left\{ qs^2-s(2n_{qh}(q-1)+q+1)\right\}\right],
\end{align}
independent of which of the two possibilities for $\alpha_0$ we choose. 
Now we need to maximize over $s$, i.e. maximize 
\begin{align}
\bar{\omega}_{\xi}^s \exp\left[-\frac{(\orbdis q s)^2}{q\ell^2}\right]
W(\alpha_{0},s)\sim\exp\left[-\frac{q\orbdis^2}{2\ell^2}\left\{s^2-2s\left(\frac{\tau_{\xi}}{q\orbdis}-\frac{2n_{qh}(q-1)+q+1}{2q}\right)\right\}\right],
\end{align}
which happens for $s_0=\frac{\tau_{\xi}}{q\orbdis}-\frac{2n_{qh}(q-1)+q+1}{2q}$. 
In order to find the approximate quasielectron position we reinsert $s_0$ and both choices of $\alpha_0$ into $\mu_{\alpha}^\alpha$ and take their mean to get 
\begin{align}
\orbdis \mu_{\alpha}^\alpha&= \tau_\xi-2\orbdis(q-1)n_{qh}-(q+1)\orbdis. 
\end{align}
Thus, the position of the quasielectron is shifted by $2(q-1)n_{qh}$ orbitals (up to the constant shift of $(q+1)\orbdis$) towards the quasiholes at position $\omega_0$.  

We now redo the calculation using  screened quasiparticle operators. 
Since we have already removed the non-zero modes, screening the quasielectron operators amounts to removing the $\tilde\varphi$ field, so that {\it e.g.} $H(\omega_0)=e^{i\frac{1}{\sqrt{q}}(\varphi_0+i\pi_0\frac{2\pi}{L}(ix_0-\tau_0))}$, and similarly for the quasielectron. Since the operators no longer carry any $\tilde Q$ charge, they are trivially screened. Here, however, we face a complication. 
The original operators were constructed so that the wave function was analytic in all the electron coordinates, corresponding to a LLL wave function. 
This was actually the original rationale for introducing the $\tilde\varphi$ field. 
Removing $\tilde\varphi$ unavoidably introduces non-analytic factors $\omega^{n/q}$, thus making the integrals Eq.~\eqref{eq:slaters} ill-defined. 
Note that this difficulty only appears for the coordinates related to a quasielectron operator. 
The wave functions are also non-analytic in the quasihole coordinates, $\eta$, but this poses no problem since $\eta$ is not integrated over. The minimal way out of this conundrum is  to enlarge the integration range as follows
\be{extension}
\int_0^L \frac {dx} L \rightarrow \int_0^{qL} \frac {dx}  {qL}
\ee
which does not effect the integrals over integer powers of $\omega$, but makes integrals over  $\omega^{n/q}$ well defined. 
To be consistent, we should also modify the projection kernel, 
\be{modproj}
K(\omega_1,\omega_2) \rightarrow \sum_{r=0}^{q-1}\sum_{s\in Z} 
(\bar\omega_1 \omega_2)^{s+r/q} e^{-\frac{1}{2q\ell^2}(\tau_{1}^2+\tau_{2}^2+2\orbdis^2 (q s+r)^2)}.
\ee

Taking this into account, we get the following expression for the quasielectron wave function
\begin{align}
\label{eq:corr2}
\Psi_{qe}=&\sum_{\alpha,s,r}  (\bar\omega_{\xi}w_\alpha)^{s+r/q} e^{-\frac{1}{2q\ell^2}(\tau_{\xi}^2+\tau_{\alpha}^2+2\orbdis^2 (q s+r)^2)}
 \\
\nonumber
& \times \langle \mathcal{O}_{\rm bg} V(\omega_{N_e}) \ldots V(\omega_{\alpha+1})\tilde V(\omega_\alpha) V(\omega_{\alpha-1}) \ldots V(\omega_{1}) H(\omega_0)^{n_{qh}}\rangle \, .
\end{align}
Again, this is nothing but a product of Gaussians with weights that depend on $\alpha$ and $s$, except that the $\mu_j^\alpha$ are now given by 
\begin{align}
\mu_j^\alpha&=\left\{\begin{array}{cc}
q(j-1)+n_{qh} & \mbox{ for } j<\alpha\\
(q-1)(j-1)+\frac{(q-1)}{q}n_{qh}+s+\frac{r}{q}-1 & \mbox{ for } j=\alpha\\
q(j-1)+n_{qh}-1 & \mbox{ for } j>\alpha
\end{array}\right. \ .
\end{align}
The constant $r$ is chosen such that $\mu_{\alpha}^\alpha$ is an integer, i.e. $r\equiv n_{qh}-qn_0$, where $n_0$ is the appropriate integer such that $r\in[0,q-1]$. We proceed in the same way as above, by first maximizing over $\alpha$ for any given $s$ (again, overall factors that do not depend on $\alpha$ or $s$ are ignored in the following):
\begin{align}
W(\alpha,s)&\sim \exp\left[-(q-1)\frac{\orbdis^2}{2\ell^2}\left\{\alpha^2-2\alpha(s-n_0+\frac{1}{2})\right\}\right]\exp\left[\frac{\orbdis^2}{2\ell^2}\left\{
s^2+2s(n_{qh}-n_0-q)\right\}\right].
\end{align}
The maximum occurs  at $\alpha_0=s-n_0$ or $\alpha_0=s-n_0+1$, and the corresponding weight is given by 
\begin{align}
W(\alpha_{0},s)&\sim\exp\left[\frac{\orbdis^2}{2\ell^2}\{qs^2+s(2r-q-1)\}\right] \ .
\end{align}
We now proceed to maximize the sum over $s$, i.e. the expression
\begin{align}
\bar{\omega}_{\xi}^{s+r/q}e^{-\frac{\delta\tau^2(qs+r)^2}{q\ell^2}}W(\alpha_0,s)
\sim\exp\left[-\frac{q\orbdis^2}{2\ell^2}\left\{s^2-2s(\frac{\tau_{\xi}}{q\orbdis}-\frac{2r+q+1}{2q})\right\}\right].
\end{align}   
This is maximized by  $s_0=\frac{\tau_{\xi}}{q\orbdis}-\frac{2r+q+1}{2q}$, which fixes the approximate quasielectron position to  
\begin{align}
\orbdis\mu_{\alpha}^\alpha &=\tau_{\xi}-\orbdis(q+1).
\end{align}
The shift in the quasielectron position is a constant that is \emph{independent} on the number of quasiholes. Such a shift can easily be compensated for by changing the details of how the screening charges are introduced.


\subsection{Properties of screened quasielectrons}

Above we discussed the role of fluctuating charges for the localization of the quasielectrons in their desired positions. In the TT limit the $\tilde{\varphi}$-field can be completely removed and therefore the lack of fluctuations does no longer pose a problem for the positions of the quasielectrons. Below we use these insights and investigate ways to improve the quasielectron wave functions. These changes can easily be implemented in the MPS description from Section \ref{sec:mps-qes}, which describes the states with the shifted quasielectrons.

The $\tilde\varphi$ field only enters the wave functions through the quasiparticle operators, so any fluctuations would have to come from an added background field. There is no obvious recipe on how to create this fluctuating background field. 
One constraint is that the $\tilde{Q}$ quantum number only can take integer values on the bonds. With opposite $\tilde{\varphi}$-charge on the quasihole and quasielectron, the background field needs to accommodate both signs of the charges. Numerically we have tried several different versions where the neutralizing charge is allowed to fluctuate on every orbital. However, we could not find any background prescription with promising behavior.

Having had no success with introducing a fluctuating background field, we next consider another way of `screening' the operators, inspired by our screening prescription in the TT-limit, which is by
removing the part of the quasiparticle operators that creates the $\tilde Q$ charge, namely $e^{\pm i (q-1)\tilde\varphi_0/\sqrt{q(q-1)}}$.
For quasiparticles on a cylinder far apart in $\tau$, only the $\tilde{\varphi}_0$ component of the $\tilde{\varphi}$ field contributes to the wave function and it is indeed this contribution that gives rise to the observed shift.

Numerically, we observe that the non-zero modes of the $\tilde{\varphi}$-field are irrelevant for quasiparticles that are separated by $\Delta \tau\gtrsim 15\ell$, meaning that $\tilde{P}_\text{max}=0$ can be used in those situations. The only part left of the $\tilde{\varphi}$-field is the zero mode, which can be screened. Updated quasiparticle operators on a cylinder with a finite circumference $L$, that give the correct asymptotic behavior, can be written as the screened (or neutralized) operators
\begin{align}
\label{sc-qe}
H_l(\eta_\alpha)&\rightarrow e^{-i(q-1)/\sqrt{q(q-1)}\tilde{\varphi}_0/2}H_l(\eta_\alpha)e^{-i(q-1)/\sqrt{q(q-1)}\tilde{\varphi}_0/2}\\
E_l(\xi_a)&\rightarrow e^{i(q-1)/\sqrt{q(q-1)}\tilde{\varphi}_0/2}E_l(\xi_a)e^{i(q-1)/\sqrt{q(q-1)}\tilde{\varphi}_0/2},\nonumber
\end{align}
with $H_l(\eta)$ given in Eq.~\eqref{eq:two-fields-qhole} and $E_l(\xi)$ in Eq.~\eqref{qe-matrix}. This screening prescription essentially amounts to having a screening charge smeared evenly around the circumference of the cylinder. The choice of putting half of the screening charge on each side of the operator rather than dividing it in some other way only amounts to a microscopic change of the quasielectron position. However, we note that the shift of all quasielectrons when putting the full screening operator on one side can be canceled by a corresponding shift in the in and out charge of the $\tilde{\varphi}$-field.

\begin{figure}[t]
	\begin{center}
		\includegraphics[width=7.5cm]{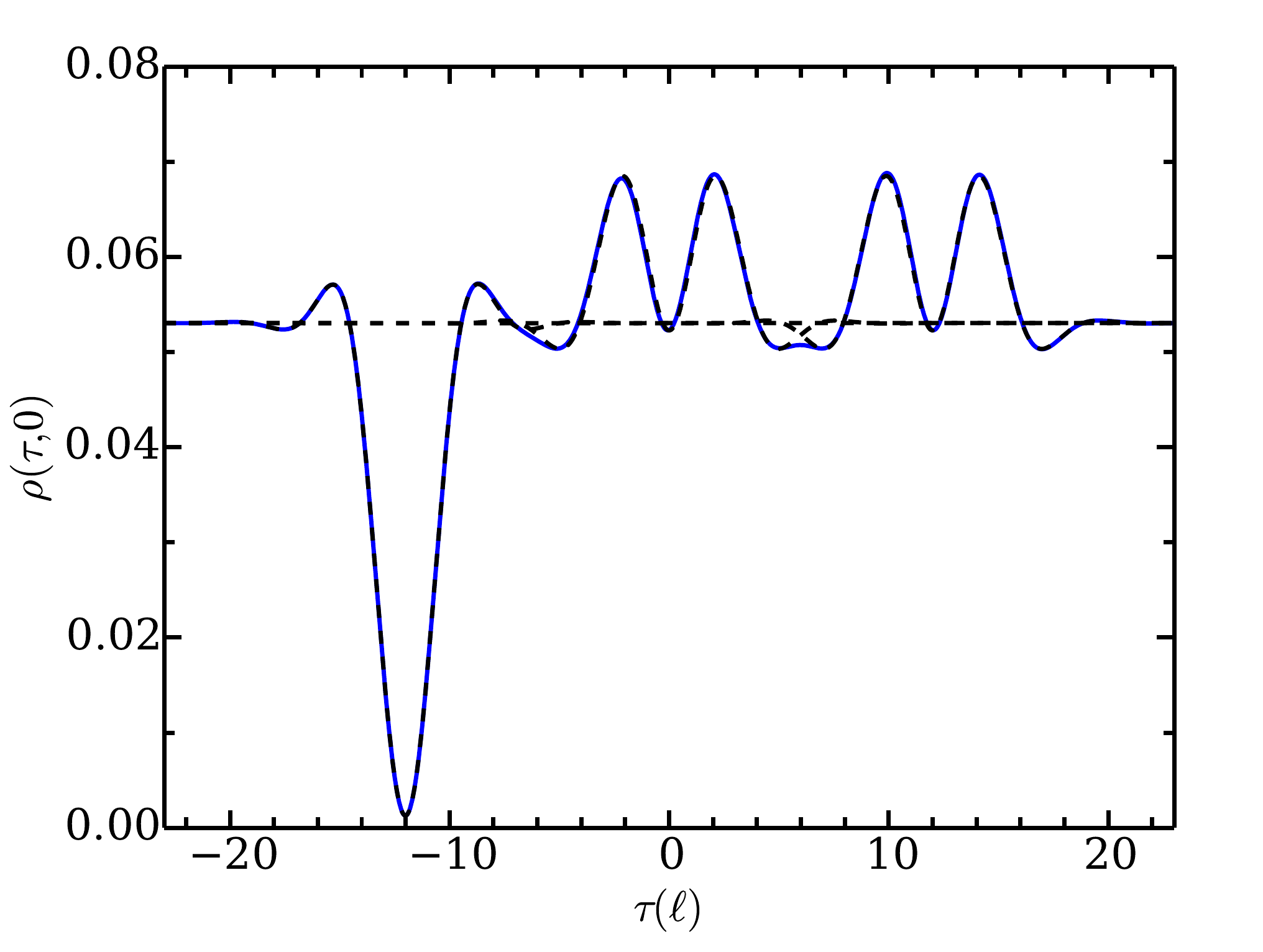}
		\includegraphics[width=7.5cm]{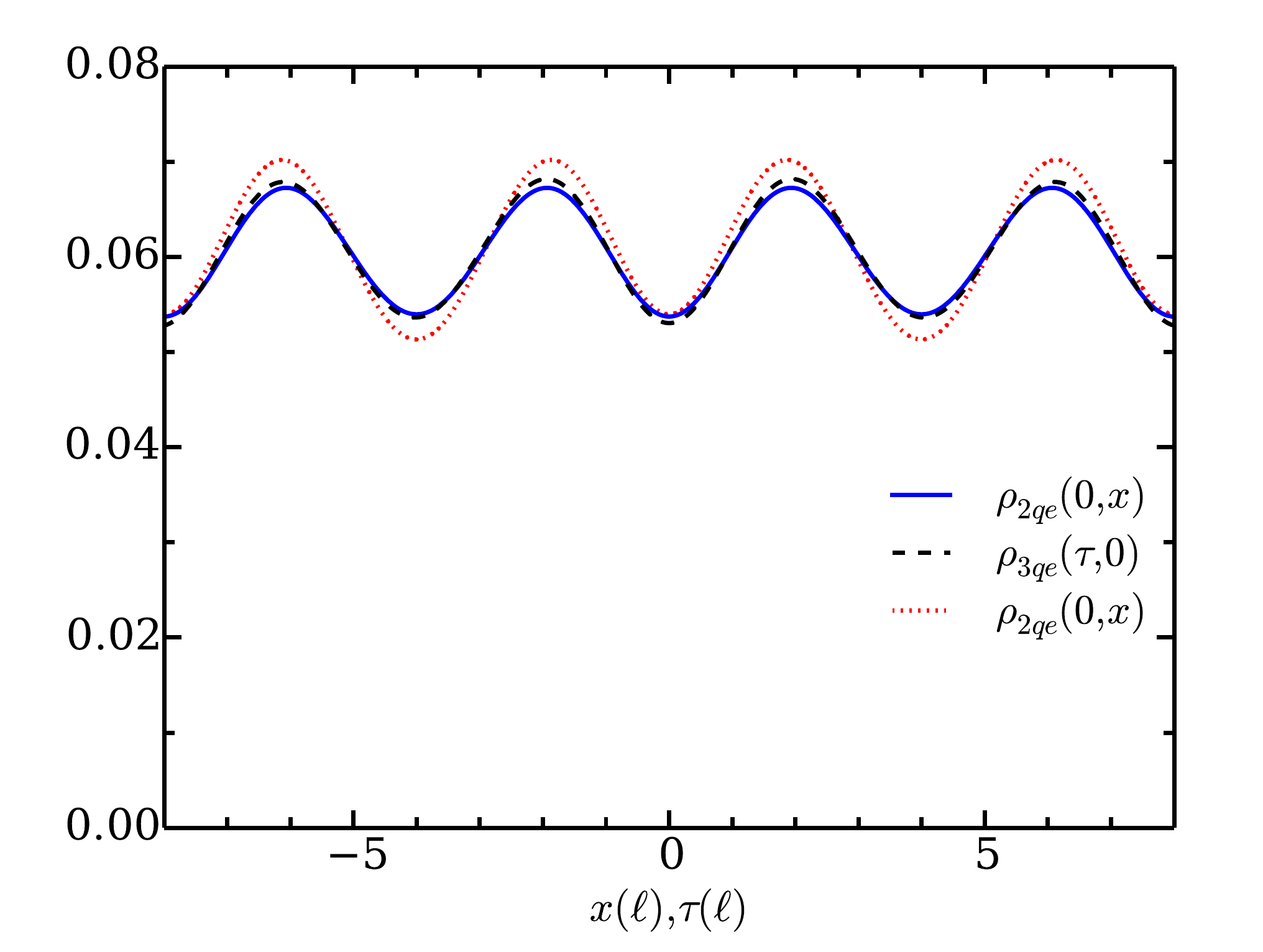}\\
		\vspace{.1cm}
		\includegraphics[height=3.5cm]{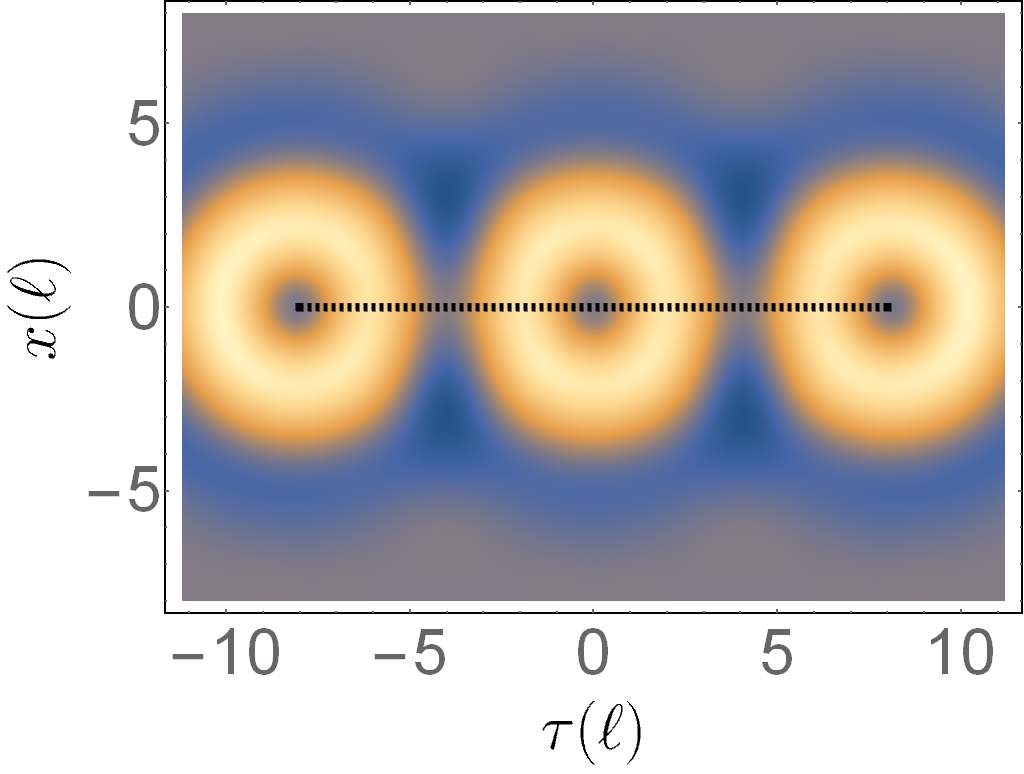}
		\includegraphics[height=3.5cm]{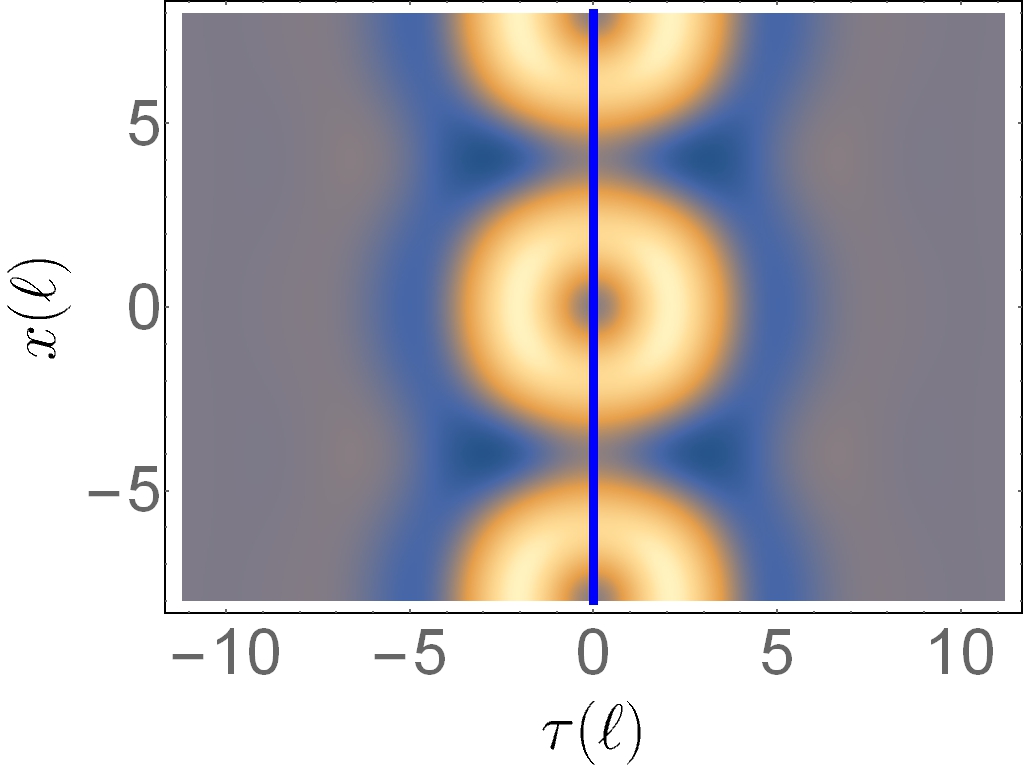}
		\includegraphics[height=3.5cm]{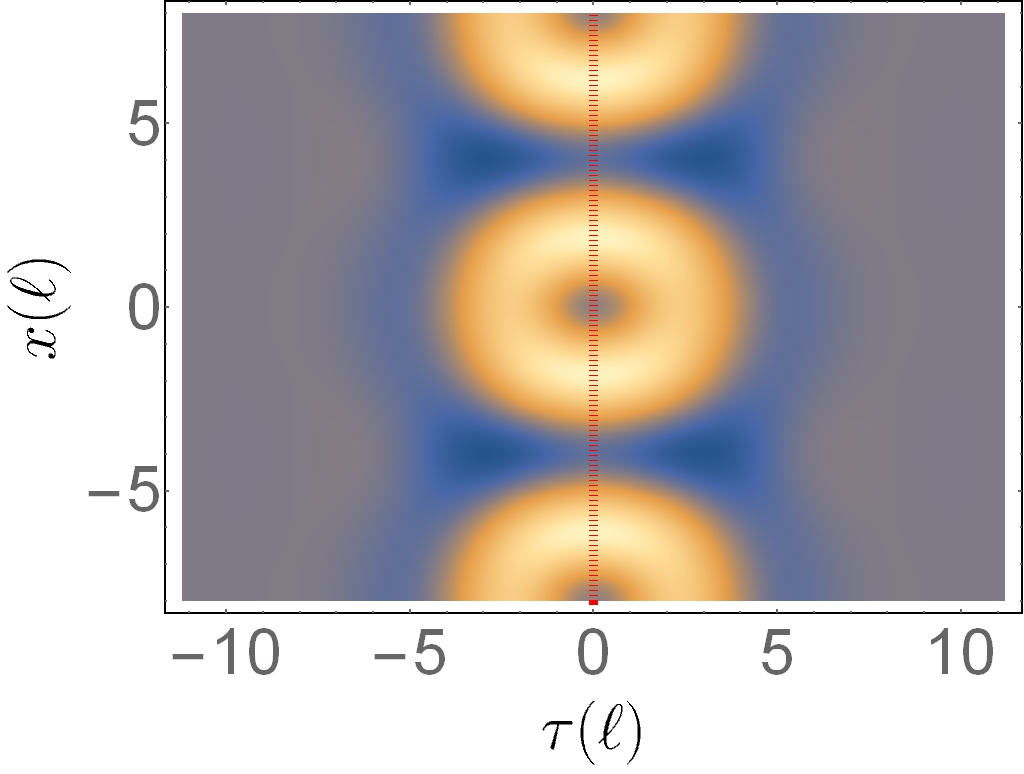}
		\includegraphics[height=3.7cm]{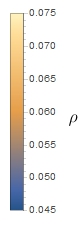}		
	\end{center}
	\caption{Cross-section of the $q=3$ Laughlin state on an infinite cylinder with the quasiparticles at the expected positions. Upper left panel: two quasielectrons and one quasihole, all at their expected position (blue solid line). Single unshifted quasiparticles are shown as a reference (black dashed lines). The parameters used are $L=20\ell$ and $\tilde{P}_\text{max}=1$. Upper right and lower panels: two quasielectrons at $\tau=0$ and $x_{\xi_1,\xi_2}=0,L/2$ using the screened operators with added phases (blue solid line in the upper right panel and lower middle panel) and using the unscreened operators (red dotted line in the upper right panel and lower right panel). As a comparison (see the main text), we plot the profile of three quasielectrons at $x=0$ and $\tau_{\xi_1,\xi_2,\xi_3}=-L/2,0,L/2$ (black dashed line  in the upper right panel and left lower panel). The parameters used are $L=16\ell$ and $\tilde{P}_\text{max}=3$. }
	\label{fig:Noshift}
\end{figure}

An example showing that the screened operators give the desired result is presented in the upper left panel of Fig.~\ref{fig:Noshift}. 
The density profile as a function of $\tau$ at $x=0$ is shown for two quasielectrons with coordinates $(\tau_{\xi_1},x_{\xi_1})=(0,0)$ and $(\tau_{\xi_2},x_{\xi_2})=(12\ell,0)$ and a quasihole at $(\tau_{\eta},x_{\eta})=(-12\ell,0)$ for the $q=3$ Laughlin state. 
All three quasiparticles are located at the expected positions.
In fact, we confirmed that with this prescription, when the $\tau$ separations between the quasiparticles are large, the quasielectrons are always at the expected positions, regardless of how many other quasielectrons or quasiholes are present. 
In addition, the density profiles are also as expected (that is, equal to the density in the absence of other quasiparticles) as long as they are all widely separated in the $\tau$ direction.

However, when quasiparticles are close in $\tau$, even if they are well separated in $x$, they do not have the expected density profile for some configurations. 
In fact, in some cases they are not even cylindrically symmetric, because the screening is only in the $\tau$-direction and the quasielectrons are non-local in the description we use. 
A surprising observation is that the unscreened operators, Eqs.~\eqref{eq:two-fields-qhole} and~\eqref{qe-matrix}, give rise to quasiparticles with reasonably symmetric density profiles, localized at the expected positions, provided that their relative coordinates fulfill $\Delta\tau \approx 0$.

We use this observation to make an {\em ad hoc} modification of the screened quasiparticle operators, so that the resulting operators give rise to quasielectrons with the expected density profile, localized at the expected position, independent of the position of the other quasiparticles.
The original, unscreened quasiparticle operators give rise to the phases
\be{ph-qe}
H_l(\eta)\propto e^{-i\frac{(q-1)(n_{qe}-n_{qh})}{q}\frac{2\pi x_\eta}{L}}\\
E_l(\xi)\propto e^{i\frac{(q-1)(n_{qe}-n_{qh})}{q} \frac{2\pi x_\xi}{L}}\nonumber,
\ee
where $n_{qh}$ and $n_{qe}$ are the number of quasihole and quasielectron operators
at smaller $\tau$. 
These phases are necessary for giving the quasiparticles the expected shape, when other quasiparticles are close by in $\tau$, but their contribution for quasiparticles well separated in $\tau$ amounts to an overall phase of the wave function. 
Note that these phases are absent in the screened quasiparticle operators.
An \emph{ad hoc} addition of them to Eq.~\eqref{sc-qe} localizes the quasielectrons at the expected positions for all configurations  we could test, even when $\Delta\tau\approx 0$. 
It is interesting to note that the only difference between the original, unscreened quasiparticle operators and the screened operators with the ad-hoc phases added, lies in the $\tau$ dependence. The dependence on $x$, via the phases that localize the quasielectrons, is the same. 

To summarize, our prescription for a general Laughlin state, containing quasielectron excitations, requires screening of the quasiparticle operators and an {\em ad hoc} addition of phase factors.
We argue that this prescription gives the correct result for well separated quasiparticles, i.e., $\Delta\tau\gtrsim 15\ell$, because the results are converged already for $\tilde{P}_\text{max}=0$, such that the quasiparticles behave as they do in the TT-limit. For reasonably short $\tau$ separations, $4\ell\lesssim\Delta\tau\lesssim 15\ell$, we can numerically reach convergence in $\tilde{P}_\text{max}$ and find that the quasiparticles have the desired asymptotic shape if they are also well separated in $x$. If they overlap, their shape is, as expected, distorted due to finite distance effects.
For small $\tau$ separations, $\Delta\tau\lesssim 4\ell$, we are not able to fully reach convergence in all of our simulation parameters simultaneously, because one needs rather large circumferences, which require large cut-offs $P_{max}$ and $\tilde{P}_{max}$.
However, for all configurations we simulated, the (non-converged) results were consistent with the expected density profiles.

The simplest such configuration we simulate consists of two quasielectrons, one located at (0,0) and the other at $(0,\frac{L}{2})$. 
A contour plot of the density is shown in the lower middle panel of Fig.~\ref{fig:Noshift}, together
with a cut along the $x$ direction in the upper right panel (solid blue line). 
To verify that the obtained density is reasonable, we compare it with the density profile of an equivalent configuration that is easy to simulate. 
Because of rotational invariance of the underlying quantum Hall liquid and the periodicity along the $x$ direction, this configuration should most closely resemble an (infinite) chain of quasielectrons located at the same $x$ and separated in the $\tau$ direction by $\Delta\tau=\frac L 2$. 
Since we can not simulate an infinite chain of quasielectrons, we instead simulate a chain of three quasielectrons at $x_{\xi_a}=0$ and $\tau_{\xi_1},\tau_{\xi_2},\tau_{\xi_3}=-L/2,0,L/2$, focusing our analysis on the quasielectron in the middle, i.e. at (0,0). 
The later configuration is comparably simple to simulate and the data converges for the $P_{max}$ and $\tilde{P}_{max}$ that we can reach. 
The corresponding density profile is plotted in the lower left panel of Fig.~\ref{fig:Noshift} and a cut
along the $\tau$ direction (from $\tau=-L/2$ to $\tau=+L/2$) in the upper right panel (dashed black line).
For sake of completeness, we also show the density profile for two \emph{unscreened} quasielectrons, located at (0,0) and $(0,L/2)$ respectively, in the lower right panel of Fig.~\ref{fig:Noshift}, together with a cut along the $x$ direction in the upper right panel (dashed red line). 
As can be seen, the density profile for all three simulations compare reasonably well around $(\tau,x)=(0,0)$ but it is hard to draw strong conclusions (since it is hard to obtain convergence for larger values of $L$). Although the ad-hoc addition of the phases is not satisfactory from a theoretical point of view, we note
that it turned out to be hard to find a description of the quasielectrons with both the correct topological properties, and  the expected density profile in the case of quasiparticles with similar $\tau$ coordinates. 

\begin{figure}[t]
	\begin{center}
		\includegraphics[width=7.5cm]{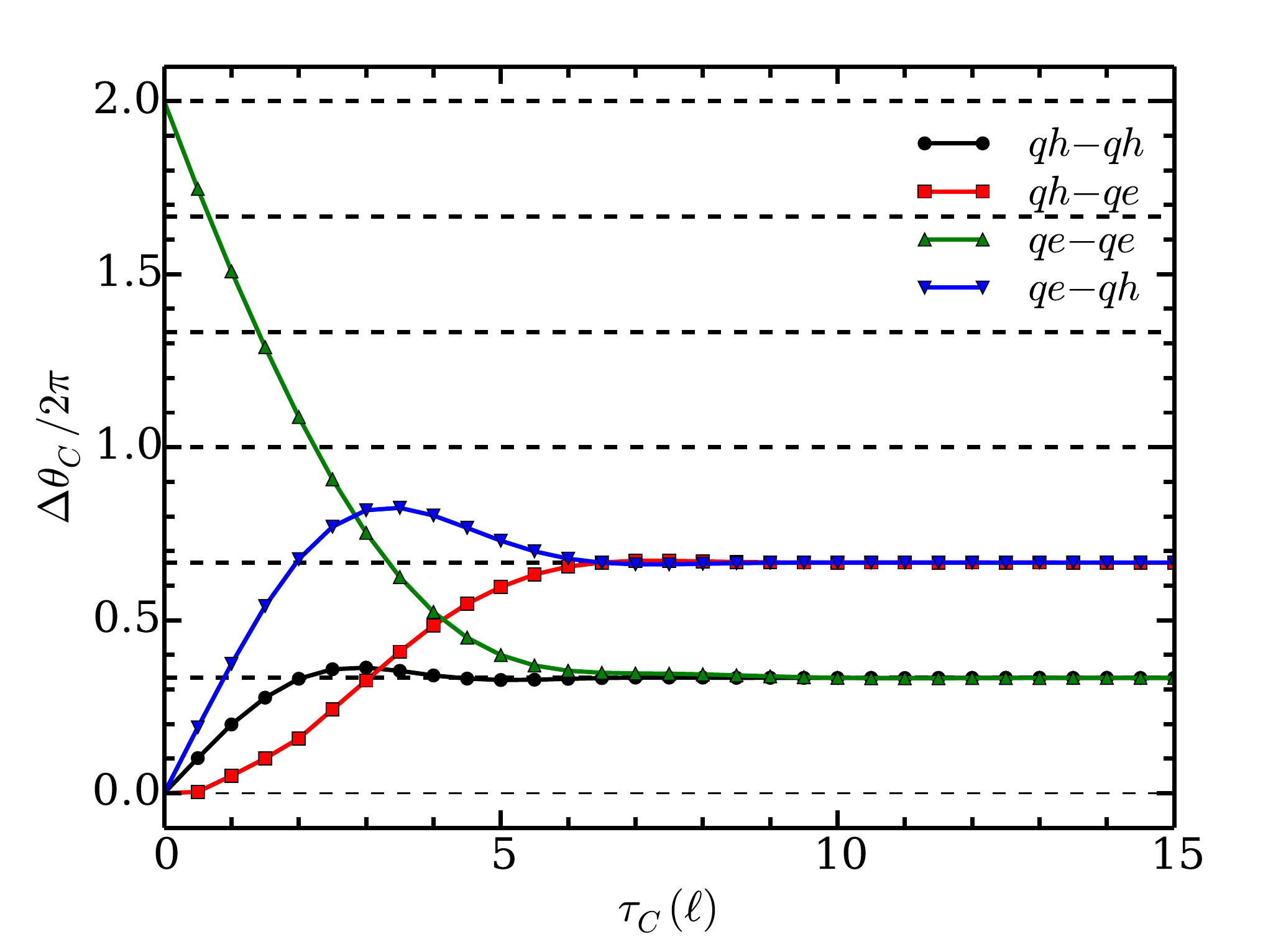}
	\end{center}
	\caption{The statistical phases $\Delta\theta_C/(2\pi)$ for the four ways $q=3$ Laughlin quasielectrons and quasiholes can be braided around each other as a function of $\tau_C$, for the screened quasiparticle operators with the additional phase. The calculations are performed on an infinite cylinder with circumference $L=16\ell$. The data is converged in $P_\text{max}$ and $\tilde{P}_\text{max}$ for all data points, except those with quasielectron(s) and $\tau_C\lesssim 4\ell$, where $\tilde{P}_\text{max}=5$ is used ($\tilde{P}_\text{max}=3$ for the qe-qe case).
	}
	\label{fig:Phase}
\end{figure}

We conclude this subsection by calculating the statistical phases for the quasiparticles with the operators Eq.~\eqref{sc-qe} and the ad hoc phases Eq.~\eqref{ph-qe} added.
In Fig.~\ref{fig:Phase} we show the statistical phase as a function of $\tau_C$, along the closed curve discussed in Section~\ref{sec:numimp}, for the four different ways quasiparticles can be braided around each other. As for the density profiles, we have been able to reach convergence in $P_\text{max}$ and $\tilde{P}_\text{max}$ for all data points, except for the cases with quasielectron(s) at $\tau_C\lesssim 4\ell$. All curves are smooth and for sufficiently large $\tau_C$, all four braiding phases converge to the expected statistical phases, obtained from analytical arguments.

\subsection{Discussion and a proposal}

The reason for spending so much time on the problem related to the shift in the quasielectron positions is it actually 
pinpoints a conceptual problem with the proposed composite fermion/CFT fermion quasielectron wave functions.
In the CFT incarnation, the  clearest way to state the  problem is that the naive notion of localization implied by the exponential function in the kernel Eq.~\eqref{eq:kernel-cylinder-real-space}, or equivalently, the projection with the holomorphic delta function centered around $\xi$,
is simply not correct. 
In the composite fermion picture, it  means that one can not think of the composite fermions as weakly interacting particles%
\footnote{%
There are other, more well-known difficulties with interpreting composite fermions in the QH states as quasiparticles; superficially they are, in accordance with their name, fermions, but the long range statistical interactions in fact makes them into anyons.
}.

As we saw in the previous section, we can solve the problem by using screened operators, which yield quasiparticles with the correct topological properties, and the expected density profiles when they are far apart. We did not find a CFT description that also gives `good' density profiles for the quasiparticles when they are close.
Having said that, we should also stress that as long as we consider only ground states and quasihole states, the CFT description is fully consistent for the chiral hierarchy. 

Given this shortcoming of the composite fermion/CFT quasielectrons, it is worthwhile to try to find other approaches to the wave functions with few quasielectrons. 
For this, we propose  a variation of the quasielectron wave functions originally proposed by Laughlin~\cite{laugh83}. 
His suggestion was to create a quasielectron at position $\xi$ in a $\nu = 1/q$ Laughlin state by inserting the factor
\be{laughlinqel}
\prod_i (2\partial_{z_i} - \bar\xi) \, .
\ee
For $\xi = 0$, \ie for a quasielectron at the origin, it is easy to see that this amounts to moving all electrons one orbital towards the center, thus creating an excess charge of $e/q$. 
If boundary effects can be ignored, the same is true for a quasielectron at an arbitrary position. 
The Laughlin quasielectron wave functions have been studied both analytically~\cite{kjonsbergM1999} and numerically~\cite{jeon03,jeon04,kjonsbergL1999,Jeon2010thermodynamic}, and show two important features that \emph{differ} from the composite fermion/CFT wave functions.  
First, the Laughlin quasielectron is localized at the correct position independent of the presence of other quasiparticles, i.e. there is no spurious shift of the position.
Secondly, care has to be taken when calculating the statistics, in that the braiding phases are well defined only for paths that encircle an area that is small in the sense of enclosing $O(1)$ of the $N_e$ particles in the system~\cite{Jeon2010thermodynamic}.
Thus, this version of the quasielectron also does not have the right topological properties.
The braiding phases of the {\em properly screened} composite fermion/CFT quasielectrons, on the other hand, converge to the
expected value for any path that keeps the particles sufficiently far apart, as we showed in the previous section. 

We now consider the methods developed in Refs.~\onlinecite{svh11a,svh11b} to find a CFT version of the Laughlin quasielectron wave functions. To do so, we use the following modified wave function for the $\nu = 1/q$ Laughlin state%
\footnote{On the disk, the modified Gaussian factor
amounts to a choice, but it corresponds to the natural form
for the wave functions on the sphere and the torus.}:%
\be{laughlinmod}
\Psi_{L,qh}(z_1\dots z_N) &=& {\mathcal P}_{LLL} \prod_{i<j}^N |z_i-z_j|^2 (z_i-z_j)^q  e^{-\frac{1}{4\ell^2}\frac{q+2}{q} \sum_j |z_j|^2}  \\
&=&  {\mathcal P}_{LLL} \av{  {\mathcal O}_{\rm bg}  \prod_{i=1}^N  V(z_{i},\bar z_i)} \nonumber \, ,
\ee
where ${\mathcal P}_{LLL}$ projects on the lowest Landau level, and the modified vertex operator describing the electrons is given by,
\be{antiholop}
V(z_{i},\bar z_i) = :e^{i \sqrt{q+1} \varphi (z)}: :e^{i \bar\chi (\bar z)}: \ ,
\ee
with $\bar\chi$ an anti-holomorphic scalar field. 
States of this type were originally proposed by Girvin and Jach~\cite{girvin}.
Note that the only difference between Eq.~\eqref{laughlinmod} and the original Laughlin state is that the electrons are pushed further away from each other while all phases remain the same.
Wave functions constructed from correlators of vertex operators with both holomorphic and anti-holomorphic components were discussed in some detail in Refs.~\onlinecite{svh11a,svh11b,hhsv} where it was proposed that they have good topological properties in the sense that braiding phases can be read directly from the monodromies just as for fully chiral states. The present case is the simplest example of such a wave function. 

It is  easy to construct a CFT version of a Laughlin quasielectron by inserting the \emph{local} operator 
$$P_{loc} (\xi)= :e^{i\bar\chi (\bar \xi) }: $$
in the correlator in \eqref{laughlinmod}. 
After projection it will yield the Jastrow-type factor \eqref{laughlinqel}.
Note that the field $\bar\chi$ is not properly screened. 
Even though the charge associated with $\bar\chi$ does fluctuate, these fluctuations are not independent of the charge fluctuations of the field $\varphi$: they are coupled via the electron operator~\eqref{antiholop}.
Nevertheless, we still expect the density profile of the quasielectron created by $P_{loc}(\xi)$ to be centered around $\xi$.
The argument is the same as the one used in Sec.~\ref{sec:plasma} to explain why the quasihole operator creates a quasihole that is located at the expected position, despite the fact that it involves the unscreened field $\tilde\varphi$. 
Namely, the operator is both local, and does not involve operators describing the  itinerant electrons. 
Thus, it seems likely that the corresponding quasielectron shows the expected braiding phase for any braiding path that keeps the quasiparticles sufficiently far apart. 
In other words, it should not show the $O(1/N_e)$ contributions to the braiding phase that plague Laughlin's original construction~\cite{kjonsbergL1999}.  

Note that we pay a rather high price for being able to write a local operator for the quasielectron. We had to introduce a new independent field $\bar\chi$. One should also be aware that it is highly nontrivial to carry out the LLL projection required in Eq.~\eqref{laughlinmod}, although some promising methods have been developed recently by Fremling \emph{et al.}~\cite{fremling}.


\section{Summary and outlook}
\label{sec:summary}
In this article we derived and implemented quasielectron excitations of the Laughlin state in the MPS formalism~\cite{zm}.
The presence of quasielectrons leads to several complications compared to the case with quasiholes only.
First, due to the non-local nature of the quasielectron operator, the MPS description of quasielectrons cannot be made
site-independent. Since the quasielectrons are obtained by a modification of the electron operators, even the matrices
corresponding to the electrons become site-dependent.
The additional chiral boson field $\tilde\varphi$ that is needed to describe systems with multiple quasiparticles substantially increases the
matrix size, in particular when quasiparticles are in close proximity.
In addition, for several technical reasons such as the need to anti-symmetrize the quasielectrons, the bond dimension 
has to be increased even further.
Regardless of all these complications, the MPS formalism is still the best numerical method available for systems with
quasielectrons, both in terms of  accuracy and accessible system sizes.

When studying the properties of the quasielectrons we observe shifts in their position depending on whether other
quasiparticles are present or not, even when these quasiparticles are far from the quasielectrons.
These shifts are due to a fundamental problem in the construction of the quasielectron, that is, the lack of screening
of the charge of the additional field. Due to these shifts, the calculated statistical phases are incorrect.
With insights from the thin-cylinder limit, we found a screening prescription for the quasihole and quasielectron
operators that works for quasiparticles that are well separated in the direction along the cylinder.
With this modification, the correct statistical phases are obtained. 
However, with the screening in place, the quasielectrons do not have the right shape in the presence of other
quasiparticles that are separated only in the direction around the cylinder.
This problem can be cured by an ad-hoc modification that numerically works well for all configurations we
could test. This modification does not affect the states with quasiparticles that have a large separation along
the cylinder, so also the modified quasielectron has the correct topological properties.

Our MPS description for states with quasielectrons in the Laughlin state opens up various possibilities for
future studies.
First of all, it stresses the importance of having a full analytic understanding of how to 
screen the quasiparticles in arbitrary configurations. It would be interesting to try to extend the
MPS formalism to describe the Girvin-Jach modification of the Laughlin state, since we argued that 
quasielectrons in this state should have the correct topological properties. This can be properly tested with an MPS description at hand, 
and it would also be very interesting to investigate and compare the entanglement spectra of both formulations in detail. 
The quasielectron states also form a stepping stone towards the description of the hierarchical/Jain composite
fermion states. Since in this case, the charge associated with the additional field is properly fluctuating, the
quasiholes over these state will be located at the correct  position. This will
make it possible to check the statistical properties of quasiparticles over the $\nu=2/5$ composite fermion state with high numerical accuracy. 
\\
\\
{\em Acknowledgements --}
We would like to thank J.~Dubail, B.~Estienne, M.~Fremling, T.~Kvorning, R.~Mong, N.~Regnault and M.~Zaletel for
stimulating and insightful discussions. The research of E.A., T.H.H. and
J.K. was sponsored, in part, by the Swedish research council.
V.D. was funded by the German Research Foundation (DFG) within the CRC
network TR 183 (project B03).
M.H. was partially funded by Emmy Noether Grant HE 7267/1-1 of the
German Research Foundation (DFG).


\appendix

\section{The chiral boson CFT}
\label{app:chiralboson}

We collect the properties of the chiral boson CFT that we use in the paper. 
We refer to Refs.~\onlinecite{bpz84,book:fms99} for more information about
CFT. 
The Laughlin wave function can be written as a correlator of the chiral part, $\varphi(z)$, of a massless bosonic field $\varphi$,
defined by the action
\begin{equation}
S = \frac{1}{8\pi} \int d^2 x \partial_\mu \varphi \partial^\mu \varphi \ .
\end{equation}
The two-point function is given by
\begin{equation}
\langle \varphi (z_1) \varphi (z_2) \rangle = - \log (z_1 - z_2) \ .
\end{equation}
The chiral boson can be expanded in modes as
\begin{equation}
\varphi(z) = \varphi_0 - i \pi_0 \ln (z)  + i \sum_{n\neq 0} \frac{a_n}{n} z^{-n} \ ,
\end{equation}
with the non-trivial commutators
\begin{align}\label{eq:App_com}
[ \varphi_0 , \pi_0 ] &= i & [ a_n, a_m ] = n \delta_{n+m,0} \ , 
\end{align}
while all the other commutators vanish. 
The modes fulfill $a_{-n} = a^\dagger_n$,  and we define $a_{-n}$ with $n>0$ as the creation operators of the non-zero modes.

For a compactified boson with radius $R = \sqrt{q}$, we define charge states as
\begin{align}
\pi_0 | Q \rangle &= \frac{Q}{\sqrt{q}} | Q \rangle &
a_n | Q \rangle &= 0 \qquad (n>0) \ .
\end{align}
The operator $e^{i \beta \varphi_0}$ adds a charge $\beta \sqrt{q}$,
\begin{equation}\label{eq:appCharge}
e^{i \beta \varphi_0} | Q \rangle = | Q + \beta\sqrt{q} \rangle \ ,
\end{equation}
so in particular $| Q \rangle = e^{i \frac{Q}{\sqrt{q}} \varphi_0} | 0 \rangle$, where
$Q$ is an integer and $| 0 \rangle$ denotes the vacuum, i.e. $\pi_0 | 0 \rangle = a_n | 0 \rangle = 0$ for $n>0$.

General states are obtained by acting with the non-zero modes on the states $| Q \rangle$. 
These states are labeled by the integer charge $Q$ and the occupation numbers of the non-zero modes $m_j$, with $j>0$. 
Alternatively, the occupation numbers can be combined into a partition.
A partition $\mu$ of a positive integer $P = |\mu| = \sum_{j>0} j m_j$ is a set of  weakly decreasing integers, $\mu = (\mu_1,\mu_2,\ldots,\mu_l)$, such that $\sum_{i = 1}^{l} \mu_i = |\mu|$. 
The $\mu_i$ are called the parts of the partition, and the orbital occupation number $m_j$ is the number of parts of $\mu$ that are equal to $j$. 
Because the momentum $P$ plays an important role, we label the states as $| Q, P,\mu \rangle$, even though $P$ is fixed by $\mu$, and write
\begin{equation}
| Q , P , \mu \rangle = \frac{1}{\sqrt{z_\mu}} \prod_{j=1}^{\infty} a_{-j}^{m_j} | Q \rangle \ .
\end{equation}
The normalization, $z_\mu = \prod_{j = 1}^{\infty} (j)^{m_j} (m_j !)$, can be computed straightforwardly using the commutators~\eqref{eq:App_com}.
In the MPS matrices, we often need to evaluate
$L_0 = \frac{1}{2}\pi_0^2 + \sum_{j>0} a_{-j} a_{j}$, 
whose action on the states only depends on $Q$ and $P$, namely $L_0 | Q, P, \mu \rangle = (\frac{Q^2}{2q} + P) | Q, P, \mu \rangle$.

Finally, we state the matrix elements of the normal ordered vertex operators $:e^{i \beta \varphi (z)}:$.
We define the normal ordering as
\begin{equation}
:e^{i \beta \varphi (z)}: =
e^{-\beta \sum_{n<0} \frac{a_n}{n}z^{-n}}
e^{i \beta \varphi_0}
e^{\beta \pi_0 \ln (z)}
e^{-\beta \sum_{n>0} \frac{a_n}{n}z^{-n}} \ .
\end{equation}
Again using~\eqref{eq:App_com}, it is straightforward to evaluate the matrix elements
$\mathcal{M} = \langle Q', P', \mu' | :e^{i \beta \varphi (z)}: | Q, P, \mu \rangle$, 
\begin{align}
\mathcal{M} &= \langle Q', P', \mu' | :e^{i \beta \varphi (z)}: | Q, P, \mu \rangle =
\delta_{Q',Q + \sqrt{q} \beta} z^{\frac{\beta Q}{\sqrt{q}}+P'-P} A^{\beta}_{\mu',\mu} \ ,
\end{align}
where 
\begin{align}
A^{\beta}_{\mu',\mu} &=
\prod_{j=1}^{\infty} \sum_{s=0}^{m'_j} \sum_{r=0}^{m_j}
\delta_{m_j-r,m'_j-s}\frac{(-1)^r}{\sqrt{r! s!}} \Bigl(\frac{\beta}{\sqrt{j}}\Bigr)^{r+s} \sqrt{\binom{m'_j}{s}\binom{m_j}{r}} \ .
\label{eq:Avalues}
\end{align}


\section{The matrices for the `polynomial' version of the states}
\label{app:matrices-polynomial}

It is useful to have an MPS representation of the polynomial part of the Laughlin state, in particular
to check the MPS description against the explicit Slater coefficients, which can be obtained by
expanding the wave function for a small number of electrons.
The matrix elements can be obtained from those for the cylinder by taking the limit $L \rightarrow \infty$,
but it is instructive to derive the MPS
description of the polynomial part of the Laughlin state directly. In this appendix, we follow Ref.~\onlinecite{estienne-long}.

Our starting point is the observation that the polynomial part of the Laughlin
wave function can be written as the CFT correlator
\begin{equation}
\label{eq:laughlin}
\Psi_{L,{\rm pol}} = \prod_{i<j} (z_i - z_j)^q = \langle 0 | e^{-i N_e \sqrt{q} \varphi_0} V(z_{N_e}) \cdots V(z_1) | 0 \rangle \ ,
\end{equation}
where the operator $V(z) = :e^{i \sqrt{q} \varphi (z)}:$ creates an electron at position $z$. The
operator $e^{-i N_e \sqrt{q} \varphi_0}$ ensures that the correlator is charge neutral. In
appendix~\ref{app:chiralboson}, we provide the details of the CFT associated with the chiral
boson field $\varphi(z)$.

The MPS technique provides an expression for the expansion coefficients $c_{\lambda}$ of the Laughlin
wave function in terms of Slater determinants, $\Psi_{L,{\rm pol}} = \sum_{\lambda} c_{\lambda, {\rm pol}} \, {\rm sl}_\lambda$,
where the partitions $\lambda = (l_{N_e},\ldots,l_2,l_1)$ encode which single-particle orbitals are
occupied for a given Slater determinant. Thus, the $l_i$ are all distinct, and are ordered as
$0\leq l_1 < l_2 < \cdots < l_{N_e} \leq N_\phi = q (N_e -1)$,
where $N_\phi$ is the highest power of any of the $z_{i}$ in Eq.~\eqref{eq:laughlin}.
In the MPS formulation, there is a matrix associated with each orbital. One therefore needs
to express the correlator in Eq.~\eqref{eq:laughlin} in terms of orbitals, rather than the
positions of the electrons. This can be achieved by (Fourier) expanding the operators $V(z)$ in modes,
\begin{align}
\label{eq:modes}
V (z) &= \sum_{l \in \mathbb{Z}} z^{l} V_{-h-l} &
V_{-h-l} = \frac{1}{2\pi i} \oint \frac{dz}{z} z^{-l} V(z) \ ,
\end{align}
where $h$ is the scaling dimension of $V(z)$, namely $h = \frac{q}{2}$ and the contour is around $z=0$.
This results in the following
expression for the $c_{\lambda, {\rm pol}}$
\begin{equation}
c_{\lambda,{\rm pol}} =  \oint \prod_{j=1}^{N_e} \frac{dz_j}{(2\pi i) z_j} z_j^{-l_j} \Psi_{L,{\rm pol}} =
\oint \prod_{j=1}^{N_e} \frac{dz_j}{(2\pi i) z_j} z_j^{-l_j}
\langle 0 | e^{-i N_e \sqrt{q} \varphi_0} V_{-h-l_{N_e}} \cdots V_{-h-l_1} | 0 \rangle .
\end{equation}
By inserting a complete set of states between the modes, one obtains an MPS expression for
the coefficients $c_{\lambda, {\rm pol}}$. This complete set of states forms a basis for the Hilbert space
associated with the chiral boson CFT (see Appendix~\ref{app:chiralboson}).

In the present form, the matrices one would obtain for the electrons depend on the orbital the electron in question
occupies.
As was observed by Zaletel and Mong~\cite{zm} (see also Estienne et al.~\onlinecite{estienne-long}), it is however
possible to obtain a site-independent MPS. We start by noting that
$V_{-h-l} = e^{i l/\sqrt{q} \varphi_0} V_{-h} e^{-i l/\sqrt{q} \varphi_0}$, which follows from the
commutation relation
$e^{i \beta \varphi_0} :\! e^{i \alpha \varphi(z)}\!\!: \, = z^{-\alpha \beta} :\!e^{i \alpha \varphi(z)}\!\!: e^{i \beta \varphi_0}$
and the definition of the mode expansion Eq.~\eqref{eq:modes}.
The resulting expression for the $c_{\lambda,{\rm pol}}$ is
\begin{equation}
c_{\lambda,{\rm pol}} =
\langle q-1 |
e^{-i(N_\phi+1-l_{N_e})\varphi_0/\sqrt{q}} V_{-h} e^{-i(l_{N_e}-l_{N_e-1})\varphi_0/\sqrt{q}} 
\cdots
e^{-i(l_2-l_1)\varphi_0/\sqrt{q}} V_{-h} e^{-il_1\varphi_0/\sqrt{q}} | 0 \rangle \ ,
\end{equation}
where $\langle q-1| = \langle 0| e^{-i(q-1)\varphi_0/\sqrt{q}}$.
We can now `spread out' the operators of the form $e^{-i(N_\phi+1-l_{N_e})\varphi_0/\sqrt{q}}$ in such
a way that we associate the operator $e^{-i\varphi_0/\sqrt{q}}$ with each empty orbital and the operators $e^{-i\varphi_0/\sqrt{q}} V_{-h}$ with each filled orbital. The charge quantum number of the out state ensures that there are exactly $N_e$ occupied orbitals.
What remains to be done is to calculate the matrix elements of the
operators associated with empty and occupied orbitals.

By using the relations stated in Appendix~\ref{app:chiralboson}, we find the matrix elements
for an empty orbital
\begin{equation}
\label{eq:b0pol}
B_{\rm pol}^{[0]} = 
\langle Q',P',\mu' | e^{-i\varphi_0/\sqrt{q}} | Q,P,\mu\rangle = \delta_{Q',Q-1}\delta_{P',P}
\delta_{\mu',\mu} \ .
\end{equation}
The matrix elements for an occupied orbital are obtained as follows
\begin{align}
\label{eq:b1pol}
B_{\rm pol}^{[1]} &=  \langle Q',P',\mu' | e^{-i\varphi_0/\sqrt{q}} V_{-h} | Q,P,\mu\rangle \nonumber \\
&=
\frac{1}{2\pi i}\oint \frac{dz}{z}  \langle Q',P',\mu' | e^{-i\varphi_0/\sqrt{q}} V(z) | Q,P,\mu \rangle 
\nonumber \\
&=
\frac{1}{2\pi i}\oint \frac{dz}{z}  \delta_{Q'+1,Q+q}\, z^{Q+P'-P} A^{\sqrt{q}}_{\mu',\mu} =
\delta_{Q',Q+q-1} \delta_{P',P-Q} A^{\sqrt{q}}_{\mu',\mu} \ ,
\end{align}
with $A^{\sqrt{q}}_{\mu',\mu}$ given by Eq.~\eqref{eq:Avalues}.
We can now calculate the Slater coefficients $c_{\lambda,{\rm pol}}$ for the polynomial part of the Laughlin state
in the MPS formulation, for instance%
\footnote{We remark that finding explicit expressions for $c_{\lambda,{\rm pol}}$ with general $\lambda$ is a hard problem!}
\begin{equation}
c_{(N_\phi,\ldots,3,0),{\rm pol}} =  \langle q-1| B_{\rm pol}^{[1]}\cdots B_{\rm pol}^{[1]}B_{\rm pol}^{[0]}B_{\rm pol}^{[0]}B_{\rm pol}^{[1]} | 0 \rangle = 1 \ .
\end{equation}

Before discussing quasiholes, we first comment on the connection between the
MPS for the polynomial and cylinder wave functions. From the single-particle wave functions
on the cylinder, we find that the Slater coefficients $c_{\lambda}$ of the cylinder wave functions are
related to the ones for the polynomial part of the wave functions, $c_{\lambda,{\rm pol}}$ as
\begin{equation}
c_{\lambda} = c_{\lambda,{\rm pol}} \prod_{j} \mathcal{N}_{l_j} \ ,
\end{equation}
where $\mathcal{N}_l = \sqrt{L \ell \sqrt{\pi}} e^{+\frac{1}{2\ell^2}\tau_l^2}$, with
$\tau_l = \frac{2\pi \ell^2 l}{L}$ being the location of the center of the $l$th orbital.

By comparing the matrix elements for the polynomial part of the wave function,
Eqs.~\eqref{eq:b0pol} and \eqref{eq:b1pol} with the ones relevant for the cylinder,
Eqs.~\eqref{noel} and \eqref{elpres}, one observes that the only difference
lies in the exponential factors present in the cylinder matrices, which
take into account the background charge.
It is a useful exercise to explicitly
calculate the total effect of these exponential factors
$e^{- \frac{2\pi }{L} \delta\tau \bigl( L_0 + \frac{1}{2\sqrt{q}} \pi_0 + \frac{1}{6q} \bigr)}$,
which are present at each orbital. 
One finds that 
\begin{equation}
-\sum_{j = 1}^{N_\phi+1} \Bigl( L_{0,j} + \frac{1}{2\sqrt{q}}\pi_{0,j} +\frac{1}{6q} \Bigr) =
-\frac{n^3}{6q} +\frac{n}{2}(n-q) \sum_{l=0}^{N_\phi} p_l
-\frac{1}{2} (2n-q) \sum_{l=0}^{N_\phi} l p_l
+\frac{1}{2} \sum_{l=0}^{N_\phi} l^2 p_l \ .
\end{equation} 
Here, the $p_i$'s are the fermion occupation numbers of the orbitals and
$n=N_\phi +1$ denotes the number of orbitals. 
We have $\sum_{l=0}^{N_\phi} p_l = N_e$, as well as
$\sum_{l=0}^{N_\phi} l p_l = \frac{q}{2} N_e (N_e-1)$, which is the total angular momentum
of the droplet (which is constant, because we do not consider quasiholes).
Apart from constant factors and terms that only depend on the number of electrons, we indeed find
that the effect of the exponential factors in the cylinder matrices gives the right
contribution, namely $e^{\frac{1}{2\ell^2}(2\pi \ell/L)^2 \sum_l l^2 p_l}$. This factor
comes solely from the $L_0$ part of the exponential in the evolution operator $U''$, see
Eq.~\eqref{eq:time-evolution}.

As already pointed out by Zaletel and Mong, states with
quasiholes can also be written as a matrix product state.
For completeness, we give the form of the Laughlin wave function in the presence of
$N_{qh}$ quasiholes,
\begin{equation}
\label{eq:Lqh}
\Psi_{L,qh,{\rm pol}} =
\prod_{\alpha<\beta} (\eta_\alpha-\eta_\beta)^\frac{1}{q}
\prod_{\alpha,i}(\eta_\alpha-z_i)
\prod_{i<j}(z_i-z_j)^q \ .
\end{equation}

To obtain the CFT correlator that describes a state with quasiholes, one inserts
the quasihole operators, $H (\eta) = :\! e^{i/\sqrt{q}\varphi(\eta)}\!\! :$, where $\eta$ is the
location of the quasihole, into the correlator. The radial ordering of the correlator fixes
the effective position of the quasihole operator. At the level of the MPS, one can actually
choose the points at which  one inserts the matrix corresponding to the quasihole and as we see below,
the matrix elements depend on this choice.

The matrix elements of the quasihole operators are easily obtained.
The labels of the orbitals are $0,1,\ldots, N_\phi$, and we denote the position of the quasihole
operator by $l$ if it is inserted between the operators corresponding to orbitals $l-1$ and $l$.
Because of the spread-out back-ground charge, we find that we actually need to
calculate the following matrix elements
\begin{align}
\langle Q',P',\mu' | e^{-i l/\sqrt{q}\varphi_0} :\! e^{i/\sqrt{q} \varphi(\eta)} \! \! : e^{i l/\sqrt{q}\varphi_0} | Q,P,\mu\rangle &= 
\eta^{l/q} \langle Q',P',\mu' | :\! e^{i/\sqrt{q} \varphi(\eta)} \! \! :  | Q,P,\mu\rangle  \nonumber \\ &
= \eta^{(Q+l)/q + P' -P}  \delta_{Q',Q+1} A^{(1/\sqrt{q})}_{\mu',\mu} \ .
\end{align}
We note that there is no $\delta$-function relating $P'$ and $P$, which for the electron operators arises
from the contour integration that picks up the appropriate mode.

The electron operator $V(z)$ and the quasihole operator $H (\eta)$ anti-commute, which is reflected in
the anti-symmetric factor $(z-\eta)$ that is present in the wave function, see Eq.~\eqref{eq:Lqh}.
Therefore, we have to introduce an additional sign in the
matrices for the quasiholes. This sign keeps track of how many matrices corresponding to occupied orbitals
already acted at the point where one acts with the matrix corresponding to the quasihole.
We denote this sign by $(-1)^{\#V}$. This information can be obtained from the quantum number $Q$ and the
position of the quasihole operator.
The quantum number $Q$, at the position of the matrix for quasihole with number $\alpha$ (i.e.,
$\alpha -1$ quasiholes are already inserted) is given by 
$Q = - l + q (\#V) + (\alpha - 1)$, where we assumed that the charge of the in-state is zero. The term $-l$ comes from the
spread-out background charge. This leads to the sign $(-1)^{(Q+l-(\alpha-1))/q}$.
Putting everything together, the elements of the matrix describing the $\alpha^{\rm th}$ quasihole, inserted at
position $l$, read
\begin{equation}
\label{eq:qh-one-field}
H_{l,\alpha,{\rm pol}} (\eta_\alpha) = (-1)^{(Q+l-(\alpha-1))/q} \eta_\alpha^{(Q+l)/q + P' -P}  \delta_{Q',Q+1} A^{(1/\sqrt{q})}_{\mu',\mu}
\ .
\end{equation}


\section{MPS for the polynomial part of the angular momentum quasielectrons}
\label{app:mps-pol-ang-qe}

In this appendix, we show how to obtain the matrix elements appearing in the
MPS expression for the polynomial version of the angular momentum quasielectron states.
We start by considering the wave functions for an arbitrary number of quasiholes
and angular momentum quasielectrons. In the case of finite systems, which we also consider,
it is important to use states that are valid on the sphere (up to the single-particle normalization
factors). These states can be obtained from the techniques presented in Ref.~\onlinecite{kvorning13},
which we review in App.~\ref{app:sphereqe}.

We assume that there are $N_e$ electrons, $N_{qh}$ quasiholes, and $N_{qe}$ quasielectrons.
We see below that the number of quasielectrons that is possible depends on both $N_e$ and $N_{qh}$.
The number of modified electron operators is $N_{qe}$, which already implies that $N_{qe} \leq N_e$,
but there are more constraints.
We start by defining the `relative' part of the wave function, that one obtains by simply using
the primary vertex operators i.e., without taking the derivatives into account (see below)
\begin{equation}
\label{full-rel}
\psi_{\rm rel} =
\prod_{\alpha<\beta} (\eta_\alpha - \eta_\beta)
\prod_{i<j} (z_i-z_j)^{q}
\prod_{a<b} (z_a-z_b)^{q-1}
\prod_{i,a} (z_i - z_a)^{q-1}
\prod_{i,\alpha}(z_i - \eta_\alpha) \ ,
\end{equation}
where the coordinates of the `modified' electron operators are $z_1,\ldots, z_{N_{qe}}$, while the remaining coordinates
$z_{N_{qe}+1}, \ldots, z_{N_e}$ correspond to `normal' electron operators. The locations of the quasiholes are
$\eta_1,\ldots,\eta_{N_{qh}}$. For the electron coordinates $z$, the indices $a,b$ run over $1,\ldots,N_{qe}$, while $i,j$
run over $N_{qe}+1,\ldots,N_e$. For quasiholes, the indices $\alpha,\beta$ run over $1,\ldots,N_{qh}$.

The wave function for the angular momentum quasielectrons 
depends on the $N_{qe}$ distinct angular momenta
$k_a$, which satisfy $k_a \leq N_e + 1 + N_{qh} - N_{qe} = k_{\rm max}$.
In terms of these, the polynomial part of the wave function is given by
\begin{equation}
\label{eq:angmomqe}
\psi^{(k_a)}_{qh,qe,{\rm pol}} (z;\eta) =
\mathcal{A}
\Biggl[
\prod_{a} \Bigl( z_a^{k_a} \partial_{z_a} - \frac{k_a}{k_{\rm max}} (q-1)(N_e-1) z_a^{k_a-1} \Bigr)
\psi_{\rm rel} 
\Biggr]
\ , 
\end{equation}
where the anti-symmetrization is over all electron coordinates.

The wave functions $\psi^{(k_a)}_{qh,qe,{\rm pol}} (z;\eta)$ were in fact obtained from  a CFT correlator,
just as in the case for the Laughlin states without quasielectrons. The operators corresponding
to the electrons, modified electrons and quasiholes are, using the notation
$f_k = \frac{k}{k_{\rm max}} (q-1)(N_e-1)$,
\begin{align}
\label{eq:sphere-vo-el}
V(z) &=\, : e^{i \sqrt{q} \varphi(z)}:\\
\label{eq:sphere-vo-mod-el}
\tilde{V}^k (z) &= (z^k \partial_z - f_k z^{k-1}) \tilde{V}(z), &
\tilde{V}(z) &=\, :e^{i (q-1)/\sqrt{q} \varphi(z)}: \, :e^{-i(q-1)/\sqrt{q(q-1)}\tilde{\varphi}(z)}:\\
\label{eq:sphere-vo-qh}
H (\eta) &=\, :e^{i/\sqrt{q}\varphi(\eta)}: \, :e^{i(q-1)/\sqrt{q(q-1)}\tilde{\varphi}(\eta)}: \ .
\end{align}

Because of the field $\tilde{\varphi} (z)$, the matrix elements of the matrices
corresponding to empty orbitals, and orbitals occupied by `normal' electrons, have to be modified
slightly in comparison to the ones given in App.~\ref{app:matrices-polynomial}.
We allow for arbitrary `in' charges $Q_0$ and $\tilde{Q}_0$ of the `in state', but
note that the matrix elements considered here do not depend on the latter. The matrix elements read
\begin{align}
B^{[0]}_{\rm pol} &= 
\delta_{Q',Q-1}
\delta_{\tilde{Q}',\tilde{Q}}
\delta_{P',P}
\delta_{\tilde{P}',\tilde{P}}
\delta_{\mu',\mu} \delta_{\tilde{\mu}',\tilde{\mu}}
\\
B^{[1]}_{\rm pol} &=
\delta_{Q',Q+q-1}
\delta_{\tilde{Q}',\tilde{Q}}
\delta_{P',P-(Q-Q_0)}
\delta_{\tilde{P}',\tilde{P}}
A^{\sqrt{q}}_{\mu',\mu} 
\delta_{\tilde{\mu}',\tilde{\mu}} \ .
\end{align}

To calculate the matrix corresponding to the modified electron operator, we proceed in the
same way as for the Laughlin state discussed in App.~\ref{app:matrices-polynomial}, but
we have to take into account a few differences.
First of all, in calculating the matrix elements, we need to consider the mode expansions of the
operators. From the definition of the mode expansion, we find that the relation between the modes of
$\tilde{V}^{k}(z)$ and $\tilde{V} (z)$ is given by
$\tilde{V}^{k}_{-h-l} = (l+1-k - f_k) \tilde{V}_{h-l+k-1}$, where in both cases $h$ refers to the scaling
dimension of the corresponding operator. Thus, in calculating the matrix elements of the modified electron
operator, we can replace the modes of the operator $\tilde{V}^k (z)$, which includes the derivative, by the
modes of $\tilde{V} (z)$ without the derivative, provided we include the factor $(l+1-k-f_k)$, where $l$ is the mode
index (i.e., the orbital associated with the operator).

Secondly, we need to consider the effect of spreading out the
background charge, which for the Laughlin state led to site independent matrices.
This followed from the relation $V_{-h-l} = e^{i l/\sqrt{q} \varphi_0} V_{-h} e^{-i l/\sqrt{q} \varphi_0}$
for the modes of $V(z)$. In the case at hand, we consider the modes of $\tilde{V}^{k}(z)$, which read
$\tilde{V}^{k}_{-h-l} = (l+1-k-f_k) \tilde{V}_{h-l+k-1}$.
Because of the difference between the vertex operator for $\varphi(z)$ in $V(z)$ and $\tilde{V}(z)$,
we now obtain the following relation instead
\begin{equation}
\tilde{V}^{k}_{-h-l} =
e^{i l/\sqrt{q}\varphi_0} \tilde{V}^{k}_{-h-l/q} e^{-i l/\sqrt{q}\varphi_0}=
(l+1-k-f_k) e^{i l/\sqrt{q}\varphi_0} \tilde{V}_{-h-l/q+k-1} e^{-i l/\sqrt{q}\varphi_0} \ .
\end{equation}
This means that in order to calculate the matrix elements, we should not use the integral
$\frac{1}{2\pi i} \oint \frac{dz}{z}$ (see Eq.~\eqref{eq:b1pol}), but
$\frac{1}{2\pi i} \oint \frac{dz}{z} z^{-l/q+k-1}$ in order
to pick up the mode of the modified electron at orbital $l$.
The additional factor $z^{-l/q+k-1}$ will change the delta function for the total momentum.
Although the exponent of $z^{-l/q+k-1}$ is in general fractional, combining it with the
factors we obtain from calculating the expectation values of the vertex operators, we find
that the total exponent of $z$ in the integrand is an integer.
The (putative) matrix elements of the modified electron operator at orbital $l$, and for a momentum
$k$ state, become
\begin{align}
&\frac{1}{2\pi i}\oint \frac{dz}{z} (l+1-k-f_k)z^{-l/q+k-1}
\langle Q',P',\mu',\tilde{Q}',\tilde{P}',\tilde{\mu}' |
e^{-i\varphi_0/\sqrt{q}} \tilde{V} (z)
| Q,P,\mu, \tilde{Q}',\tilde{P}',\tilde{\mu}' \rangle 
\\\nonumber
=&
(l+1-k-f_k)
\delta_{Q'+1,Q+(q-1)}
\delta_{\tilde{Q}',\tilde{Q}-(q-1)}
A^{\frac{q-1}{\sqrt{q}}}_{\mu',\mu}
A^{-\frac{q-1}{\sqrt{q(q-1)}}}_{\tilde{\mu}',\tilde{\mu}}  \\\nonumber&
\times \frac{1}{2\pi i}\oint \frac{dz}{z}
z^{P'+\tilde{P}'-P-\tilde{P} + Q - Q_0 + k - 1 - \frac{1}{q} (Q-Q_0 +\tilde{Q} - \tilde{Q}_0 + l)}
\\\nonumber 
=&
(l+1-k-f_k) \delta_{Q',Q+(q-2)}
\delta_{\tilde{Q}',\tilde{Q}-(q-1)}
\delta_{P'+\tilde{P}',P+\tilde{P} -Q+Q_0 - k + 1 + \frac{1}{q} (Q-Q_0 +\tilde{Q}-\tilde{Q}_0 + l)}
A^{\frac{q-1}{\sqrt{q}}}_{\mu',\mu}
A^{-\frac{q-1}{\sqrt{q(q-1)}}}_{\tilde{\mu}',\tilde{\mu}} \ .
\end{align}

So far, we have not yet taken into account that we need to anti-symmetrize the
modified electrons, both with respect to the ordinary electrons and amongst themselves.
In addition, as explained in Sec.~\ref{sec:mps-qes}, we need to make sure that
in the MPS expansion, each term contains one and only one matrix corresponding
to each modified electron operator.
These problems were solved together in Sec.~\ref{sec:mps-qes}
by enlarging the auxiliary Hilbert space (including signs for the anti-symmetrization).
Here, we only discuss the signs that are necessary to take the anti-symmetrization
between the normal and modified electron operators into account.

This can be done by introducing the factor $(-1)^{\#V}$.
Here, $\#V$ denotes the number of ordinary electrons that
were inserted before the modified electron. Alternatively, one can also
use the factor $(-1)^{\#V+\#\tilde{V}}$, where $\#\tilde{V}$ denotes the
number of modified electrons that were already inserted. The difference
between these two prescriptions is merely an overall sign.

To calculate the factors $(-1)^{\#V}$ and $(-1)^{\#V+\#\tilde{V}}$ at orbital $l$, we
consider the quantum numbers $Q$ and $\tilde{Q}$ at that point. They are given by
\begin{align}
Q - Q_0 &= - l + q \#V + (q-1) \#\tilde{V} + \#{\rm qh} ,\\
\tilde{Q} - \tilde{Q}_0 &=  -(q-1) \#\tilde{V} + (q-1)\#{\rm qh} \ ,
\end{align}
where $\#{\rm qh}$ is the number of quasiholes matrices that acted
before orbital $l$. This leads to
\begin{align}
\#V &= \frac{1}{q} \bigl( Q-Q_0 + \tilde{Q}-\tilde{Q}_0 + l \bigr) - \#{\rm qh}, \\
\#V+\#\tilde{V} &= \frac{1}{q}\left( Q-Q_0 + l -\frac{\tilde{Q}-\tilde{Q}_0}{q-1} \right) \ .
\end{align}
We see that it is slightly easier to use $(-1)^{\#V+\#\tilde{V}}$ to perform the
anti-symmetrization of the modified electrons with respect to the ordinary electrons,
because this expression does not depend on the number of quasihole matrices that
already acted.
Putting the results together, we find the following form of the matrix elements
$E_{k,l,{\rm pol}}$ for the modified electron operators at orbital $l$ and angular momentum $k$,
that appear in the enlarged structure for the
total electron operators (compare Eqs.~\eqref{B1-1qe} and \eqref{B1-2qe}) 
\begin{align}
E_{k,l,{\rm pol}} &=
(-1)^{\frac{1}{q}\bigl( Q-Q_0 + l -(\tilde{Q}-\tilde{Q}_0)/(q-1) \bigr)}
(l+1-k-f_k)
\delta_{Q',Q+(q-2)}
\delta_{\tilde{Q}',\tilde{Q}-(q-1)} 
\\\nonumber&\times
\delta_{P'+\tilde{P}',P+\tilde{P} -Q+Q_0 - k + 1 + \frac{1}{q} (Q-Q_0 +\tilde{Q}-\tilde{Q}_0 + l)}
A^{\frac{q-1}{\sqrt{q}}}_{\mu',\mu}
A^{-\frac{q-1}{\sqrt{q(q-1)}}}_{\tilde{\mu}',\tilde{\mu}} \ .
\end{align}

We end this appendix by giving the 
matrix elements for the quasiholes in the formulation that uses the field $\tilde{\varphi}$.
The changes in comparison to Eq.~\eqref{eq:qh-one-field} are as follows:
the exponent of the quasihole position
$\eta$ changes, and the sign necessary to anti-symmetrize the quasiholes with respect to the
`ordinary' electrons is now given by
$(-1)^{\frac{1}{q} \bigl( Q-Q_0 + \tilde{Q}-\tilde{Q}_0 + l \bigr) - (\alpha-1)}$,
for quasihole number $\alpha$. 
The matrix elements for the $\alpha^{\rm th}$ quasihole, with coordinate $\eta$,
inserted at position $l$  (i.e., in between orbitals $l-1$ and $l$) are given by
\begin{align}
\label{eq:qh-two-fields}
H_{l,\alpha,{\rm pol}} (\eta_\alpha) =&
(-1)^{\frac{1}{q} \bigl( Q-Q_0 + \tilde{Q}-\tilde{Q}_0 + l \bigr) - (\alpha-1)} 
\eta_{\alpha}^{P'+\tilde{P}'-P-\tilde{P} + \frac{1}{q}\bigl( Q -Q_0 +\tilde{Q} - \tilde{Q}_0 + l \bigr)}
 \\
\nonumber & \times
\delta_{Q',Q+1}
\delta_{\tilde{Q}',\tilde{Q}+(q-1)}
A^{\frac{1}{\sqrt{q}}}_{\mu',\mu}
A^{\frac{q-1}{\sqrt{q(q-1)}}}_{\tilde{\mu}',\tilde{\mu}} \ .
\end{align}
This concludes the description of the matrix elements for the polynomial part of the wave functions.


\section{The matrices for the full cylinder  states} \label{app:matrices-cylinder}

In this appendix, we discuss the matrix elements for the quasielectron states on the cylinder.
In the main text, we gave the matrix elements for the operators corresponding to empty and
occupied orbitals, in the case where the operators only depended on the field $\varphi$.
Because of the presence of $\tilde{\varphi}$, the matrix elements $B^{[0]}$ and 
$B^{[1]}$ have to be modified. In particular, the $\delta$-functions should be modified, as well
as the time evolution matrix elements, which now
depend not only on $Q'$ and $P'$, but also on the quantum numbers associated with
$\tilde{\varphi}$, i.e. $\tilde{Q}'$ and $\tilde{P}'$, though we note that there is no smeared out
background charge associated with $\tilde{\varphi}$. The matrix elements of the time
evolution become
\begin{equation}
\label{eq:time-evolution-tilde}
U'' = 
e^{-\frac{2\pi \delta\tau}{L} \bigl(\frac{1}{2q} Q'^2 + P' + \frac{1}{2q} Q' + \frac{1}{6q} \bigr)}
e^{-\frac{2\pi \delta\tau}{L} \bigl(\frac{1}{2q(q-1)} \tilde{Q}'^2 +\tilde{P}' \bigr)} \ .
\end{equation}
For the matrix elements corresponding to the empty orbitals and orbitals occupied by the
unmodified electrons, these are the only differences, so that we obtain
\begin{align}
\label{eq:two-fields-empty}
B^{[0]} &=
e^{-\frac{2\pi\delta\tau}{L} \bigl( \frac{(Q')^2}{2q} + P' + \frac{Q'}{2q} +\frac{1}{6q}
+ \frac{(\tilde{Q}')^2}{2q(q-1)} + \tilde{P}' \bigr)}
\delta_{Q',Q-1}\delta_{P',P} \delta_{\mu',\mu}
\delta_{\tilde{Q}',\tilde{Q}} \delta_{\tilde{P}',\tilde{P}} \delta_{\tilde{\mu}',\tilde{\mu}} 
\\
\label{eq:two-fields-electron}
B^{[1]} &=
e^{-\frac{2\pi\delta\tau}{L} \bigl( \frac{(Q')^2}{2q} + P' + \frac{Q'}{2q} +\frac{1}{6q}
+ \frac{(\tilde{Q}')^2}{2q(q-1)} + \tilde{P}' \bigr)}
\delta_{Q',Q+q-1} \delta_{P',P-(Q-Q_0)} A^{\sqrt{q}}_{\mu',\mu}
\delta_{\tilde{Q}',\tilde{Q}} \delta_{\tilde{P}',\tilde{P}} \delta_{\tilde{\mu}',\tilde{\mu}} \ .
\end{align}
Here, we allowed for arbitrary `in' charges $Q_0$ and $\tilde{Q}_0$, though the latter does not
appear in these matrix elements.

To derive the matrix elements for the modified electron operators, a few issues need to be taken
care of. We consider a quasielectron with angular momentum $k$, and assume that the associated
matrix acts at orbital $l$. The associated operator is
$\bigl(\omega^k \partial_{\omega} - f_k \omega^{k-1}\bigr) \tilde{V}(\omega)$. To deal with the term with the derivative,
it is easiest to consider the expressions for the Slater coefficient Eqs.~\eqref{eq:slaters} and
\eqref{eq:c-lambda-corr}. Because the $\varphi$ charge of the operator $\tilde{V}$ is $q-1$ instead of $q$, we
find that we have to evaluate an expression of the form (we concentrate on the modified electron here, and recall
that the integral is to be performed at fixed $\tau$)
\begin{align}
&\int_{-L/2}^{L/2} \frac{dx}{L} e^{i x \frac{2\pi}{L}l}  (\omega^{k} \partial_\omega - f_k \omega^{k-1})
e^{-i x \frac{2\pi}{L}l (1-\frac{1}{q})}
\av{\mathcal{O}_{\rm bg} \cdots \tilde{V}(\omega) \cdots} \\\nonumber
&=
\int_{-L/2}^{L/2} \frac{dx}{L} e^{i x \frac{2\pi}{L}\frac{l}{q}}  (l + 1- k - f_k )\omega^{k-1}
\av{\mathcal{O}_{\rm bg} \cdots \tilde{V}(\omega) \cdots} \ ,
\end{align}
where we have  integrated the first term by parts. In comparison to the case without quasielectrons, we find that there
is an exponent $e^{i x \frac{2\pi}{L}\frac{l}{q}}$ left in the integral. This exponent is necessary, because
when combined with the factors coming from the matrix elements of the operator $\tilde{V}(\omega)$, 
it  ensures that the integral over $x$ is well defined. It is interesting to see that a similar factor appeared in the
calculation of the matrix elements for the polynomial part of the wave function, though there it had a rather
different origin. We note that in the expression above, we dropped a Gaussian factor
$e^{-\tau_l^2/(2 q \ell^2)}$, which is associated with the normalization of the single-particle orbitals. This factor naturally
arises as part of the localizing kernel when we consider localized quasielectrons, as discussed in Sec.~\ref{qeo}.

We now turn our attention to the change in the factors coming from the free time evolution $U''$.
In App.~\ref{app:matrices-polynomial}, we showed explicitly that the contribution from $U''$
precisely gave rise to the exponential factors that are needed for the cylinder wave functions.
Here, we are dealing with the matrix elements for the modified electron operators, which depend
both on the orbital $l$ and the angular momentum $k$. We should thus check if the free time
evolution gives rise to $l$ and $k$ dependent factors, apart from the necessary Gaussian factors
for the cylinder normalization. In the case of the localized quasielectrons, we need to perform
a sum over the angular momenta $k$, so we must make sure that we do not introduce any
spurious $k$ (and $l$) dependent normalization factors. If such factors are present, they should
be corrected for.

For ease of calculation, we consider the case without quasiholes, and find that the time-evolution
does indeed give rise to the Gaussian factors that are necessary for the cylinder normalization. There are,
however, additional contributions and we drop an unimportant factor that only depends on $q$ and the
number of electrons $N_e$. In case the modified electron operator acts at orbital $l$ and
describes a quasielectron with angular momentum $k$, the additional $l$ and $k$ dependent
factor is given by
\begin{equation}
e^{
\frac{2\pi\delta\tau}{L}
\bigl( \frac{l^2}{2q} + \frac{l}{q} (\frac{q}{2} - Q_0 - \tilde{Q}_0) 
-  k \bigl(l - \frac{q}{2} - Q_0)\bigr)} \ .
\end{equation}
In the matrix elements for the modified electron operators, we need to correct for this 
factor. Putting all the pieces together, we obtain the following matrix elements for
the modified electron operators at orbital $l$ and with angular momentum $k$
\begin{align}
\label{eq:btilde}
E_{k,l} &=
(l+1-k-f_{k})
(-1)^{\frac{1}{q}\bigl( Q-Q_0 + l -(\tilde{Q}-\tilde{Q}_0)/(q-1) \bigr)}
\\\nonumber&\times
e^{-\frac{2\pi\delta\tau}{L} \bigl( \frac{(Q')^2}{2q} + P' + \frac{Q'}{2q} +\frac{1}{6q}
+ \frac{(\tilde{Q}')^2}{2q(q-1)} + \tilde{P}' \bigr)}
e^{-\frac{2\pi\delta\tau}{L} \bigl( 
\frac{l^2}{2q} + \frac{l}{q} (\frac{q}{2} - Q_0 - \tilde{Q}_0) - k (l - \frac{q}{2} - Q_0)
\bigr)}
\\\nonumber&\times
\delta_{Q',Q+(q-2)}
\delta_{\tilde{Q}',\tilde{Q}-(q-1)}
\delta_{P'+\tilde{P}',P+\tilde{P} -Q+Q_0 - k + 1 + \frac{1}{q} (Q-Q_0 +\tilde{Q}-\tilde{Q}_0 + l)}
A^{\frac{q-1}{\sqrt{q}}}_{\mu',\mu}
A^{-\frac{q-1}{\sqrt{q(q-1)}}}_{\tilde{\mu}',\tilde{\mu}} \ .
\end{align}
These are the matrix elements appearing in the enlarged structure for the matrices describing all
total electron operators $\tilde{B}^{[p_l]}$, which for a system with three quasielectrons takes the form
\begin{align}
\tilde{B}^{[p_l=1]}=
\begin{pmatrix}
B^{[1]}&0&0&0&0&0&0&0\\
E_{k_1,l}&B^{[1]}&0&0&0&0&0&0\\
-E_{k_2,l}&0&B^{[1]}&0&0&0&0&0\\
0&E_{k_2,l}&E_{k_1,l}&B^{[1]}&0&0&0&0\\
E_{k_3,l}&0&0&0&B^{[1]}&0&0&0\\
0&-E_{k_3,l}&0&0&E_{k_1,l}&B^{[1]}&0&0\\
0&0&-E_{k_3,l}&0&-E_{k_2,l}&0&B^{[1]}&0\\
0&0&0&E_{k_3,l}&0&E_{k_2,l}&E_{k_1,l}&B^{[1]}
\end{pmatrix}.
\label{B_3qe}
\end{align}

This form, as discussed in  Sec.~\ref{sec:mps-qes}, ensures that there is a contribution from each modified electron operator once and only once in the MPS, and the explicit minus signs take care of the anti-symmetrization between the quasielectrons. The structure for systems with one or two quasielectrons were presented in the main text and addition of more quasielectrons is straightforward. 
The corresponding matrices necessary for the localized quasielectron can be obtained as described in
Sec.~\ref{sec:mps-qes} in the main text, and are given by
\begin{equation}
E_l(\xi_a) = e^{-\frac{\tau_\xi^2}{2 q \ell^2}} \sum_{k_a} 
e^{-\bigl(\frac{2 \pi}{L}\bigr)^2 q \ell^2 k_a^2} e^{\frac{2 \pi}{L} k_a (i x_\xi + \tau_\xi)} E_{k_a,l} \ ,
\end{equation}
with $E_l(\xi_a)$ replacing $E_{k_a,l}$ in Eq.~\eqref{B_3qe}.

Finally, we discuss the matrix elements for the quasihole operators. The differences with the
corresponding expression in the case where one only uses the field $\varphi(\omega_\eta)$ are straightforward,
and one obtains, for the $\alpha^{\rm th}$ quasihole  inserted between orbitals
$l-1$ and $l$
\begin{align}
\label{eq:two-fields-qhole}
H_l(\eta_\alpha) =&
(-1)^{(Q-Q_0+\tilde{Q}-\tilde{Q}_0+l-(\alpha-1))/q}
e^{+\frac{2\pi}{L} (l \delta\tau - \tau_{\eta_\alpha})
\bigl( \frac{(Q)^2}{2q} + P + \frac{Q}{2q} +\frac{1}{6q} +
\frac{(\tilde{Q})^2}{2q(q-1)} + \tilde{P} \bigr)}
\\\nonumber&\times
e^{-\frac{2\pi}{L} (l \delta\tau - \tau_{\eta_\alpha})
\bigl( \frac{(Q')^2}{2q} + P' + \frac{Q'}{2q} +\frac{1}{6q} +
\frac{(\tilde{Q}')^2}{2q(q-1)} + \tilde{P}'  \bigr)}
e^{- \frac{2\pi }{L} (i x_{\eta_\alpha}) \bigl( P' - P + \tilde{P}' - \tilde{P} + (Q-Q_0+\tilde{Q}-\tilde{Q}_0)/q
+ \tilde\tau_{\eta_\alpha}/q \bigr)}
\\\nonumber&\times
\delta_{Q',Q+1} A^{(1/\sqrt{q})}_{\mu',\mu}
\delta_{\tilde{Q}',Q+(q-1)} A^{(1/\sqrt{q(q-1)})}_{\tilde{\mu}',\tilde{\mu}} \ ,
\end{align}
where we recall that $\tilde\tau_{\eta_\alpha} = \tau_{\eta_\alpha}/(\delta\tau)$, i.e. the $\tau$ coordinate
of the quasihole, in terms of the distance between two orbitals.

We explicitly checked the contribution from the time evolution to the wave function in the case of a
single quasihole. In this case, the free-evolution should provide the correct normalization
of the single-particle orbitals, in addition to the `$\tau_\eta$ part' of the quasihole coordinate
that is explicitly present in the wave function, through the product $\prod_{i} (\omega_i - \omega_\eta)$.
Namely, in the wave function, the quasihole position comes in through factors
$\omega_\eta^s = e^{-\bigl(\frac{2\pi i}{L}(x_\eta+i \tau_\eta)\bigr) s}$, where $s$ is an integer, that
depends on which orbitals are occupied. In addition, there is the dependence through the
overall Gaussian factor
$e^{-\frac{1}{2q\ell^2}\tau_\eta^2}$. The dependence on $x_\eta$ is taken
into account explicitly in the MPS formulation, while the $\tau_\eta$ dependence
comes from the free time evolution. An explicit calculation of the free evolution
results in the Gaussian factor
$e^{-\frac{1}{2q\ell^2}(\tau_\eta-\delta\tau(\frac{q}{2}+Q_0+\tilde{Q}_0))^2}$ and 
the $\tau_\eta$ part of the $\omega_\eta^s$ dependence is
$e^{\frac{2\pi}{L} \bigl(\tau_\eta- \delta\tau(\frac{q}{2}+Q_0)\bigr) s}$.
We find that there is a shift of $q/2$ orbitals in the position of the quasihole. This shift is because
we are working on the cylinder, and therefore should have picked up a factor
coming from the conformal transformation from the plane to the cylinder. We did not do this
explicitly, because it merely leads to a shift in the position of the droplet on the cylinder.
This position can further be influenced by choosing a different value of $Q_0$, the charge of
the in-state associated with $\varphi(\omega)$, as can be seen explicitly in the two contributions above.
Finally, we see that there is a dependence on the charge $\tilde{Q}_0$ in the Gaussian factor, but not in the
other contribution from the free evolution. Because there is no spread out background charge
associated with $\tilde{\varphi}(\omega)$, it is not expected that changing the value $\tilde{Q}_0$ will change
the position of the quasihole. Indeed, a change in $\tilde{Q}_0$ only changes is the overall normalization
of the wave function.


\section{The quasielectron wave functions on the sphere.}
\label{app:sphereqe}

We here give the recipe for obtaining the angular momentum quasielectron
wave functions Eq.~\eqref{eq:angmomqe}, using the coherent state
approach developed in Ref.~\onlinecite{girvin} (see also Ref.~\onlinecite{svh11a})
and the techniques for calculating
correlators on the sphere as presented in Ref.~\onlinecite{kvorning13}.
For general information about quantum Hall state on
the sphere, we refer the reader to Ref.~\onlinecite{haldane83}.
We discuss the case of one quasielectron without any quasiholes,
the general case can be obtained in a similar way. The vertex operators
for the electron and modified electron are give in Eqs.~\eqref{eq:sphere-vo-el} and
\eqref{eq:sphere-vo-mod-el}. The strategy is to first obtain a localized quasielectron
on the sphere, from which the angular momentum quasielectron states are obtained by
expanding the result in angular momentum states.

For a fixed number of flux quanta $N_\phi$, there are $N_\phi+1$ lowest Landau level
single-particle states on the sphere, given by 
$z^m (1 + |z|^2/(4R^2))^{-N_\phi/2}$. 
In the following,   we   set the radius $R$ of the sphere to $R=\ell/2$.
The magnetic length $\ell$ is set to 1 in the remainder of this appendix. 
The factor $(1 + |z|^2)^{-N_\phi/2}$ is the analog of the
Gaussian factor $e^{-|z|^2/4}$ on the disk.

The sphere, in contrast to the disc and cylinder,
has non-zero curvature, which affects the two-point function of the chiral boson field,
which becomes
\begin{align}
\langle\varphi(z_{1})\varphi(z_{2})\rangle=-\log(z_{1}-z_{2})-\frac{1}{2}\Omega(z_{1},\bar{z}_{1})-\frac{1}{2}\Omega(z_{2},\bar{z}_{2}),
\end{align}
where $\Omega =-\log (1+\bar z z)+\log(2R)$ and $z=x+iy$ is related to the more familiar spherical coordinates by $z=e^{i\phi}\tan\left(\frac{\theta}{2}\right)$. 
To get a quasielectron operator with the correct geometric properties we have to also change the derivatives to~\cite{kvorning13}
\begin{align}
\frac{\partial^{n}}{\partial z^{n}}\rightarrow e^{n\Omega}\left(e^{-2\Omega}\frac{\partial}{\partial z}\right)^{n}\ .
\end{align}
We take the operator for the first electron to be the modified electron operator~\eqref{eq:sphere-vo-mod-el}, while the remaining ones
are ordinary electron operators~\eqref{eq:sphere-vo-el}.
Thus the starting correlator, to be calculated on the sphere, is given by
\begin{align}
 \Psi &= \langle\mathcal{O}_{\rm bg}\tilde{V} (\zeta_1) V(\zeta_2)\cdots V(\zeta_{N_e})\rangle \nonumber\\
 &=e^{-\Omega(\zeta_{1},\bar{\zeta}_{1})} \partial_{\zeta_{1}}\prod_{j}\!^{\prime}(\zeta_{1}-\zeta_{j})^{q-1}e^{\frac{q-1}{2}\left(\Omega(\zeta_{1},\bar{\zeta}_{1})+\Omega(\zeta_{j},\bar{\zeta}_{j})\right)}\prod_{j<k}\!^{\prime}(\zeta_{j}-\zeta_{k})^{q}e^{\frac q 2\left(\Omega(\zeta_{j},\bar{\zeta}_{j})+\Omega(\zeta_{k},\bar{\zeta}_{k})\right)}\ .
 \end{align}
Commuting the derivative through the factors $e^{\Omega}$, we arrive at 
 \begin{align}
 \Psi &= 
 (1+|\zeta_1|^2)^{-(q-1)(N_e -1)/2}
\prod_{j}\!^{\prime} (1+|\zeta_j|^2)^{(-q(N_e -1)+1)/2}
\, \left[(1+|\zeta_1|^2) \partial_{\zeta_1} + C \bar\zeta_1 \right] \Psi_{\rm rel}
\end{align}
where $C= -(q-1)(N_e-1)/2$, $\Psi_{\rm rel}$ is defined as  
\begin{align}
\Psi_{\rm rel}= \prod_{j<k}\!^\prime (\zeta_j - \zeta_k)^{q} \prod_{j}\!^\prime (\zeta_1 - \zeta_j)^{q-1} 
\end{align}
and the prime indicates that the indices run only over $2,3,\ldots, N_{e}$. In the following, we denote the power of
$(1+|\zeta_1|^2)$ by $B = -(q-1)(N_e-1)/2$.

Before projecting to the lowest Landau level, we need to insert a coherent state factor
that localizes the quasielectron at position $\xi$ and ensures the homogeneity of the quantum Hall droplet.
In the case of one quasielectron, this factor is
$$
(1+ \zeta_1 \bar\xi)^{N_e} (1 + |\zeta_1|^2)^{-N_e/2} \ .
$$
For completeness, we also state the coherent state factor for the case of $N_{qh}$ quasiholes and
$N_{qe}$ quasielectrons, localized at positions $\xi_a$
$$
\prod_a (1+ \zeta_a \bar\xi_a)^{N_e + 1 + N_{qh} - N_{qe}}
(1 + |\zeta_a|^2)^{-(N_e + 1 + N_{qh} - N_{qe})/2} \ .
$$

Finally, to obtain the quasielectron wave function on the sphere, we need to
project onto the lowest Landau level, by means of the operator
\begin{equation}
\label{eq:pLLL}
\mathcal{P}_{LLL} = \int \prod_j \frac{d^2\zeta_j}{(1+|\zeta_j|^2)^2}
\frac{
	(1 + \bar\zeta_j z_j)^{N_\phi}
}
{
	(1 + |\zeta_j|^2)^{N_\phi/2}
	(1 + |z_j|^2)^{N_\phi/2}
} \ .
\end{equation}
The factor $(1+|\zeta_j|^2)^{-2}$ is the integration measure on the sphere.
In general,
$N_\phi = q (N_e-1)+N_{qh} - N_{qe}$, so $N_\phi = q (N_e-1)-1$ in the present case.
In the following, we drop the `Gaussian' factor $(1 + |z_j|^2)^{-(N_\phi/2)}$ associated with the
single-particle orbitals.
To obtain the wave function of a localized quasielectron on the sphere, we need to calculate
\begin{multline}
\label{eq:sphere-int}
\psi_{qe} (z_j; \xi) = \mathcal{A}
\Bigl[
\mathcal{P}_{LLL} 
(1+ \zeta_1 \bar\xi)^{N_e} (1 + |\zeta_1|^2)^{-N_e/2}
\prod_{j}\!^\prime (1+|\zeta_j|^2)^{(-q(N_e -1)+1)/2} 
\\
\times
\bigl(
(1+|\zeta_1|^2) \partial_{\zeta_1} + C \bar\zeta_1
\bigr)
(1 + |\zeta_1|^2)^{B}
\Psi_{\rm rel}
\Bigr]
\ ,
\end{multline}
where $\mathcal{A}$ denotes anti-symmetrization over the electron coordinates.
It turns out that one can explicitly evaluate the integrals, because the various contributions
of the integrand combine to delta functions on the sphere
$$
\delta^2 (\zeta-z)= \frac{1}{4\pi (N_\phi+1)}
\frac{
	(1 + \bar\zeta z)^{N_\phi}
}
{
	(1 + |\zeta|^2)^{N_\phi}
} \ ,
$$
which includes the `Gaussian' part $(1 + |\zeta|^2)^{-N_\phi/2}$ of the
wave function one acts on, but not the measure.
We start by collecting the weight factors $(1+ |\zeta_j|^2)$ in 
$\psi_{qe} (z_j; \xi)$ in Eq.~\eqref{eq:sphere-int}.
For $j = 2,\ldots, N_e$, the total factor indeed becomes
$(1+ |\zeta_j|^2)^{-(N_\phi+2)}$.
After evaluating $\partial_{\zeta_1}$ on the factor $(1+|\zeta_1|^2)^B$
the integrand becomes
$$
\prod_j (1 + \bar\zeta_j z_j)^{N_\phi}
\prod_j\!^\prime (1 + |\zeta_j|^2)^{-N_\phi-2}
(1+ \zeta_1 \bar\xi)^{N_e} (1+|\zeta_1|^2)^{-N_e/2+B}
\bigl(
(1+|\zeta_1|^2) \partial_{\zeta_1} + (B+C)\bar{\zeta_1}
\bigr) \Psi_{\rm rel} \ .
$$
The factor containing the derivative $\partial_{\zeta_1}$ is proportional
to $(1+|\zeta_1|^2)^{-N_e/2 + B +1} = (1+|\zeta_1|^2)^{-(N_\phi+2)}$, which
is the desired result. The other factor is proportional to
$\bar{\zeta_1} (1+|\zeta_1|^2)^{-(N_\phi+3)}$, which we write as
$-1/(N_\phi+2) \bigl(\partial_{\zeta_1}(1+|\zeta_1|^2)^{-(N_\phi+2)}\bigr)$.
Integrating this last term by parts, the total integrand becomes
$$
\prod_j
\frac{(1 + \bar\zeta_j z_j)^{N_\phi}}{(1 + |\zeta_j|^2)^{N_\phi+2}}
\left[
(1+\zeta_1 \bar\xi)^{N_e} \partial_{\zeta_1} \Psi_{\rm rel}
+
\frac{B+C}{N_\phi+2} \partial_{\zeta_1} 
\bigl(
(1+\zeta_1 \bar\xi)^{N_e} \Psi_{\rm rel}
\bigr)
\right]\ .
$$
The second factor is holomorphic in the $\zeta_j$, while the first factor is the holomorphic
delta function, including the measure. Therefore, we can perform the integrals over the $\zeta_j$, to obtain
$$
\psi_{qe} (z_j;\xi) = 4\pi (N_\phi+1)
\mathcal{A}
\Bigl[
(1+ z_1 \bar\xi)^{N_e} \partial_{z_1} \Psi_{\rm rel} + \frac{B+C}{N_\phi+2} \partial_{z_1}
\bigl(
(1+z_1 \bar\xi)^{N_e} \Psi_{\rm rel}
\bigr)
\Bigr] \ ,
$$
where from now on $\Psi_{\rm rel}$ is in terms of the variables $z_j$ instead of
$\zeta_j$.
After evaluating the derivative in the second term, grouping terms and ignoring an
unimportant overall factor, one obtains the wave function for a quasielectron
localized at $\xi$ on the sphere as
\begin{equation}
\psi_{qe} (z_j;\xi) =
\mathcal{A}
\Bigl[
\bigl(
(1+ z_1 \bar\xi)^{N_e} \partial_{z_1}  - \bar\xi (q-1)(N_e-1) 
(1+ z_1 \bar\xi)^{N_e-1}
\bigr) \Psi_{\rm rel} \Bigr] \ .
\end{equation}

The angular momentum quasielectron wave functions on the sphere are obtained
by expanding the factors $(1+ z_1 \bar\xi)^{N_e}$ and $(1+ z_1 \bar\xi)^{N_e-1}$
in powers of $z_1$. This gives the following result for the quasielectron with
angular momentum $k$ (after dropping a $k$-dependent overall factor) 
\begin{equation}\label{eq:qespherefinal}
\psi_{qe}^k (z_j) =
\mathcal{A}
\Bigl[
\bigl(
z_1^k \partial_{z_1}  - k (q-1)\frac{(N_e-1)}{N_e} z_1^{k-1}
\bigr) \Psi_{\rm rel} \Bigr]
\ .
\end{equation}
We note that we dropped the sphere normalization factors
$(1+|z_j|^2)^{-N_\phi/2}$ present in $\mathcal{P}_{LLL}$, Eq.~\eqref{eq:pLLL},
which means that Eq.~\eqref{eq:qespherefinal} is really the `polynomial part' of
the wave function. One can check that the states $\psi_{qe}^k (z_j)$ with
$k=0,1,\ldots, N_e$ form an angular momentum multiplet with $k=0$ for the
lowest weight state, and $k=N_e$ for the highest. 

The generalization of this example  to the general case with
several quasielectrons and quasiholes is now straightforward. The correlator one starts with
changes in the expected way, namely, one uses
$N_{qe}$ operators $\tilde{V}$ and one inserts $N_{qh}$ operators $H$. The other changes
are described above. In the end, one obtains the expression given in Eq.~\eqref{eq:angmomqe}.

We used the results of this appendix to study quasielectron states (both
localized and angular momentum) on the sphere with MPS, by taking the sphere normalization
factors into account `by hand' when calculating observables (for systems up to 40 electrons).
We note that the observed `shift' in the position of the quasielectron (see Sec.~\ref{sec:results}) is also present on the sphere.


\section{Details of the Tao-Thouless limit}
\label{app:TTlimit}

In this appendix, we analyze the trial wave functions studied in this article in the Tao-Thouless (TT) limit. This corresponds to defining the wave functions on a cylinder with circumference $\lenx \to 0$. Starting from the wave function for the Laughlin state with quasiparticle excitations, we seek to analytically compute the orbitals that are occupied, and hence infer the ``position'' of the excitation. 

A generic trial wave function consists of a totally anti-symmetric polynomial in $\coord_j = e^{-2\pi i \, z_j /\lenx}$ (with $z_j = x_j + i \tau_j$) times an exponential factor which restricts the wave function to the lowest Landau level. Schematically, 
\beq
   \Psi ( \coll{\coord} ) 
   \sim \sum_{\bmu} \left( \prod_{i=1}^{N_e} \coord_i^{\mu_i} \right) e^{-\sum_{i=1}^{N_e} \frac{\tau_i^2}{2\ell^2}}
   = \sum_{\bmu} \left( \prod_{i=1}^{N_e} \coord_i^{\mu_i} e^{-\tau_i^2/2\ell^2} \right),
\eeq
where $\coll{\coord} = \left( \coord_1, \dots \coord_{N_e} \right) \in \cmplx^{N_e}$ denotes the coordinates of the $N_e$ electrons and $\bmu = \left( \mu_1, \dots \mu_{N_e} \right) \in \intg^{N_e}$ denotes the set of occupied orbitals. Further analysis hinges on the observation that 
\beq 
  \coord^{\mu_i} e^{-\mu_i^2/2\ell^2}  = \expn{ -\frac{2\pi i \, \mu_i \, x}{\lenx} - \frac{(\tau - \const \mu_i)^2}{2\ell^2} + \frac{2\pi^2 \ell^2}{\lenx^2} \mu_i^2} = e^{\frac{2\pi^2 \ell^2}{\lenx^2} \mu_i^2} \phi_{\mu_i} (z),   \label{eq:beta_sub}
\eeq 
where $\const = \frac{2\pi\ell^2}{\lenx}$ is the separation of the single-particle orbitals and $\phi_{\mu}(z)$ denotes the single electron lowest Landau level wave function on a cylinder with ``momentum'' $k$, which is exponentially localized at the $\mu^{\text{th}}$ orbital, i.e., around $\tau = \delta\tau \mu$. Thus, 
\beq 
   \Psi( \coll{\coord} ) 
 \sim \sum_{\bmu} \left( \prod_{i=1}^{N} e^{\frac{2\pi^2 \ell^2}{\lenx^2} \mu_i^2} \phi_{\mu_i} (z_i) \right) 
 = \sum_{\bmu} e^{\frac{2\pi^2 \ell^2}{\lenx^2} \norm{\bmu}_2^2} \Phi_\bmu( \coll{\coord} ),   
\eeq 
where $\norm{\bmu}_2$ is the $L_2$ norm of $\bmu$. Recall that the $L_p$ norms are defined as 
\[
\norm{\boldsymbol{\xi}}_p = \left[ \sum_{i=1}^n |\xi_i|^p \right]^{1/p} \!\!\!\!\!\!\!\!; \quad\quad\quad \boldsymbol{\xi} = \left( \xi_1, \xi_2, \dots \xi_n \right) \in \cmplx^n.
\]
Furthermore, we have defined a set of many-body wave functions (indexed by $\bmu$) as $\Phi_\bmu ( \coll{\coord} ) = \prod_{i} \phi_{\mu_i} (z_i)$, which is simply a product of single-electron wave functions localized along $\tau$ at $\tau = \const\mu_i$ and delocalized along $x$. 
In the TT limit, this wave function is exponentially dominated by terms which maximize $\norm{\bmu}_2$, i.e., 
\beq 
  \Psi( \coll{\coord} ) 
 = e^{\frac{2\pi^2 \ell^2}{\lenx^2} M}  \sum_{\norm{\bmu}_2=M} \Phi_\bmu( \coll{\coord} ) + O\left( e^{-1/\lenx^2} \right),   \label{eq:config_sch}
\eeq 
where $M = \max{\norm{\bmu}^2_2}$, and the remaining sum is over the `degenerate' configurations, i.e., $\bmu$ such that $\norm{\bmu}_2^2 = M$, which includes all permutations of $\mu_i$'s. We shall term this dominant contribution $\bmu_{\max}$ (up to permutations) the \emph{orbital configuration} corresponding to the given wave function. 

The exact form of the wave function imposes constraints on $\bmu$'s, thereby allowing only certain occupation patterns. Thus, to find the orbital configuration, we seek to find $\bmu_{\max}$ that maximizes $\norm{\bmu}_2$ subject to all the constrains imposed by the form of the wave function. We shall derive this maximal configuration $\bmu_{\max}$ iteratively, following a \emph{greedy algorithm}. Denoting the step count by $s$, this algorithm can be stated as follows: 
\begin{enumerate}
 \item Start with $s = 1$.
 \item Set $\mu_s$ to the value that locally maximizes $F[\bmu] = \norm{\bmu}_2^2$, subject to the constraints. 
 \item Update the constraint for remaining $\mu_i, i > s$. 
 \item $s \to s+1$, go to step 2 if $s \leq N_e$. 
\end{enumerate}
The constraints on $\bmu$ as well as the update of step 3 will be obtained from a given wave function by inspection. In the following, we analytically compute this configuration for the Laughlin wave function for the ground state as well as the states with a few quasiparticle excitations.

\subsection{TT limit for Laughlin wave function}
The Laughlin wave function for $N_e$ electrons on a cylinder with filling fraction $\nu = 1/q$  is 
\beq 
  \Psi_{L} ( \coll{\coord} ) 
  = \prod_{i<j} \left( \coord_i - \coord_j \right)^q \;  \expn{-\sum_{i=1}^{N_e} \frac{\tau_i^2}{2\ell^2}} 
  =  \sum_{\bmu} \left( \prod_{i=1}^{N_e} \coord_i^{\mu_i} e^{-\tau_i^2/2\ell^2} \right),   \label{eq:wf_cyl}
\eeq 
where in the second step, we have expanded out the polynomial part of the wave function as a sum over monomials indexed by $\bmu$. From the form of the polynomial part $\prod \left( \coord_i - \coord_j \right)^q$, we deduce that the exponent $\mu_i$ of each $\coord_i$ is bounded from below by 0 and from above by $q(N_e-1)$. Furthermore, since the polynomial is homogeneous, the sum of all exponents is the same for all monomials. Thus, $\bmu$ is subject to the constraints
\beq
  \norm{\bmu}_{1} = \frac{1}{2}q N_e(N_e-1), \quad 0 \leq \mu_i \leq q(N_e-1).   \label{eq:mu_cond}
\eeq 
Physically, these express the fact that the many-body wave function has a fixed orbital angular momentum distributed among the single electron orbitals, and furthermore, each of those states has an upper bound on the angular momentum, corresponding to the size of the quantum Hall droplet.

\begin{figure}[h]
	\begin{center}
		\includegraphics[width=0.6\columnwidth]{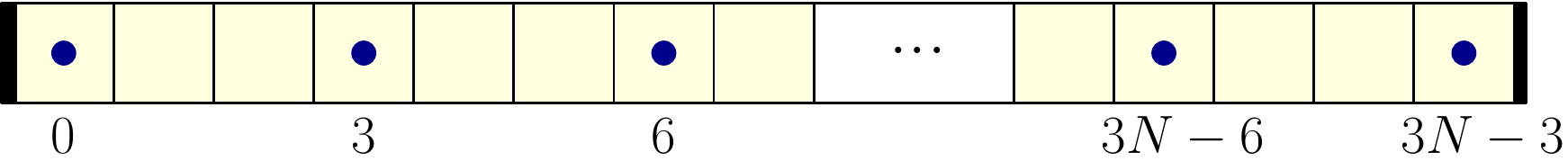} 
		\caption{Orbital configuration for the ground state of $\nu=1/3$ Laughlin state for $N_e$ electrons. The blue dots represent the occupied orbitals, whose number (counting the leftmost as zero) is shown underneath.}
		\label{fig:occ_pattern_GS}
	\end{center}
\end{figure}

We next maximize $\norm{\bmu}_2$ subject to these constraints, using the algorithm outlined earlier. At the first iteration, 
\[ 
 s=1 \!: \quad 0 \leq \mu_i \leq q(N_e-1),
\]
and since we seek to maximize $\sum_i \mu_i^2$, we set $\mu_1 = q(N_e-1)$.
This reduces by $q$ the maximum exponent that can be attained by any other $\coord_j$. Thus, the constraint for the remaining orbitals is modified to $0 \leq \mu_i \leq q(N_e-2)$. Iterating this procedure $N$ times, we get 
\[ 
  \mu_1 = q(N_e-1), \;\; \mu_2 =  q(N_e-2), \; \dots \mu_s = q(N_e-s), \;\; \dots, \;  \mu_{N_e} = 0 \implies \bmu_{\max} = \left\{ nq \right\}_{n = 1, \dots N_e-1}.
\] 
This is the well known orbital configuration for the Laughlin state in the TT limit~\cite{bergholtz08}, as depicted in Fig~\ref{fig:occ_pattern_GS}. Note that any other permutation of this set $\{\mu_i\}$ would lead to the same value of $\norm{\bmu}_{2}$, and these constitute the set of ``degenerate'' configurations referred to in \eq{eq:config_sch}. 

\subsection{Excitations} 
The derivation of the orbital configuration in the presence of excitations follows a similar strategy, although the iterations become more complicated. 
In the following, we discuss a few analytically tractable cases for Laughlin wave functions in presence of excitations.  

\subsubsection{Quasiholes}
The Laughlin wave function for $N_{qh}$ quasiholes at positions $\omega_{\eta_\alpha} =  e^{-2\pi i \, \eta_\alpha /\lenx}$ is given by 
\beq 
  \Psi_{L, qh} \left( \coord_j; \omega_{\eta_k} \right) = \prod_{i<j} \left( \coord_i - \coord_j \right)^q \;  \prod_{j,k} \left( \coord_j - \omega_{\eta_k} \right) \;  \expn{-\sum_{i=1}^{N_e} \frac{\tau_i^2}{2\ell^2}}.     \label{eq:wf_qh}
\eeq 
Expanding into monomials and using \eq{eq:beta_sub}, we get  
\begin{align}
  \Psi_{L, qh}\left( \coord_j; \omega_{\eta_k} \right) 
  = & \; \sum_{\bmu} \expn{ \frac{2\pi^2 \ell^2}{\lenx^2} \left( \norm{\bmu}_2^2 + \frac{\lenx}{\pi \ell^2} \sum_{k=1}^{N_{qh}} \wt{\mu}_k \tau_{\eta_k} \right) } 
  e^{-i\frac{2\pi}{\lenx} \sum_{k=1}^{N_{qh}} \wt{\mu}_k x_{\eta_k}} \Phi_\bmu( \coll{\coord} ).
\end{align}
We can again derive the constraints on $\bmu$ by inspection, as
\beq
  \norm{\bmu}_{1} + \norm{\wt{\bmu}}_{1} = \frac{1}{2}q N_e(N_e-1) + N_{qh} N_e \equiv K, \quad 0 \leq \mu_i \leq q(N_e-1) + N_{qh},  \label{eq:mu_cond_qh}
\eeq 
where the increase in the upper bound on $\mu_i$'s can be physically interpreted as expansion of the (finite) quantum Hall droplet on addition of $N_{qh}$ quasiholes. 
To find the TT configuration, we seek to maximize $ \norm{\bmu}_2^2 + \frac{\lenx}{\pi \ell^2} \sum_{k=1}^{N_{qh}} \wt{\mu}_k \tau_{\eta_k}$ subject to these constraints. 
This could be accomplished numerically following the greedy algorithm outlined earlier, but there is no direct analytical solution in general. 

We thus restrict ourselves to an analytically tractable special case, when all quasiholes lie at the same position $x_{\eta_k}=0$ and $\tau_{\eta_k}=\wt{\tau}$. We can then use the first constraint of \eq{eq:mu_cond_qh} to get 
\begin{align}\label{eq:normmu}
\norm{\bmu}_2^2 + \frac{\lenx}{\pi \ell^2} \sum_{k=1}^{N_{qh}} \wt{\mu}_k \tau_{\eta_k} =  \sum_{i=1}^{N_e} (\mu_i - \wt{\Delta})^2  + (2K-\wt{\Delta}) \wt{\Delta} 
\end{align}
with $\wt{\Delta} = \frac{\wt{\tau}}{\const}$.
Thus,  the quasihole is expected to be localized around $\wt{\Delta}$. The last term in Eq. \eqref{eq:normmu} can be ignored since it is independent of $\bmu$, so we are left to to maximize 
$\norm{\bmu - \wt{\Delta} \onev}_2, \;  \onev = \left( 1, \dots 1 \right) \in \intg^{N_e}$, 
subject to only the second constraint of \eq{eq:mu_cond_qh}.

\begin{figure}[h]
	\begin{center}
		\includegraphics[width=0.75\columnwidth]{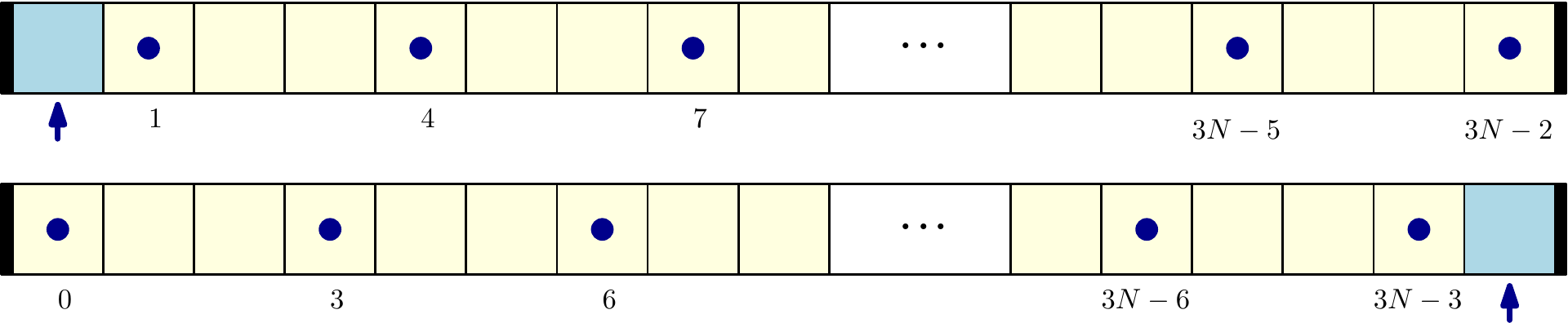} 
		\caption{Orbital configuration for the $\nu=1/3$ Laughlin state for $N_e$ electrons, with a single quasihole to the left (top) and the right (bottom). The orbital shaded in blue denotes the expected position of the quasihole, and the blue arrow denotes the actual position. Clearly, the quasihole is localized at the expected position.}
		\label{fig:occ_pattern_QH}
	\end{center}
\end{figure}

For $\wt{\Delta} \leq 0$, i.e, a quasihole at the left end,  we need to choose the largest possible $\mu_i$ at each step in order to maximize $(\mu_i - \wt{\Delta} )^2$.
Thus, this case is almost identical to the case without a quasihole, except for a rightward shift by $N_{qh}$, i.e, $\mu_s = N_{qh} + qs$, 
as depicted in Fig~\ref{fig:occ_pattern_QH}.

For $\wt{\Delta} \geq q(N_e+1)+N_{qh}$, i.e, a quasihole at the right end,  we need to choose the \emph{smallest} possible $\mu_i$ at each step, in order to maximize  $(\mu_i - \wt{\Delta} )^2$. The iteration proceeds in a fashion analogous to the case without a quasihole, except that we now update only the lower bound. To wit, after $s$ steps, we have 
\[
  \mu_s = qs, \quad q(s+1) \leq \mu_i \leq q(N_e-1)+ N_{qh},
\]
for the remaining orbitals $i>s$. Thus, the resulting occupation pattern is identical to that for the ground state.

For $0 < \wt{\Delta} < q(N_e-1)+N_{qh}$, the iteration is more complicated now, since at each step, we must decide whether to pick the lowest or highest possible value. We omit the details of this iteration since they are  not very illuminating, but the result is as expected, i.e.  the quasihole is localized at the position $\wt{\Delta}$.

\subsubsection{Quasielectrons}
The most general expression for a multiple-quasielectron Laughlin wave function can be rather complicated, so we shall instead start with the case of a single quasielectron wave function, which can be written as 
\beq 
    \Psi_{L, qe} \left( \coll{\coord}, \omega_{\xi} \right) =  \sum_a \left[ \proj(\coord_{\xi},\coord_a) \prod_{\mathclap{\substack{i<j \\ i \neq j\neq a}}} \left( \coord_i - \coord_j \right)^q 
    \; \partial_a \left( \prod_{i \neq a} \left( \coord_i - \coord_a \right)^{q-1} \right)  \expn{-\sum_{i=1}^{N_e} \frac{\tau_i^2}{2\ell^2} + \frac{\tau_a^2}{2q\ell^2} } \right],   \label{eq:wf_qe} 
\eeq 
where $\proj(\coord_{\xi},\coord_a)$ is the kernel Eq.~\eqref{eq:kernel-cylinder}. 
We rewrite the two single-particle wave functions in the kernel as
\begin{align}\label{eq:kernel_app}
 \phi_k(\coord_a) 
 = & \coord_a^k \; \expn{- \frac{\tau_a^2}{2q\ell^2} - \frac{2\pi^2\ell^2}{\lenx^2} q k^2 } \nonumber \\ 
\bar{\phi}_k(\coord_{\xi}) 
  = & \expn{\frac{2\pi i k x_\xi}{\lenx} - \frac{2\pi^2\ell^2}{\lenx^2} q \left( k - \frac{\tau_{\xi}}{q\const} \right)^2 },
\end{align}
and then expand the wave function \eqref{eq:wf_qe} into monomials to get 
\beq 
  \Psi_{L, qe} \left( \coll{\coord}, \eta \right) = \sum_k \expn{ - \frac{2\pi^2\ell^2}{\lenx^2} q \left[ k^2 + \left( k - \frac{\tau_{\xi}}{q\const} \right)^2 \right] } e^{ i \frac{2\pi k x_\xi}{\lenx}} \left[ \sum_{\bmu} e^{\frac{2\pi^2 \ell^2}{\lenx^2} \norm{\bmu}_2^2} \Phi_\bmu( \coll{\coord} ) \right].  \label{eq:expn_qe_proj}
\eeq 
In order to obtain the occupation pattern of the orbitals for this wave function, we first determine $\bmu_{\max}(k)$ that maximizes the sum over $\bmu$ for a given value of $k$ following our iteration, and then choose the $k$ that maximizes the overall coefficient of the monomials. 

For a given $k$, we need to derive the constraints on $\bmu$.
Without loss of generality, we choose the term $a=N_e$ in \eqref{eq:wf_qe}, which leads to:
\beq \label{eq:qeconstraint}
  k \leq \mu_{N_e} \leq k + (q-1)(N_e-1) -1,  \quad\quad 
0 \leq \mu_i \leq q(N_e-1) - 1, \quad  i = 1, \dots N_e-1.         
\eeq 	
The pre-factor $\coord_{N_e}^k$ in the first line of \eqref{eq:kernel_app} leads to a nonzero lower bound for $\mu_{N_e}$, while the derivative reduces the upper bound from what one would expect from the Jastrow factor involving $\coord_{N_e}$. The derivative should also affect the upper bound for \emph{one of the} $\coord_i$'s, but that turns out to be unimportant for the rest of this calculation. Thus, we seek to maximize $\norm{\bmu}_2$ subject to these constraints. 

At each step of the iteration, we have a choice between $\mu_{N_e}$ and $\mu_i, \; i \neq {N_e}$, depending on the highest allowed value for them at each stage. If we choose $\mu_{N_e}$, the upper bound on $\mu_i$'s decreases by $q-1$. On the other hand, if we choose $\mu_i$, the upper bound on $\mu_j, \, j \neq i, {N_e}$ decreases by $q$ and on $\mu_{N_e}$ by $q-1$. It is precisely this interplay of updates that leads to interesting shifts in the quasielectron positions. 

Explicitly, let $\mu_i$ be chosen for the first $s$ steps, so that $\mu_s = q({N_e}-s)-1$, and the constraint on $\bmu$ becomes 
\[  
    k \leq \mu_{N_e} \leq k + (q-1)({N_e}-s-1) -1,  \quad\quad 0 \leq \mu_i \leq q({N_e}-s-1) - 1, \quad  s < i < {N_e}.
\]
Thus, we must choose $\mu_{N_e}$ when 
\[ 
 k + (q-1)({N_e}-s-1) -1 > q({N_e}-s-1) - 1 \implies s > {N_e} - k - 1.
\]
Define $s_0 = {N_e} - k$, which is the smallest integer to satisfy the above condition. Then,  at the $(s_0 + 1)^{\text{th}}$ step we must set $\mu_{N_e} = q(k - 1)$, and the condition on the remaining $\mu$'s is simply $0 \leq \mu_i \leq q(k - 2), \;  s_0 < i < {N_e}$. The rest of the iteration proceeds as in the case of the ground state wave function, and the set of occupied orbitals becomes 
\[
  \bmu_{\max}(k) = \left\{  0, q, \dots q(k - 2), q(k - 1), qk - 1, q(k + 1) - 1, \dots q({N_e}-1)-1 \right\}.    \label{eq:config_qe}
\]
This is the occupation pattern for an \emph{angular momentum quasielectron}, which is expected to be localized at the $qk^{\text{th}}$ orbital. However, from the occupation pattern, we compute its position as 
\beq
  \frac{1}{2} \left[ qk -1 + q(k-1) \right] = q (k - 1) + \frac{q-1}{2} = q k - \frac{q+1}{2}.    \label{eq:qe_pos}
\eeq 
Thus, for delocalized quasielectrons, we see a shift \emph{to the left} by $\frac{1}{2}(q+1)$ orbitals, inherent in the construction of the wave functions. 

Finally, we can compute the occupation pattern for the \emph{localized} quasielectron, whose wave function can be written, using \eq{eq:expn_qe_proj}, as 
\beq 
  \Psi_{L, qe} \left( \coll{\coord}, \eta \right) = \sum_k \expn{ \frac{2\pi^2\ell^2}{\lenx^2} q \left[ \frac{1}{q} M(k) - k^2 - \left( k - \frac{\tau_{\xi}}{q\const} \right)^2  \right] } e^{i \frac{2\pi k x_\xi}{\lenx}} \sum_{\norm{\bmu}_2=M(k)} \Phi_\bmu( \coll{\coord} ) + O\left( e^{-1/\lenx^2} \right),
\eeq 
where 
\begin{align}
   M(k) \equiv \norm{\bmu_{\max}(k)}_2^2 = \sum_{n=0}^{k-1} (qn)^2 + \sum_{n=k}^{N_e-1} (qn-1)^2 = F(N) + q k^2 - (q+1) k ,   \label{eq:sum_qe}
\end{align}
with $F(N_e)$ is a (unimportant) constant independent of $k$. Thus, we need to maximize 
\[
 \frac{1}{q} \left[ q k^2 - (1+q) k \right] - k^2 - \left( k - \frac{\tau_{\xi}}{q\const} \right)^2 = - \left[ k - \left( \frac{\tau_{\xi}}{q\const} - \frac{q+1}{2q} \right) \right]^2 + \text{constants},
\]
over $k$, so that we choose $k$ as the nearest integer to  $\frac{\tau_{\xi}}{q\const} - \frac{q+1}{ 2q}$. 
Using \eq{eq:qe_pos}, we compute the position of the quasielectron as 
$
  \frac{\tau_{\xi}}{\const} - (q+1), 
$
with an error of up to $\pm q/2$. Thus, the localized quasielectron is shifted \emph{to the left} by $(q+1)$ orbitals.

\begin{figure}[h]
	\begin{center}
		\includegraphics[width=0.95\columnwidth]{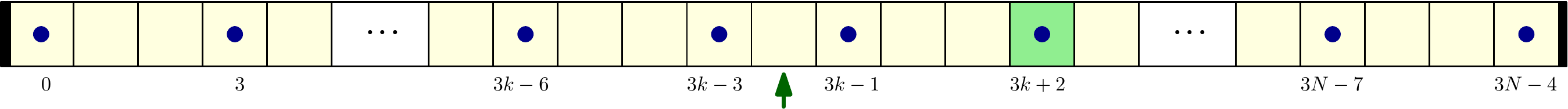} 
		\caption{Orbital configuration for the $\nu=1/3$ Laughlin state for $N_e$ electrons, with a quasielectron in the middle. The orbital shaded in green (corresponding to $\tau = (3k+2) \const$) denotes the expected position of the quasielectron, and the green arrow denotes the actual position. We clearly notice a shift to the left by $q+1=4$ orbitals.}
	\end{center}
\end{figure}

Let us now consider a single localized quasielectron in the presence of $N_{qh}$ quasiholes. In order to make this problem tractable, we assume that the quasiholes are again localized at the same point $\omega_{\eta}$ with  $x_{\eta_k}=0$ and $\tau_{\eta_k}=\wt{\tau}$, and that the quasielectron  at $\tau_{\xi}\gg \wt{\tau}$ is far away from the quasiholes.
The wave function for this setup is 

\begin{multline}
\label{eq:qe+qh} 
\Psi_{L, qe+qh} \left( \coll{\coord}, \omega_{\xi} \right) =  \sum_a \left[ \proj(\coord_{\xi},\coord_a) 
\prod_{j\neq a}\left(\coord_j-\omega_\eta\right)^{N_{qh}}
\prod_{\mathclap{\substack{i<j \\ i \neq j\neq a}}}  \left( \coord_i - \coord_j \right)^q \right.
\nonumber\\
\left. \times\partial_a \left( \prod_{i \neq a} \left( \coord_i - \coord_a \right)^{q-1} \right)  \expn{-\sum_{i=1}^{N_e} \frac{\tau_i^2}{2\ell^2} + \frac{\tau_a^2}{2q\ell^2} } \right].   
\end{multline} 
The extra factor enlarges the quantum Hall droplet, and the constraints in Eq. \eqref{eq:qeconstraint} change to become
\begin{align}
k \leq \mu_{N_e} \leq k + (q-1)(N_e-1) - 1 ,\qquad 0 \leq \mu_i \leq q(N_e-1) - 1 + \delta_{ex},  \;  i = 1, \dots N_e-1 
\end{align}
with $\delta_{ex}\equiv N_{qh}$. 
We now proceed along the same lines as earlier: we pick $\mu_i$ for the first $s$ step, and choose $\mu_{N_e}$ when 
\[ 
k + (q-1)(N_e-s-1) -1 > q(N_e-s-1) - 1 + \delta_{ex} \implies s > N_e - k - 1 + \delta_{ex}.
\]
Thus, $\mu_{N_e} = q (k - 1) - (q-1) \delta_{ex}$, and the occupation pattern is 
\beq 
\bmu_{\max} = \left\{ \dots q (k - 1) - (q-1) \delta_{ex}, q k_0 - (q-1) \delta_{ex} - 2, \dots q(N_e-1) - 1 + \delta_{ex} \right\}. 
\eeq
The delocalized quasielectron is localized at the orbital 
\[
q k - \frac{q+1}{2} - (q-1) \delta_{ex},
\]
and experiences a charge-dependent shift \emph{to the left} by $\frac{q+1}{2} - (q-1) \delta_{ex}$ orbitals. 

Finally, for a localized quasielectron, 
\beq 
M(k) = F(N_e, \delta_{ex}) + qk^2 - \left[(q+1) + 2 \delta_{ex} (q-1) \right] k,
\eeq 
so that we need to maximize 
\beq 
\frac{1}{q} M(k) - k^2 - \left( k - \frac{\tau_\xi}{q\const} \right)^2 = - \left[ k - \left( \frac{\tau_\xi}{q\const} - \frac{(q+1) + 2\delta_{ex} (q-1)}{2q} \right) \right]^2 + \text{constants}. 
\eeq 
We need to choose $k$ as the integer nearest to
\[  
  \frac{\tau_\xi}{q\const} - \frac{(q+1) + 2\delta_{ex} (q-1)}{2q} 
= \delta_{ex} +  \frac{\tau_\xi}{q\const} - \frac{q+1 - 2\delta_{ex}}{2q} ,
\]
so that the quasielectron is localized near $ \frac{\tau_{\xi}}{\delta\tau} - (q+1) - 2 (q-1)  \delta_{ex}$. 
Thus, in the presence of additional excitations, the quasielectron is shifted to the left by an additional $2 (q-1) \delta_{ex}$ orbitals, where $\delta_{ex} = N_{qh}$ corresponds to the number of quasihole to the left of the quasielectron in question.

\begin{figure}[h]
	\begin{center}
		\includegraphics[width=\columnwidth]{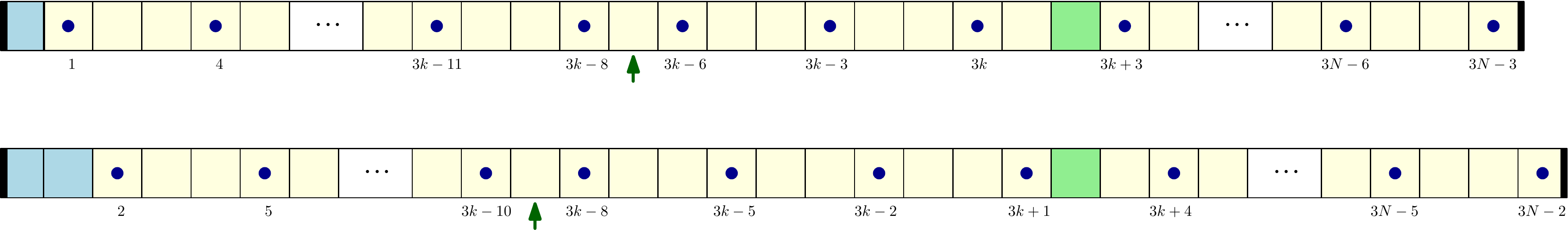} 
		\caption{Orbital configuration for the $\nu=1/3$ Laughlin state for $N_e$ electrons, with a quasielectron in the middle, and (top) 1 and (bottom) 2 quasiholes to the left. The orbital shaded in green denotes the expected position of the quasielectron, and the green arrow denotes the actual position. We clearly notice a shift to the left proportional to the number of quasiholes to the left.}
	\end{center}
\end{figure}

For more general setups --- i.e. multiple quasiparticles at arbitrary positions --- this calculation is no longer analytically tractable. However, as long as the quasiparticles are far apart, one can verify numerically that the quasielectron positions are shifted to the left by $ (q+1) + 2 (q-1)  \delta_{ex}$ orbitals, where $\delta_{ex}=n_{qh}-n_{qe}$ is the total charge to the left of the quasielectron, i.e. $n_{qh}/n_{qe}$ is the number of quasiholes/quasielectrons at a smaller $\tau$.


\begin{thebibliography}{99}


\bibitem{fqhe} 
D.C.~Tsui, H.L.~St\"ormer, A.C.~Gossard,
{\it Two-Dimensional Magnetotransport in the Extreme Quantum Limit},
Phys. Rev. Lett. {\bf 48}, 1559 (1982),
\doi{10.1103/PhysRevLett.48.1559}.


\bibitem{wen}
X.-G.~Wen,
{\it Topological orders and Edge excitations in FQH states},
Adv. Phys. {\bf 44}, 405 (1995),
\doi{10.1080/00018739500101566}.


\bibitem{willet87}
R.L.~Willett, J.P.~Eisenstein, H.L.~Stormer, D.C. ~Tsui, A.C.~Gossard, J.H.~English,
{\it Observation of an even-denominator quantum number in the fractional quantum Hall effect},
Phys. Rev. Lett. {\bf 59}, 1776 (1987),
\doi{10.1103/PhysRevLett.59.1776}.


\bibitem{pan99}
W.~Pan, J.-S.~Xia, V. ~Shvarts, E.D.~Adams, H.L.~St\"ormer, D.C.~Tsui, L.N.~Pfeiffer,
K.W.~Baldwin, K.W.~West,
{\it Exact quantization of the even-denominator fractional quantum Hall state at $\nu=5/2$ Landau level filling factor},
Phys. Rev. Lett. {\bf 83}, 3530 (1999),
\doi{10.1103/PhysRevLett.83.3530}.


\bibitem{laugh83}
R.B.~Laughlin,
{\it Anomalous quantum Hall effect: an incompressible quantum fluid with fractionally charged excitations},
Phys. Rev. Lett. {\bf 50}, 1395 (1983),
\doi{10.1103/PhysRevLett.50.1395}.


\bibitem{MR}
G.~Moore, N.~Read,
{\it Nonabelions in the fractional quantum Hall effect},
Nucl. Phys. B {\bf 360}, 362 (1991),
\doi{10.1016/0550-3213(91)90407-O}.


\bibitem{jaincf}
J.K.~Jain,
{\it Composite-fermion approach for the fractional quantum Hall effect},
Phys. Rev. Lett. {\bf 63}, 199 (1989),
\doi{10.1103/PhysRevLett.63.199}.


\bibitem{jainbook}
J.K.~Jain,
{\it Composite fermions},
(Cambridge University Press, 2007),
\doi{10.1017/CBO9780511607561}.


\bibitem{pandata}
W.~Pan, H.L.~St\"ormer, D.C.~Tsui, L.N.~Pfeiffer, K.W.~Baldwin, K.W.~West
{\it Fractional quantum hall effect of composite fermions},
Phys. Rev. Lett. {\bf 90}, 016801 (2003),
\doi{10.1103/PhysRevLett.90.016801}.


\bibitem{witten}
E.~Witten,
{\it Quantum field theory and the Jones polynomial},
Comm. Math. Phys. {\bf 121}, 351 (1989),
\doi{10.1007/BF01217730}.


\bibitem{wen91}
X.-G.~Wen,
{\it Non-Abelian Statistics in the Fractional Quantum Hall States},
Phys. Rev. Lett. {\bf 66}, 802 (1991).
\doi{10.1103/PhysRevLett.66.802}.


\bibitem{blokwen}
B.~Blok, X.-G.~Wen,
{\it Many-body systems with non-Abelian statistics},
Nucl. Phys. B {\bf 374}, 615 (1992),
\doi{10.1016/0550-3213(92)90402-W}.


\bibitem{haldane83}
F.D.M.~Haldane,
{\it Fractional quantization of the Hall effect: a hierarchy of incompressible quantum fluid states},
Phys. Rev. Lett. {\bf 51}, 605 (1983),
\doi{10.1103/PhysRevLett.51.605}.


\bibitem{halperinhierarchy}
B.I.~Halperin,
{\it Statistics of Quasiparticles and the Hierarchy of Fractional Quantized Hall States},
Phys. Rev. Lett. {\bf 52}, 1583 (1984),
\doi{10.1103/PhysRevLett.52.1583}.


\bibitem{bondsling}
P.~Bonderson, J.K.~Slingerland,
{\it Fractional quantum Hall hierarchy and the second Landau level},
Phys. Rev. B {\bf 78}, 125323 (2008),
\doi{10.1103/PhysRevB.78.125323}.


\bibitem{levhalp}
M.~Levin, B.I.~Halperin,
{\it Collective states of non-Abelian quasiparticles in a magnetic field},
Phys. Rev. B {\bf 79}, 205301 (2009),
\doi{10.1103/PhysRevB.79.205301}.


\bibitem{hermanns}
M.~Hermanns,
{\it Condensing Non-Abelian Quasiparticles},
Phys. Rev. Lett. {\bf 104}, 056803 (2010),
\doi{10.1103/PhysRevLett.104.056803}.


\bibitem{hhsv}
T.H.~Hansson, M.~Hermanns, S.H.~Simon, S.F.~Viefers,
{\it Quantum Hall physics: Hierarchies and conformal field theory techniques},
Rev. Mod. Phys. {\bf 89}, 025005 (2017),
\doi{10.1103/RevModPhys.89.025005}.


\bibitem{lopezfradkin}
A.~Lopez, E.~Fradkin,
{\it Universal structure of the edge states of the fractional quantum Hall states},
Phys. Rev. B {\bf 59}, 15323 (1999),
\doi{10.1103/PhysRevB.59.15323}.


\bibitem{arovas}
D.~Arovas, J.R.~Schrieffer, F.~Wilczek,
{\it Fractional statistics and the quantum Hall effect},
Phys. Rev. Lett. {\bf 53}, 722 (1984),
\doi{10.1103/PhysRevLett.53.722}.


\bibitem{kjonsbergM1999}
H.~Kj\o{}nsberg, J.~Myrheim,
{\it Numerical study of charge and statistics of Laughlin quasiparticles},
Int. J. Mod. Phys. A {\bf 14}, 537 (1999),
\doi{10.1142/S0217751X99000270}.


\bibitem{jeon03}
G.S.~Jeon, K.L.~Graham, J.K.~Jain,
{\it Fractional statistics in the fractional quantum Hall effect},
Phys. Rev. Lett. {\bf 91}, 036801 (2003),
\doi{10.1103/PhysRevLett.91.036801}.


\bibitem{jeon04}
G.S.~Jeon, K.L.~Graham, J.K.~Jain,
{\it Berry phases for composite fermions: Effective magnetic field and fractional statistics},
Phys. Rev. B {\bf 70}, 125316 (2004),
\doi{10.1103/PhysRevB.70.125316}.


\bibitem{MRqh}
M.~Baraban, G.~Zikos, N.~Bonesteel, S.H.~Simon,
{\it Numerical analysis of quasiholes of the Moore-Read wave function},
Phys. Rev. Lett. {\bf 103}, 076801 (2009),
\doi{10.1103/PhysRevLett.103.076801}.


\bibitem{mpsnashort}
Y.-L.~Wu, B.~Estienne, N.~Regnault, B.A.~Bernevig,
{\it Braiding Non-Abelian Quasiholes in Fractional Quantum Hall States},
Phys. Rev. Lett. {\bf 113}, 116801 (2014),
\doi{10.1103/PhysRevLett.113.116801}.


\bibitem{wu15}
Y.-L.~Wu, B.~Estienne, N.~Regnault, B.A.~Bernevig,
{\it Matrix product state representation of non-Abelian quasiholes},
Phys. Rev. B {\bf 92}, 045109 (2015),
\doi{10.1103/PhysRevB.92.045109}.


\bibitem{kitaev-preskill}
A.~Kitaev, J.~Preskill,
{\it Topological Entanglement Entropy},
Phys. Rev. Lett. {\bf 96}, 110404 (2006),
\doi{10.1103/PhysRevLett.96.110404}.


\bibitem{lw06}
M.~Levin, X.-G.~Wen,
{\it Detecting Topological Order in a Ground State Wave Function},
Phys. Rev. Lett. {\bf 96}, 110405 (2006),
\doi{10.1103/PhysRevLett.96.110405}.


\bibitem{Haque07}
M.~Haque, O.~Zozulya, K.~Schoutens,
{\it Entanglement Entropy in Fermionic Laughlin States},
Phys. Rev. Lett. {\bf 98}, 060401 (2007),
\doi{10.1103/PhysRevLett.98.060401}.


\bibitem{li08}
H.~Li, F.D.M.~Haldane,
{\it Entanglement spectrum as a generalization of entanglement entropy: Identification of topological order in non-Abelian fractional quantum Hall effect states},
Phys. Rev. Lett. {\bf 101}, 010504 (2008),
\doi{10.1103/PhysRevLett.101.010504}.


\bibitem{ostlund}
S.~\"Ostlund, S.~Rommer,
{\it Thermodynamic Limit of Density Matrix Renormalization},
Phys. Rev. Lett. {\bf 75}, 3537 (1995),
\doi{10.1103/PhysRevLett.75.3537}.


\bibitem{perez-garcia}
D.~Perez-Garcia, F.~Verstraete, M.M.~Wolf, J.I.~Cirac,
{\it Matrix Product State Representations},
Quantum Inf. Comput. {\bf 7}, 401 (2007),
arXiv:\href{https://arxiv.org/abs/quant-ph/0608197}{quant-ph/0608197}.


\bibitem{dubail12}
J.~Dubail, N.~Read, E.H.~Rezayi,
{\it Edge state inner products and real-space entanglement spectrum of trial quantum Hall states},
Phys. Rev. B {\bf 86}, 245310 (2012),
\doi{10.1103/PhysRevB.86.245310}.


\bibitem{zm}
M.P.~Zaletel, R.S.K.~Mong,
{\it Exact matrix product states for quantum Hall wave functions},
Phys. Rev. B {\bf 86}, 245305 (2012),
\doi{10.1103/PhysRevB.86.245305}.


\bibitem{ciracsierra}
J.I.~Cirac, G.~Sierra,
{\it Infinite matrix product states, conformal field theory, and the Haldane-Shastry model},
Phys. Rev. B {\bf 81}, 104431 (2010),
\doi{10.1103/PhysRevB.81.104431}.


\bibitem{white}
S.R.~White,
{\it Density matrix formulation for quantum renormalization groups},
Phys. Rev. Lett. {\bf 69}, 2863 (1992),
\doi{10.1103/PhysRevLett.69.2863}.


\bibitem{schollwoeck}
U. Schollw\"ock,
{\it The density-matrix renormalization group},
Rev. Mod. Phys. {\bf 77}, 259 (2005),
\doi{10.1103/RevModPhys.77.259}.


\bibitem{vedral}
L.~Amico, R.~Fazio, A.~Osterloh, V.~Vedral,
{\it Entanglement in many-body systems},
Rev. Mod. Phys. {\bf 80}, 517 (2008),
\doi{10.1103/RevModPhys.80.517}.


\bibitem{orus}
R.~Orus,
{\it A Practical Introduction to Tensor Networks: Matrix Product States and Projected Entangled Pair States},
Ann. Phys. {\bf 349}, 117 (2014),
\doi{10.1016/j.aop.2014.06.013}.


\bibitem{estienne-short}
B.~Estienne, Z.~Papi\'c, N.~Regnault, B.A.~Bernevig,
{\it Matrix product states for trial quantum Hall states},
Phys. Rev. B {\bf 87}, 161112(R) (2013).
\doi{10.1103/PhysRevB.87.161112}.


\bibitem{estienne-long}
B.~Estienne, N.~Regnault, B.A.~Bernevig,
{\it Fractional Quantum Hall Matrix Product States For Interacting Conformal Field Theories},
arXiv:\href{https://arxiv.org/abs/1311.2936}{1311.2936} (unpublished).


\bibitem{zmp13}
M.P.~Zaletel, R.S.K.~Mong, F.~Pollmann,
{\it Topological Characterization of Fractional Quantum Hall Ground States from Microscopic Hamiltonians},
Phys. Rev. Lett. {\bf 110}, 236801 (2013),
\doi{10.1103/PhysRevLett.110.236801}.


\bibitem{zpm15}
M.P.~Zaletel, Z.~Papi\'c, R.S.K.~Mong,
{\it Competing Abelian and non-Abelian topological orders in $\nu=1/3+1/3$ quantum Hall bilayers},
Phys. Rev. B {\bf 91}, 205139 (2015),
\doi{10.1103/PhysRevB.91.205139}.


\bibitem{mzpp17}
R.S.K.~Mong, M.P.~Zaletel, F.~Pollmann, Z.~Papi\'c,
{\it Fibonacci anyons and charge density order in the 12/5 and 13/5 quantum Hall plateaus},
Phys. Rev. B {\bf 95}, 115136 (2017),
\doi{10.1103/PhysRevB,95.115136}.


\bibitem{nielsen}
A.E.B.~Nielsen, I.~Glasser, I.D.~Rodriguez,
{\it Symmetry between quasielectrons and quasiholes for fractional quantum Hall models defined on lattices},
arXiv:\href{https://arxiv.org/abs/1609.02389}{1609.02389} (unpublished).


\bibitem{hhrv}
T.H.~Hansson, M.~Hermanns, N.~Regnault, S.~Viefers,
{\it Quantum Hall quasielectrons - Abelian and non-Abelian},
Phys. Rev. Lett. {\bf 102}, 166805 (2009),
\doi{10.1103/PhysRevLett.102.166805}.


\bibitem{hhv}
T.H.~Hansson, M.~Hermanns, S.~Viefers,
{\it Quantum Hall quasielectron operators in conformal field theory},
Phys. Rev. B {\bf 80}, 165330 (2009),
\doi{10.1103/PhysRevB.80.165330}.


\bibitem{halperin84}
B.I.~Halperin,
{\it Statistics of quasiparticles and the hierarchy of fractional quantized Hall states},
Phys. Rev. Lett. {\bf 52}, 1583 (1984),
\doi{10.1103/PhysRevLett.52.1583}.


\bibitem{hansson07a}
T.H.~Hansson, C.-C.~Chang, J.K.~Jain, S.~Viefers, 
{\it Conformal Field Theory of Composite Fermions},
Phys. Rev. Lett. {\bf 98}, 076801 (2007),
\doi{10.1103/PhysRevLett.98.076801}.


\bibitem{hansson07b}
T.H.~Hansson, C.-C.~Chang, J.K.~Jain, S.~Viefers,
{\it Composite-fermion wave functions as correlators in conformal field theory}, 
Phys. Rev. B {\bf 76}, 075347 (2007),
\doi{10.1103/PhysRevB.76.075347}.


\bibitem{kjonsbergL1999}
H.~Kj\o{}nsberg, J.M.~Leinaas,
{\it Charge and statistics of quantum Hall quasi-particles -- a numerical study of mean values and fluctuations},
Nucl. Phys. B, {\bf 559}, 705 (1999),
\doi{10.1016/S0550-3213(99)00353-3}.


\bibitem{bergholtz08}
E.J.~Bergholtz, A.~Karlhede,
{\it Quantum Hall system in Tao-Thouless limit},
Phys. Rev. B {\bf 77}, 155308 (2008),
\doi{10.1103/PhysRevB.77.155308}.


\bibitem{screeningplasma}
J.M.~Caillol, D.~Levesque, J.J.~Weis, J.P.~Hanson,
{\it A Monte Carlo study of the classical two-dimensional one-component plasma},
J. Stat. Phys. {\bf 28}, 325 (1982),
\doi{10.1007/BF01012609}.


\bibitem{deleeuw82}
S.W.~de Leeuw, J.W.~Perram,
{\it Statistical mechanics of two-dimensional coulomb systems: II. The two-dimensional one-component plasma},
Physica A {\bf 113}, 546 (1982),
\doi{10.1016/0378-4371(82)90156-X}.


\bibitem{halperin83}
B.I.~Halperin,
{\it Theory of the quantized Hall conductance},
Helv. Phys. Acta {\bf 56}, 75 (1983),
\doi{10.5169/seals-115362}.


\bibitem{TaoThouless}
R.~Tao, D.J.~Thouless,
{\it Fractional quantization of Hall conductance},
Phys. Rev. B {\bf 28}, 1142(R) (1983),
\doi{10.1103/PhysRevB.28.1142}.


\bibitem{Jeon2010thermodynamic}
G.S.~Jeon, J.K.~Jain,
{\it Thermodynamic behavior of braiding statistics for certain fractional quantum Hall quasiparticles},
Phys. Rev. B {\bf 81}, 035319 (2010),
\doi{10.1103/PhysRevB.81.035319}.


\bibitem{svh11a}
J.~Suorsa, S.~Viefers, T.H.~Hansson,
{\it Quasihole condensates in quantum Hall liquids},
Phys. Rev. B  {\bf 83}, 235130 (2011),
\doi{10.1103/PhysRevB.83.235130}.


\bibitem{svh11b}
J.~Suorsa, S.~Viefers, T.H.~Hansson,
{\it A general approach to quantum Hall hierarchies},
New Journal of Physics {\bf 13}, 075006 (2011),
\doi{10.1088/1367-2630/13/7/075006}.


\bibitem{girvin}
S.M.~Girvin, T.~Jach,
{\it Formalism for the quantum Hall effect: Hilbert space of analytic functions},
Phys. Rev. B {\bf 29}, 5617 (1984),
\doi{10.1103/PhysRevB.29.5617}.


\bibitem{fremling}
M.~Fremling, J.~Fulsebakke, N.~Moran, J.K.~Slingerland,
{\it Energy projection and modified Laughlin states},
Phys. Rev. B {\bf 93}, 235149 (2016),
\doi{10.1103/PhysRevB.93.235149}.


\bibitem{bpz84}
A.A.~Belavin, A.M.~Polyakov, A.B.~Zamolodchikov,
{\it Infinite conformal symmetry in two-dimensional quantum field theory},
Nucl. Phys. B {\bf 241}, 333 (1984),
\doi{10.1016/0550-3213(84)90052-X}.


\bibitem{book:fms99}
P.~Di~Francesco, P.~Mathieu, D.~S\'en\'echal,
{\it Conformal field theory},
Springer, New York (1999),
\doi{10.1007/978-1-4612-2256-9}.


\bibitem{kvorning13}
T.~Kvorning,
{\it Quantum Hall hierarchy in a spherical geometry},
Phys. Rev. B {\bf 87}, 195131 (2013),
\doi{10.1103/PhysRevB.87.195131}.


\end{thebibliography}
\end{document}